\documentclass[12pt]{article}
\usepackage{amsmath,amssymb}
\usepackage{graphicx,psfrag,epsf, color}
\usepackage{enumerate}
\usepackage{natbib}
\usepackage{url} 

\newcommand{\blind}{1}

\newcommand{\pr}{\mathbb{P}} 
\newcommand{\In}{\mathbb{I}} 

\DeclareMathOperator*{\argmin}{arg\,min}

\newcommand{\D}{\mathbf{D}}

\newcommand{\bH}{\mathbf{H}}

\newcommand{\T}{\mathbf{T}}
\newcommand{\U}{\mathbf{U}}
\newcommand{\V}{\mathbf{V}}

\newcommand{\0}{\mathbf{0}}

\newcommand{\bu}{\mathbf{u}}
\newcommand{\bv}{\mathbf{v}}

\newcommand{\blambda}{\boldsymbol{\lambda}}

\newcommand{\bTheta}{\boldsymbol{\Theta}}
\newcommand{\bOmega}{\boldsymbol{\Omega}}

\newcommand{\be}{\begin{equation}}
\newcommand{\ee}{\end{equation}}
\newcommand{\bes}{\begin{equation*}}
\newcommand{\ees}{\end{equation*}}
\newcommand{\bi}{\begin{itemize}}
\newcommand{\ei}{\end{itemize}}
\newcommand{\bea}{\begin{eqnarray}}
\newcommand{\eea}{\end{eqnarray}}

\begin{document}

\def\spacingset#1{\renewcommand{\baselinestretch}%
{#1}\small\normalsize} \spacingset{1}

\pagenumbering{roman}

\if1\blind
{
  \title{\bf To Wait or Not to Wait:\\ {Two-way Functional Hazards Model for Understanding Waiting in Call Centers}}
  \author{Gen Li\thanks{Corresponding author: gl2521@cumc.columbia.edu}\hspace{.2cm}\\
    Department of Biostatistics, Columbia University\\
    and \\
    Jianhua Z.\ Huang \\
    Department of Statistics, Texas A$\&$M University\\
    and \\
    Haipeng Shen\\
    Faculty of Business and Economics, University of Hong Kong}
  \maketitle
  \newpage
} \fi

\if0\blind
{
  \bigskip
  \bigskip
  \bigskip
  \begin{center}
    {\LARGE\bf To Wait or Not to Wait:\\ \vskip.05in {Two-way Functional Hazards Model for Understanding Waiting in Call Centers}}
\end{center}
  \medskip
} \fi

\bigskip
\begin{abstract}
Telephone call centers offer a convenient communication channel between businesses and their customers.  Efficient management of call centers needs accurate modeling of customer waiting behavior, 
which contains important information about customer patience (how long a customer is willing to wait) and service quality (how long a customer needs to wait to get served).
Hazard functions offer dynamic characterization of customer waiting behavior, and provide critical inputs for agent scheduling.
Motivated by this application, we develop a two-way functional hazards (tF-Hazards) model to study customer waiting behavior as a function of two timescales, waiting duration and the time of day that a customer calls in.
The model stems from a two-way piecewise constant hazard function, and imposes low-rank structure and  smoothness on the hazard rates to enhance interpretability.
We exploit an alternating direction method of multipliers (ADMM) algorithm to optimize a penalized likelihood function of the model.
We carefully analyze the data from a US bank call center, and provide informative insights about customer patience and service quality patterns along waiting time and across different times of a day. The findings provide primitive inputs for call center agent staffing and scheduling, as well as for call center practitioners to understand the effect of system protocols on customer waiting behavior.
\end{abstract}

\noindent%
\vfill

\newpage
\spacingset{1.45} 

\pagenumbering{arabic}
\setcounter{page}{1}
\section{Introduction}

Call centers have become increasingly important in today's business world.
They serve as a major channel connecting companies and their customers. Global Industry Analysts projects the global call centers market to reach \$337.8 billion by 2018.
The need for efficient call center operation has stimulated a large amount of research in stochastic modeling as surveyed by~\cite{gans2003telephone} and~\cite{aksin2007modern}, and more recently in statistical modeling using call-by-call data~\citep{brown2005statistical, shen2008forecasting, shen2008interday, gans2010service, mandelbaum2013data, ibrahim2013forecasting}.

The call center service process consists of three fundamental components: the arrival process of customer calls to the center, the service process (i.e., conversation durations between customers and agents), and customer waiting times. Efficient operation management of call centers aims at achieving optimal balance between customer demand and agent capacity (i.e., staffing cost), subject to certain service level constraints on waiting time or abandonment. One of the main issues of interests is optimal agent staffing which is closely related to customer waiting times/patience behavior.

Customer waiting is a system output that reflects the interplay among demand (customer arrivals), capacity (agent staffing), customer patience, and system protocol. When a customer gets connected to a call center, s/he is typically placed in a virtual tele-queue, waiting to be served by the next available agent.
While waiting, s/he may become impatient and abandon the queue, or is patient enough to get served by an agent.
We define the {\em patience time} as the time s/he is willing to wait, and define the {\em offered wait time} as the time the customer needs to wait in order to get served. For each customer, one observes the actual waiting time -- the minimum of the patience time and the offered wait time, one of which is right-censored depending on the outcome of the waiting. If the call is answered, the offered wait time equals the observed waiting time, while the corresponding patience time is longer than the waiting time and unobserved; if the customer abandons, the patience time is observed, while the offered wait time is right-censored.

Customer patience and abandonment is a unique and important aspect of tele-queues that involve human customers. Hazard function of patience time offers a dynamic view of customer waiting behavior, and has been incorporated in many queueing studies.~\cite{mandelbaum2009staffing} studied asymptotically optimal staffing of many-agent queues with customer patience having general hazard function.~\cite{reed2012hazard} used patience hazard function scaling for large queueing systems to obtain accurate approximation for various performance measures. {\cite{liu2011large} exploited time-varying patience-time distributions to study the asymptotics for the many-server fluid queue.}~\cite{atlason2008optimizing} optimized agent staffing via simulation and analytic methods.~\cite{gurvich2010staffing} considered uncertain demand forecasts and used chance-constraint for agent staffing. Furthermore,
~\cite{mandelbaum2000model} considered queues with rational customers, and characterized their patience behaviors via more basic primitives of waiting costs and service benefits;~\cite{zohar2002adaptive} showed that customers can adapt their patience according to their anticipated waiting time;~\cite{aktekin2012bayesian} considered Bayesian inference of queues with abandonment; ~\cite{akcsin2013structural} modeled the abandonment decision of callers as an optimal-stopping problem.


Queueing studies usually assume the hazard function as known. To be practical, one needs to estimate patience hazard as a function of waiting time.~\cite{mandelbaum2013data} pointed out that the amount of empirical work about customer waiting behavior in telephone queues lagged behind that about customer arrivals and their services. In this paper, we aim at narrowing this gap and focus on modeling customer waiting, to better understand their waiting behavior and help with agent staffing.

As far as we know, \cite{brown2005statistical}, \cite{mandelbaum2013data}, {and \cite{aktekin2014bayesian}} are the only empirical papers studying patience patterns in tele-queues. In particular, \cite{aktekin2014bayesian} proposed mixture models and piecewise linear models to estimate customer patience hazard rate.
However, all three studies focused on one-way variation of the customer patience as a function of waiting time.
In practice, customer patience varies across the time of day. To cope with it, one would like the patience hazard function to also depend on the time of day, i.e., it varies in two directions: how long a customer waits and what time s/he starts waiting.

{This paper makes a unique and important contribution by developing a two-way functional hazards model (i.e., {\em tF-Hazards}) to provide an efficient estimator of the {\em two-way patience hazard surface} as inputs to call center staffing models.
The tF-Hazards model stems from a simple yet flexible two-way piecewise constant hazard model, and incorporates low-rank and smooth structure into the discrete hazard matrix indexed by two timescales (in our application, the two timescales are waiting duration and time of day).
The model can be viewed as a generalization of the two-way functional singular value decomposition (tF-SVD) approach for Gaussian data \citep{huang2009analysis}.
Unlike tF-SVD though, our model accommodates censored time-to-event data and estimates the underlying two-way hazard surface.
The model is estimated via a regularized likelihood framework, where the low-rank constraint and the roughness penalty are incorporated simultaneously. 
We exploit an alternating direction method of multipliers (ADMM) algorithm to solve the constrained optimization problem.
We apply the proposed method to the US bank waiting times previously analyzed by \cite{mandelbaum2013data}, and offer novel insights into the customer waiting behavior across time of day, reflecting dynamics of customer patience.
We also investigate the offered wait patterns, which provides interesting findings reflecting the dynamic effect of system protocol and congestion.
Both the patience time and the offered wait time are  important components of {\em impatience} modeling from the operational viewpoint~\citep{brown2005statistical}.
}

Although primarily motivated by the application of efficient call center operations management, the proposed two-way functional hazards model is generally applicable to a range of scientific studies.
For example, it can be easily adapted to survival analysis in epidemiological research involving dual timescales \citep{efron2002two}.

{The concept of two-way hazards model was first introduced by \cite{cox1972regression} and \cite{farewell1979note}. However, the development of appropriate methods lagged behind due to the common ignorance of calendar time in survival analysis.
Only a couple of two-way hazards models have been proposed in the literature ever since.
\cite{efron2002two} developed a parametric two-way hazards model to study the survival of heart transplant recipients with respect to lifetime and calendar time, simultaneously.
Later on, \cite{iacobelli2013multiple} extended Efron's work to a multi-state model to study chronic myeloid leukemia.
However, both methods require the specification of parametric forms for the underlying hazard functions.
Without any prior knowledge, the parametric requirement may be too rigid for practical applications.
\cite{kauermann2006additive} developed a nonparametric additive two-way hazards model to study unemployment in Germany.
The model assumes the hazard function has an additive structure on the log scale, which is a rank-one special case of our proposed model  as we shall show in Section~\ref{subsec:model}.
\cite{kauermann2006additive} proposed to fit the additive model via a penalized spline approach.
However, the large number of spline basis functions imposes a tremendous computational burden to the fitting procedure.
We remark that the two-way hazards model also loosely connects to the nonparametric proportional hazards models \citep[cf.][]{fan1997local,chen2010global,chen2007local}, if one treats the second timescale as a covariate. However, most existing nonparametric hazards models exclusively focus on the estimation of the covariate pattern via a partial likelihood approach, and ignore the estimation of the baseline hazard function. This is not appropriate for the call center application where estimation of the underlying hazard function/surface is of primary interest.}

The rest of the paper is organized as follows.
Section \ref{sec:method} starts with the piecewise constant hazard model and then introduces the proposed tF-Hazards model.
Section \ref{sec:alg} elaborates on the model fitting algorithm.
Section \ref{sec:real} introduces the call center data, and illustrates the analyses of customer patience and offered wait respectively, highlighting additional insights from our two-way model and comparison with other approaches.
Discussions and future research directions are provided in Section \ref{sec:dis}.
Adaptive tuning parameter selection is in the appendix.
The Matlab code/manual and the call center data are available upon request.

\section{Model}\label{sec:method}
In this section, we first introduce piecewise constant hazard functions in Section \ref{subsec:piece},
and then describe the proposed tF-Hazards model in Section \ref{subsec:model}.
\textcolor{black}{Due to the space limit, the simulation study is contained in the supplementary material.}

\subsection{Piecewise Constant Hazard Function}\label{subsec:piece}
In survival analysis, a hazard function presents the instantaneous failure rate at time $t$ given the event happens no earlier than $t$.
Mathematically, the hazard function $h(t)$ is defined as
\bes
h(t)\triangleq\lim_{\Delta t\rightarrow 0} {\mathbb{P}(t< T\leq t+\Delta t\ | T> t) \over \Delta t},
\ees
where $T$ is a random variable of the event time.
{\cite{palm1957research} first introduced the concept of hazard rates to call center customer impatience modeling. \cite{brown2005statistical} and \cite{aktekin2014bayesian} further explored the idea to model abandonment in call center operations.}

Piecewise constant hazard functions have been widely used in survival analysis for its simplicity and flexibility \citep[see][for example]{ibrahim2005bayesian,miller2011survival}.
In a call center waiting time study, a piecewise constant hazards model is especially desirable due to the nature of the data: waiting times are usually recorded at certain resolution, e.g., 1 second in our application. 
It is thus natural to partition the waiting time axis into 1-second intervals and assume a constant hazard rate in each interval.

Consider the following piecewise constant hazard function $h(t)$ of time $t\in(0,\tau]$ with $p$ discrete values:
\bes
h(t)=h_k,\qquad \tau_{k-1}<t\leq \tau_k;\, k=1,\cdots,p;
\ees
where $0=\tau_0<\tau_1<\cdots<\tau_p=\tau$ specify $p$ time intervals in $(0,\tau]$.
Suppose that we have $n$ independent observations from the model. For the $i$th observation $(i=1,\cdots,n)$, let $t_i$ ($t_i\leq\tau$) denote the event time or the censoring time and $d_i$ denote the event indicator (i.e., when $d_i=1$, an event is observed; when $d_i=0$, the observation is right censored).

{To derive the likelihood function of the observed event times given the event indicators,
we first transform the data pair $(t_i,d_i)$ by breaking each of them down into $p$ segments.}
Specifically, let
\be\label{transf}
\begin{split}
t_{ik}=\left\{
\begin{aligned}
&0,\quad\quad\quad\quad\,\, t_i\leq\tau_{k-1}\\
&t_i-\tau_{k-1},\quad \tau_{k-1}< t_i\leq\tau_k\\
&\tau_{k}-\tau_{k-1},\quad t_i>\tau_k
\end{aligned}\right. \mbox{ and }\
d_{ik}=d_i\In(\tau_{k-1}< t_i \leq\tau_k);\ k=1,\cdots,p;
\end{split}
\ee
where $\In(\cdot)$ is the indicator function.
The vectors $(t_{i1},\cdots,t_{ip})^T$ and $(d_{i1},\cdots,d_{ip})^T$ are the decompositions of $t_i$ and $d_i$ into the $p$ time intervals.
In particular, $\sum_{k=1}^p t_{ik}=t_i$ and $\sum_{k=1}^p d_{ik}=d_i$.
The likelihood function can be expressed in terms of the new data as
\bes
\mathcal{L}(h_1,\cdots,h_p)=\prod_{i=1}^n\prod_{k=1}^p {h_k}^{d_{ik}}\exp(-h_k t_{ik}),
\ees
and the log likelihood function is
$\log\mathcal{L}(h_1,\cdots,h_p)=\sum_{i=1}^n\sum_{k=1}^p\{ d_{ik}\log h_k -t_{ik}h_k\}.$
The maximum likelihood estimate (MLE) of the hazard rates are
\bes
\widehat{h}_k={\sum_{i=1}^n d_{ik}\over \sum_{i=1}^n t_{ik}}; \quad k=1,\cdots,p.
\ees

{{\noindent\bf Two-way piecewise constant hazard function}
In the context of the call center example, we introduce one random variable $A$ as the arrival time of a call. It takes continuous clock time values.
A two-way hazard function $h(a,t)$ is defined as the instantaneous event rate at time $t$ conditional on the arrival time being $a$.
Namely, we have
\bes
h(a,t)\triangleq\lim_{\Delta t\rightarrow 0} {\pr(t<T\leq t+\Delta t\ |\ A=a,T> t) \over \Delta t}.
\ees
Piecewise constant hazard functions can be extended to the two-way hazards model \citep{keiding1990statistical,berzuini1994bayesian}.}
The hazard rates are assumed to be constant in a neighborhood of the arrival time and the waiting time.
Formally, we break the time of day into $m$ intervals, with knots denoted as $0=\alpha_0<\alpha_1<\cdots<\alpha_m=\alpha$, and the waiting time into $p$ intervals, with the knots denoted as $0=\tau_0<\tau_1<\cdots<\tau_p=\tau$ as in the one-way situation, and assume the hazard rates to be constant in each block.
Namely,
\bes
h(a,t)=h_{jk},\quad \alpha_{j-1}<a\leq\alpha_j,\tau_{k-1}<t\leq\tau_{k},
\ees
for $j=1,\cdots,m$ and $k=1,\cdots,p$.
We typically align the hazard rates in an $m\times p$ matrix $\bH$, where the $(j,k)$th entry is $h_{jk}$.

{Given $n$ observations, each being a triplet $(a_i,t_i,d_i)$ where $a_i$ is the arrival time, $t_i$ is the waiting time, and $d_i$ is the censoring indicator, we exploit a (conditional) maximum likelihood approach to estimate the piecewise constant hazard rates.} Similar to the one-way case, we first transform the data. In particular, we convert $t_i$ and $d_i$ into two $m\times p$ matrices $\T_i$ and $\D_i$, where all entries are zero other than the $j$th rows such that $\alpha_{j-1}<a_i\leq\alpha_{j}$.
Entries of the $j$th rows of $\T_i$ and $\D_i$ are defined in the same way as in \eqref{transf}, i.e., decomposing $t_i$ and $d_i$ into the $p$ waiting time intervals respectively.
We denote the $(j,k)$th entries of $\T_i$ and $\D_i$ as $t_{ijk}$ and $d_{ijk}$, respectively.
{By assuming that the waiting times are conditionally independent given the arrival times and censoring indicators, we derive the conditional log likelihood function of the observed waiting times as}
\be\label{loglik}
\log\mathcal{L}(\bH)=\sum_{j=1}^m\sum_{k=1}^p \left[\left(\sum_{i=1}^n d_{ijk} \right)\log h_{jk}-\left(\sum_{i=1}^n t_{ijk}\right)h_{jk}\right].
\ee
Hence, the MLE for each entry of the hazard-rate matrix $\bH$ has an explicit expression as
\bes
\widehat{h}_{jk}={\sum_{i=1}^n d_{ijk}\over \sum_{i=1}^n t_{ijk}};\quad j=1,\cdots,m;\quad k=1,\cdots,p.
\ees


{We remark that it is generally reasonable to assume the waiting times are conditionally independent given the arrival times or intervals. Thus the total conditional log likelihood function \eqref{loglik} is just the summation of the individual conditional log likelihood for each observation. 
Otherwise, \eqref{loglik} can be viewed as a marginal composite likelihood \citep{varin2011overview}, the maximization of which still provides good estimates and permits useful inference of the hazard rates.
The development of other estimation methods under dependence is a future research direction.}

\subsection{Two-Way Functional Hazards Model}\label{subsec:model}
The piecewise constant two-way hazards model is easy to comprehend and estimate, but it has several disadvantages.
One obvious drawback is that the estimated hazard function can be very rough.
The hazard rate in each block is estimated individually without borrowing information from adjacent blocks.
Thus even neighboring blocks may have  distinct hazard rate estimates, making the results less interpretable.
Another issue is that the MLE is sensitive to the fineness of the grids, especially along the waiting time direction.
If the grids are dense, many blocks may not have any observed events, resulting in an estimated hazard-rate matrix with many zeros;
if the grids are coarse, most blocks would have multiple events, leading to a matrix estimate with few zero entries.
Namely, the estimated hazard function has large variability.
In addition, a less obvious issue is the probable occurrence of missing values.
If there is a lack of observations with long duration at a time of day (e.g., if $\sum_{i=1}^n t_{ijk}=0$ for arrival interval $j$ and waiting time interval $k$), the MLE will be undefined for that particular block.
It leads to a hazard-rate matrix estimated with missing values.
If we want to apply an arbitrary smoother (e.g., tF-SVD) to the estimated hazard rate matrix, we have to either impute the missing values or shorten the duration range -- either way is restrictive and ad hoc.
To address all the issues, below we introduce the tF-Hazards model.

The tF-Hazards model assumes that the two-way hazard function $h(a,t)$ consists of $r$ components (or layers), each being a product of two univariate functions.
Namely,
\be\label{model_c}
h(a,t)=u_1(a)v_1(t)+\cdots+u_r(a)v_r(t),
\ee
where $u_1(a),\cdots,u_r(a)$ and $v_1(t),\cdots,v_r(t)$ are smooth univariate functions of the arrival time $0<a\leq\alpha$ and the waiting time $0<t\leq\tau$, respectively.
There is no sign restriction on the functions, other than $h(a,t)\geq0$ for any $(a,t)$ in the domain.
The multiplicative structure in each layer nicely separates the time-of-day effect from the duration effect on hazard rates.
The sum of the $r$ components makes the model flexible enough to capture complex patterns of a hazard surface in a simple way.

{We remark that when $r=1$, the tF-Hazards model degenerates to a multiplicative model in the following form
\bes
h(a,t)=u(a)v(t).
\ees
Since $h(a,t)\geq 0$, without loss of generality, we have $u(a)\geq 0$ and $v(t)\geq 0$.
In particular, if we treat $a$ as a covariate and $v(t)$ as a baseline hazard function, it corresponds to a proportional hazards model.
Furthermore, if we assume the hazard rates are strictly positive and take the logarithmic transformation on both sides of the formula, we exactly obtain the nonparametric additive two-way hazards model in \cite{kauermann2006additive}.
Clearly, their model is a special case of our tF-Hazards model.
}

Following the notations in the piecewise constant two-way hazards model, the discrete version of Model \eqref{model_c} can be expressed as
\be\label{model_d}
\bH=\bu_1\bv_1^T+\cdots+\bu_r\bv_r^T=\U\V^T
\ee
where $\bH$, an $m\times p$ matrix, is the discrete sampling of the hazard function at the $(\alpha_1,\cdots,\alpha_m)\times(\tau_1,\cdots,\tau_p)$ grid; the columns of $\U=(\bu_1,\cdots,\bu_r)$ and the columns of $\V=(\bv_1,\cdots,\bv_r)$ are discrete samplings of the smooth univariate functions respectively.
In particular, Model \eqref{model_d} can be viewed as a rank-$r$ decomposition of the discrete hazard-rate matrix $\bH$.

For identifiability, we require that both $\U$ and $\V$ have orthogonal columns when $r$ is greater than 1, and each column of $\U$ has mean 1.
The orthogonality condition guarantees that the patterns captured by different components are unrelated.
The unit-mean condition enhances the interpretability of each layer: $\bv_i$ is the average hazard function over different arrival times, and each entry of $\bu_i$ represents a proportion of the average hazard function.
Similar to the continuous version of the model, the columns of $\U$ capture the time-of-day patterns and the columns of $\V$ capture the waiting time patterns of a hazard function.
As discussed in Section \ref{subsec:piece}, in the call center waiting time study, it is natural and desirable to use the discretized version of the model.

To obtain an estimate of the hazard function from Model \eqref{model_d}, we solve the following constrained optimization problem
\be\label{penlik}
\begin{split}
\min_\bH&\quad -\log\mathcal{L}(\bH)+{\rho\over2}\mathcal{P}(\U,\V) \\
&\mbox{s.t. } \quad \bH=\U\V^T, \, \bH\geq 0,
\end{split}
\ee
where $\log\mathcal{L}(\bH)$ is the log likelihood of the discrete hazard matrix, as in \eqref{loglik}, $\mathcal{P}(\U,\V)$ is a penalty term to be definded in \eqref{pen} that ensures the smoothness of the columns of $\U$ and $\V$ (and therefore the smoothness of the hazard matrix $\bH$), $\rho$ is a tuning parameter that balances the goodness of fit term (the negative log-likelihood) and the penalty term, and $\bH\geq 0$ is the validity constraint that all entries of $\bH$ must be nonnegative.

{
The roughness of a length-$p$ vector $\bv$ can be characterized by a quadratic form $\bv^T\bOmega\bv$, where $\bOmega$ is some fixed positive semi-definite matrix.
For example, under some special setting, $\bv^T\bOmega\bv$ is a summation of the squared differences of the adjacent entries, i.e., $\sum_{i=2}^{p-1}(2v_i-v_{i-1}-v_{i+1})^2$.}
While there are many options of $\bOmega$, here we adopt the  setting from \cite{green1994nonparametric}.
For each rank-one layer (i.e., $\bu_i\bv_i^T$), we consider the following two-way roughness penalty proposed by \cite{huang2009analysis}:
\bes
\begin{split}
\mathcal{P}_{\blambda_i}(\bu_i,\bv_i)=&\lambda_{\bu_i}\bu_i^T\bOmega_{\bu}\bu_i\cdot\|\bv_i\|_F^2+
\lambda_{\bv_i}\bv_i^T\bOmega_{\bv}\bv_i\cdot\|\bu_i\|_F^2\\
&+\lambda_{\bu_i}\bu_i^T\bOmega_{\bu}\bu_i\cdot\lambda_{\bv_i}\bv_i^T\bOmega_{\bv}\bv_i,\quad i=1,\cdots,r,
\end{split}
\ees
where $\bOmega_{\bu}$ and $\bOmega_{\bv}$ are two fixed positive semi-definite matrices, $\blambda_i=(\lambda_{\bu_i},\lambda_{\bv_i})$ is a pair of smoothing parameters for the $i$th layer, and $\|\cdot\|_F$ is the Frobenius norm. {The penalty function has several desirable properties such as scale invariance and equivariance~\citep{huang2009analysis}.}
We then set the penalty term in \eqref{penlik} as
\be\label{pen}
\mathcal{P}(\U,\V)=\sum_{i=1}^r \mathcal{P}_{\blambda_i}(\bu_i,\bv_i).
\ee

There are $2r+1$ penalty parameters in the optimization problem \eqref{penlik}.
In particular, $\lambda_{\bu_i}$ and $\lambda_{\bv_i}$ control the roughness penalties imposed on $\bu_i$ and $\bv_i$; $\rho$ balances the log likelihood and the overall penalty term.
We remark that the tuning parameter $\rho$ is not redundant, and we introduce it based on two considerations:
1) equivariance under rescaling of the log likelihood; 2) computational simplicity.
In terms of the first point, suppose we have a replicate of the data at hand (i.e., $-\log\mathcal{L}(\bH)\rightarrow -2\log\mathcal{L}(\bH)$).
If we double the value of $\rho$ and maintain all the other penalty parameters, we can obtain exactly the same solution as before.
However, this cannot be achieved without $\rho$.
In terms of the second point, we shall give more details in the next section.

\section{Algorithm}\label{sec:alg}
We introduce an Alternating Direction Method of Multipliers (ADMM) algorithm to fit the tF-Hazards model via solving the optimization problem \eqref{penlik}. {For now, we assume all the tuning parameters (i.e., the $2r$ smoothing parameters, the balancing parameter $\rho$, and the rank of the hazard matrix $r$) are known.
In practice, one needs to select those parameters in a data-driven fashion.
We remark that the estimation of the smoothing parameters can be embedded in the iterative algorithm; the balancing parameter can be set to a small value initially and updated over iterations to expedite the convergence; the rank is usually set to be a small number for easy interpretation.
More details of the tuning parameter selection is discussed in the appendix.} 

Constrained optimization problems are commonly solved by the augmented Lagrange method.
It converts the original constrained problem to an unconstrained one by adding a Lagrange multiplier and an augmentation term to the optimization objective \citep{nocedal2006conjugate}.
In the context of our problem, solving (\ref{penlik}) is equivalent to minimizing the following augmented Lagrangian function:
\be\label{auglag}
\begin{split}
\mathcal{C}(\bH,\U,\V,\bTheta)\ \triangleq& \ -\log\mathcal{L}(\bH) + {\rho\over2}\mathcal{P}(\U,\V)\\
 &+ <\bTheta,\bH-\U\V^T> + {\rho\over2}\|\bH-\U\V^T\|_F^2,
\end{split}
\ee
with respect to $\U$, $\V$, and $\bH$ under the sign constraint $\bH\geq 0$.
The third term is the Lagrange multiplier with $\bTheta$ being the Lagrange parameter matrix, with $<\cdot,\cdot>$ denoting the matrix inner product induced by the Frobenius norm.
The fourth term is the augmentation term where we particularly set $\rho$ to be the augmentation parameter.

The function (\ref{auglag}) is not convex with respect to all variables and thus simultaneous optimization is formidable.
As an alternative, we propose to use the \textcolor{black}{ADMM} algorithm to solve \eqref{auglag} iteratively.
In particular, for the $i$th iteration, the iterative scheme alternates among the following three steps until convergence:
\begin{eqnarray}
\label{H}
\bH^{(l)}&\triangleq&\argmin_{\bH\geq0}\,\, \mathcal{C}(\bH,\U^{(l-1)},\V^{(l-1)},\bTheta^{(l-1)}),\\
\label{UV}
\left(\U^{(l)},\V^{(l)}\right)&\triangleq&\argmin_{(\U,\V)}\,\,\mathcal{C}(\bH^{(l)},\U,\V,\bTheta^{(l-1)}),\\
\label{Lambda}
\bTheta^{(l)}&\triangleq&\bTheta^{(l-1)}+\rho\left(\bH^{(l)}-\U^{(l)}{\V^{(l)}}^T\right),
\end{eqnarray}
where the first two steps are partial optimizations of $\bH$ given $(\U, \V)$ and of $(\U,\V)$ given $\bH$, respectively, and the third step is a dual update of the Lagrange parameter $\bTheta$. 
The tuning parameter $\rho$ serves as the step size in the dual update.
Below we give more details of each step.
When there is no confusion, we shall drop the superscript indicating iterations.

The optimization in (\ref{H}) is a convex minimization problem.
It can be well separated into $mp$ univariate optimization problems, each corresponding to an entry of $\bH$.
In particular, with respect to the $(j,k)$th entry $h_{jk}$, we have the following optimization problem:
\bes
\begin{split}
\min\limits_{h_{jk}\geq0} \quad&-\sum_{i=1}^nd_{ijk}\log h_{jk}+\sum_{i=1}^nt_{ijk} h_{jk}\\
&+ \theta_{jk}\left(h_{jk}-\bu_{(j)}^T\bv_{(k)}\right) + {\rho\over2} \left(h_{jk}-\bu_{(j)}^T\bv_{(k)}\right)^2,
\end{split}
\ees
where $\theta_{jk}$ is the $jk$th entry of $\bTheta$, and $\bu_{(j)}$ and $\bv_{(k)}$ are $r\times1$ vectors of the $j$th row of $\U$ and the $k$th row of $\V$ respectively.
We denote $t_{\cdot jk}\triangleq\sum_{i=1}^nt_{ijk}$ and $d_{\cdot jk}\triangleq\sum_{i=1}^nd_{ijk}$ for simplicity.
Equating the first order derivative of the objective function to zero, we obtain a unique closed-form solution of the optimization problem as
\be\label{thetasol}
\widetilde{h}_{jk}={-(t_{\cdot jk}-\rho\bu_{(j)}^T{\bv_{(k)}}+\theta_{jk})+\sqrt{\left(t_{\cdot jk}-\rho\bu_{(j)}^T{\bv_{(k)}}+\theta_{jk}\right)^2+4\rho d_{\cdot jk}}   \over 2\rho}.
\ee

We now compare the estimator $\widetilde{h}_{jk}$~\eqref{thetasol} with $\widehat{h}_{jk}=d_{\cdot jk}/t_{\cdot jk}$, i.e. the MLEs under the piecewise constant two-way hazards model of Section~\ref{subsec:piece}. In a nutshell, our estimator $\widetilde{h}_{jk}$ has several desirable properties and address the issues of the MLE $\widehat{h}_{jk}$ (discussed at the beginning of Section \ref{subsec:model}).
First, $\widetilde{h}_{jk}$ exploits information in all the data points through $\bu_{(j)}$, $\bv_{(k)}$ and $\theta_{jk}$, all of which are estimated from a global view.
Therefore, it is less sensitive to different choices of grids.
Secondly, the value of $\widetilde{h}_{jk}$ is similar across adjacent blocks, making it a smooth estimator.
This is because $\widetilde{h}_{jk}$ depends on $\bu_{(j)}$ and $\bv_{(k)}$ which have similar values in neighboring rows of $\U$ and $\V$ respectively.
In particular, the estimator $\widetilde{h}_{jk}$ is dominated by $\bu_{(j)}^T\bv_{(k)}$ when $d_{\cdot jk}$ and $t_{\cdot jk}$ are small (i.e., when we do not have enough data in the $jk$th block).
Finally, the estimator is well defined even when there is no observation in a block.
When $t_{\cdot jk}=d_{\cdot jk}=0$, we have $\widetilde{h}_{jk}=\max(\bu_{(j)}^T{\bv_{(k)}}-\theta_{jk}/\rho\ ,\ 0)$.


Next, to obtain an estimate of $(\U,\V)$ from \eqref{UV}, we rewrite the optimization problem equivalently as
\be\label{UV1}
\min_{(\U,\V)}\quad\|\bH+{\bTheta\over\rho}-\U\V^T\|_{F}^2+\mathcal{P}(\U,\V),
\ee
where the penalty parameter $\rho$ in front of $\mathcal{P}(\U,\V)$ is omitted together with the augmentation parameter.
Without the identifiability conditions, the above optimization problem can be exactly solved by the tF-SVD method \citep{huang2009analysis}.
In particular, tF-SVD sequentially estimates the rank-one layers of $\U\V^T$.
For the $i$th layer $\bu_i\bv_i^T$, the algorithm removes the effect of the preceding layers and alternates between two steps: 1) fix $\bu_i$ and estimate $\bv_i$; 2) fix $\bv_i$ and estimate $\bu_i$.
Each step has an analytical solution.
After convergence, the identifiability conditions on $\U$ and $\V$ can be retrieved through a post-processing step.
More specifically, we apply SVD to the final estimated $\U\V^T$ and adjust the scale of the two components in each layer to satisfy the unit-mean condition.

Once we obtain the estimates of $\bH$, $\U$, and $\V$,  we plug them into \eqref{Lambda} to update the Lagrange parameter $\bTheta$.
The updating formula is derived from the dual feasible condition of the original optimization problem \eqref{penlik} \citep{boyd2011distributed}.

{The stopping rule of the iterative algorithm is based on the primal and dual residuals.
The primal and dual residuals are defined as the Frobenius norms of  $\bH^{(l)}-\U^{(l)}{\V^{(l)}}^T$ and $\bH^{(l)}-\bH^{(l-1)}$, respectively.
At the end of each iteration, we evaluate both residuals.
If the maximum of the two residuals exceeds some preset threshold, the iteration continues; otherwise, the algorithm reaches convergence.
In all of our numerical studies, the algorithm typically converges within tens of iterations.
In particular, with the adaptive selection of the balancing parameter, the convergence is expedited.
}

\section{Call Center Application}\label{sec:real}
In this section, we apply the proposed method to a real application concerning customer waiting behavior at a US Bank call center.
The data are provided to us through the courtesy of the Technion SEELab ---  \url{http://ie.technion.ac.il/Labs/Serveng/}.
{The data have been analyzed by \cite{mandelbaum2013data} and \cite{huang2017refined} among others.}

\subsection{Waiting Time Data}\label{subsec:data}
{We focus on the over one million call records in December, 2002, as studied by \cite{mandelbaum2013data}, to facilitate comparison and reproducibility.
For each call record, we extract information relevant for the waiting component: the start time, end time, outcome of waiting (e.g., being answered, abandoned, transferred or disconnected), and  service type (e.g., retail, loans, business, etc).
Out of the 17 service types in total, each of the 5 major groups (i.e., Retail, Business, Consumer Loans, CCO, Quick\&Reilly) contains more than 5\% of the total calls, respectively.
Since different customer groups may have distinct offered wait and patience patterns, we analyze each customer group separately.
In particular, to be consistent with \cite{mandelbaum2013data}, we focus on the Quick\&Reilly group in the main paper (about 150 thousand calls), and investigate the other 4 major groups in Section D of the supplementary material.
In addition, we only consider calls either being answered or being abandoned, and ignore those being transferred, lost, or undefined.
We aim to investigate the dynamics of customer patience and offered wait along waiting time and across different times of day using the tF-Hazards model.
}

\begin{figure}[h]
\begin{center}
\includegraphics[width=2.4in]{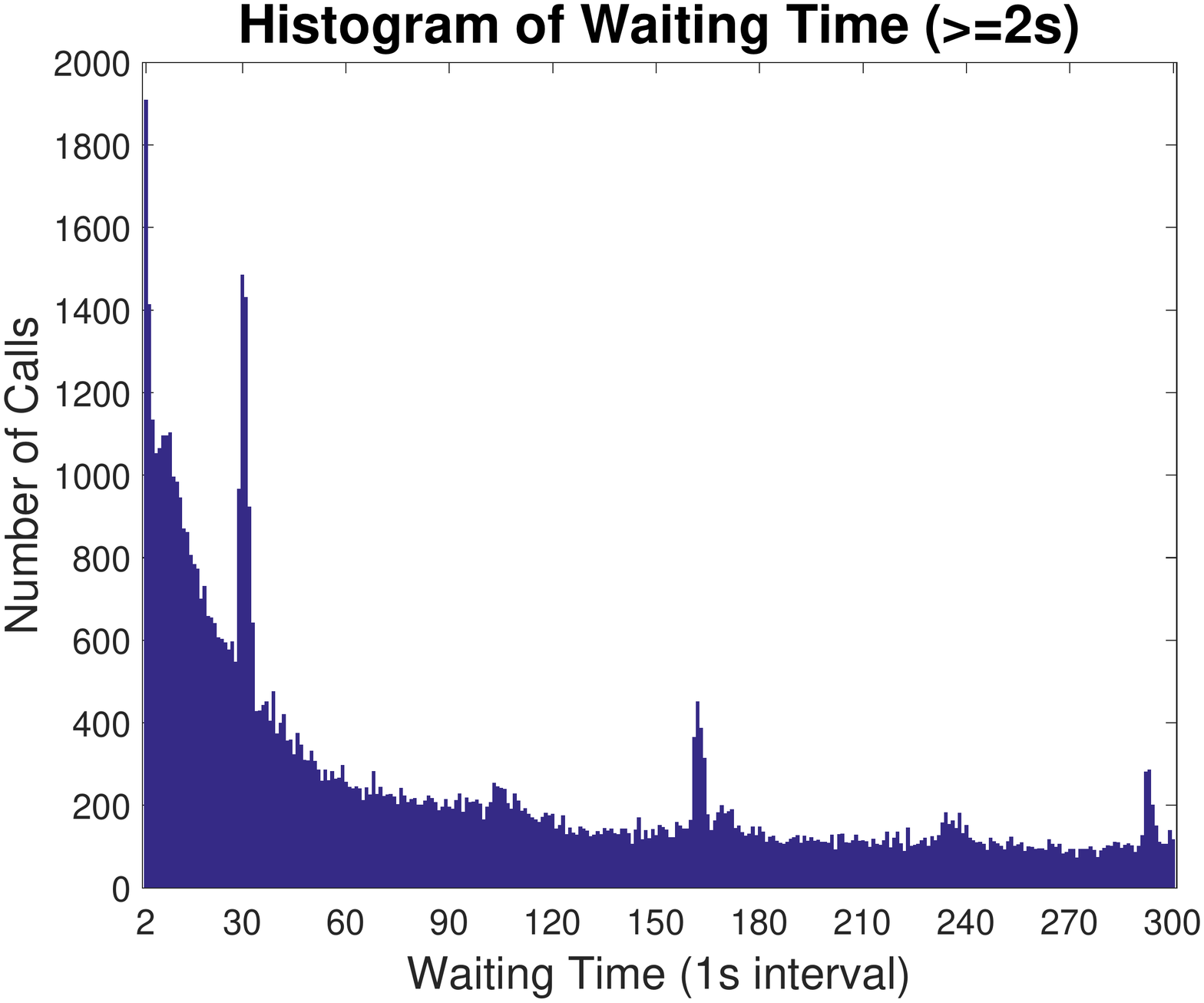}
\includegraphics[width=2.4in]{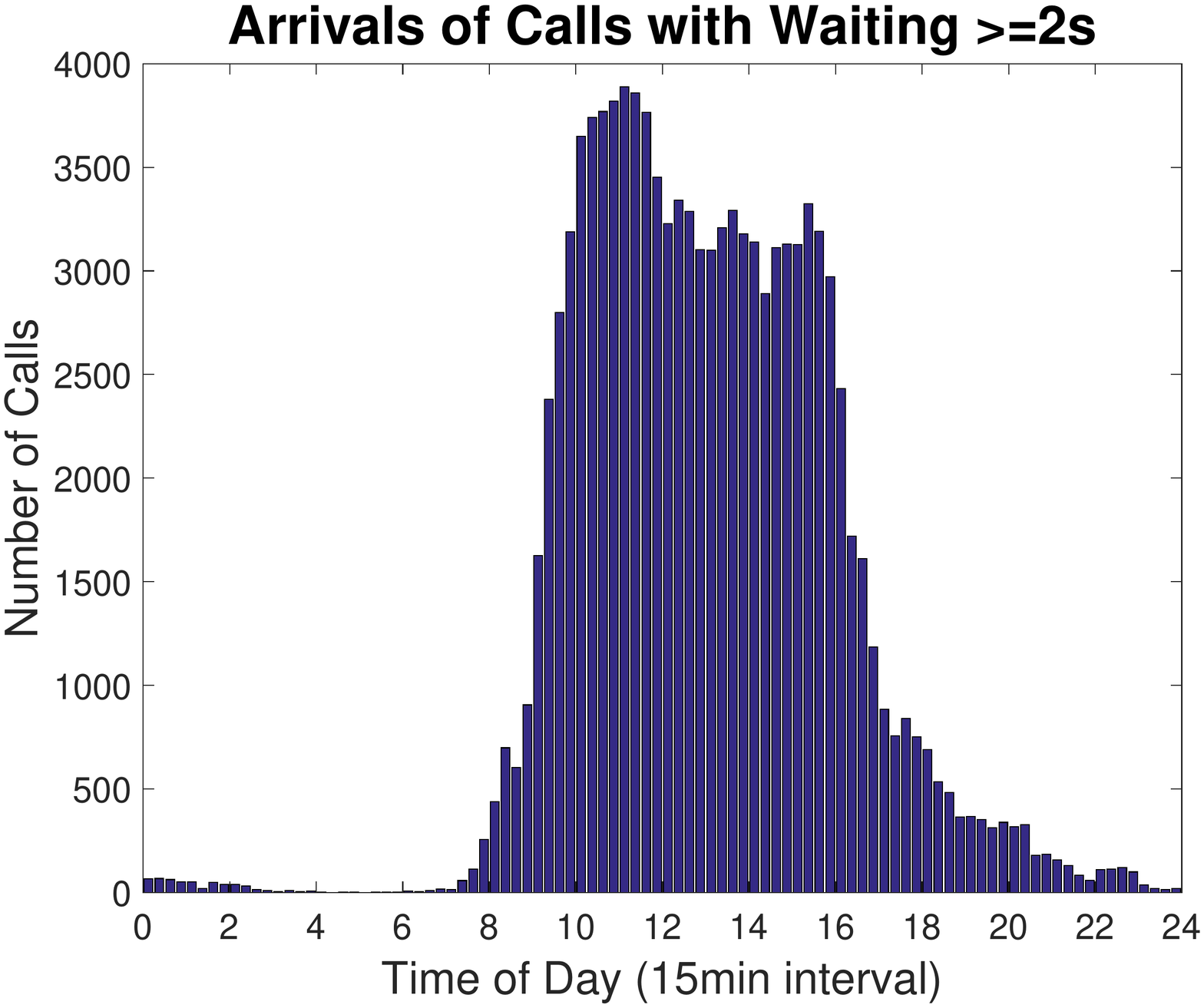}
\end{center}
\vskip-.1in
\caption{US bank call center data for the Quick\&Reilly service type in December, 2002. Left: the histogram of waiting time between 2 seconds and 300 seconds; Right: the histogram of arrival rates for call with waiting time greater or equal to 2 seconds in each 15-minute interval of a day.}
\label{fig:arrival}
\end{figure}

We first need to set domains and intervals for the waiting time and the time of the day.
For the waiting time, the resolution of the records is 1 second.
In other words, each observed waiting time has been rounded to the nearest integer.
Therefore, one second serves as a natural waiting-time interval length. To determine the domain of the waiting time, we tabulate the actual waiting times by the second.
{More than three-quarter calls have waiting time less than 5 minutes. Thus we set the upper bound of the waiting-time domain to be 300 seconds.
It is a bit trickier to determine the lower bound of the domain since the waiting time distribution is highly skewed.
In particular, there is a large number of calls with extremely short waiting times: 0 second (about 13.23\%) and 1 second (about 12.91\%). Among these calls with less than 2 seconds of waiting, over 98\% were answered immediately by an agent. The corresponding offered-wait dynamics and patience behavior are quite different from the remaining calls, and may need to be studied separately \citep{zeltyn2005call}. Waiting times greater or equal to 2 seconds are more common. The histogram is shown in the left panel of Figure \ref{fig:arrival}, for the number of calls with waiting time from 2 seconds to 300 seconds, i.e. over $p=299$ waiting time intervals, aggregated over the month.
We shall focus on the calls with waiting time greater or equal to 2 seconds, and discuss the impact of including those short waiting calls on the hazard estimation in Sections B and C of the supplementary material.}

Regarding the time of the day, call arrivals are typically aggregated into 15-minute or 30-minute intervals in forecasting studies as these are common granularity for work schedules \citep{gans2003telephone}.
Here we use 15 minutes as the time-of-day interval length.
The right panel of Figure \ref{fig:arrival} is the histogram of call arrivals with waiting time no less than 2 seconds in each of the 15-minute interval in a day, aggregated over the month.
{Apparently, most calls arrived between 8:00 am and 8:00 pm.
Therefore, we focus on the 12-hour period, which consists of $m=48$ 15-minute intervals.}

{As a result of the pruning process, we end up with slightly over 100 thousand qualified calls, out of which about $32.23\%$ are abandoned and $67.77\%$ are answered by service agents.}
In the offered wait study, the answered calls are viewed as observed events while the abandoned calls are viewed as right-censored;
in the customer patience study, it is the opposite.
In addition, all calls with waiting time longer than five minutes are truncated and viewed as right-censored.

{We apply the proposed method to the waiting time data, and estimate two-way hazard functions for customer patience and offered wait, respectively.
In particular, in the customer patience study, we assume the patience times are independent and use a full likelihood.
In the offered wait study, we exploit a marginal composite likelihood to accommodate the (potentially dependent) offered wait times.
How to better account for the dependence structure among the observations is a future research direction as discussed in Section \ref{sec:dis}.}

{As a comparison, we also implement the additive model \citep{kauermann2006additive} using the  R package \textsf{TwoWaySurvival} developed by the authors {(which for some reason has been retracted from CRAN and no longer maintained)}. The additive model is fitted via a penalized backfitting procedure, which is extremely computationally expensive.
{It cannot be directly applied to the complete data.
Even if we reduce the number of iterations from the default value 100 to just 10, the additive model fitting procedure is still burdensome.
In particular, fitting the additive model on 5\% of the data with 10 iterations takes about 13 minutes; 10\% of the data takes about 52 minutes; 15\% of the data takes over 2 hours.
As a comparison, fitting the tF-Hazards model on all data with full convergence only takes a few seconds.
In the following customer patience and offered wait studies, we randomly select one tenth of all the observations and only run 10 iterations for the additive model fitting. The results are compared to those from the tF-Hazards on the complete data.}


\subsection{Customer Patience}\label{subsec:p}
We use the call center waiting time data to study the dynamic patterns of customer patience along waiting time and across different times of a day.

{We particularly set $r$, the rank of the underlying hazard matrix in the tF-Hazards model, to be 1 for several reasons.}
First of all, the scree plot of the MLE of the two-way piecewise constant hazard rate matrix suggests the first rank explains the majority of variations in the hazard surface. Secondly, the existing additive model is rank-one; hence the results are directly comparable. Thirdly, the rank-one tF-Hazards model connects nicely to a proportional hazards model so the obtained patterns are highly interpretable. Last but not least, the estimated unit-rank tF-Hazards model can be used as inputs to queueing models~\citep{reed2012hazard, liu2012stabilizing} to have direct impact on time-varying operations.

The hazard surfaces for customer patience estimated from the two methods are shown in Figure \ref{fig:psurf}.
Both methods provide consistent results in terms of the overall patterns and the magnitude of the hazard rates.
The tF-Hazards method presents richer variations than the additive method, possibly due to the fact that the former exploits all available data while the latter only uses a small portion and the algorithm might not have converged.

\begin{figure}[h]
\begin{center}
\includegraphics[width=3in]{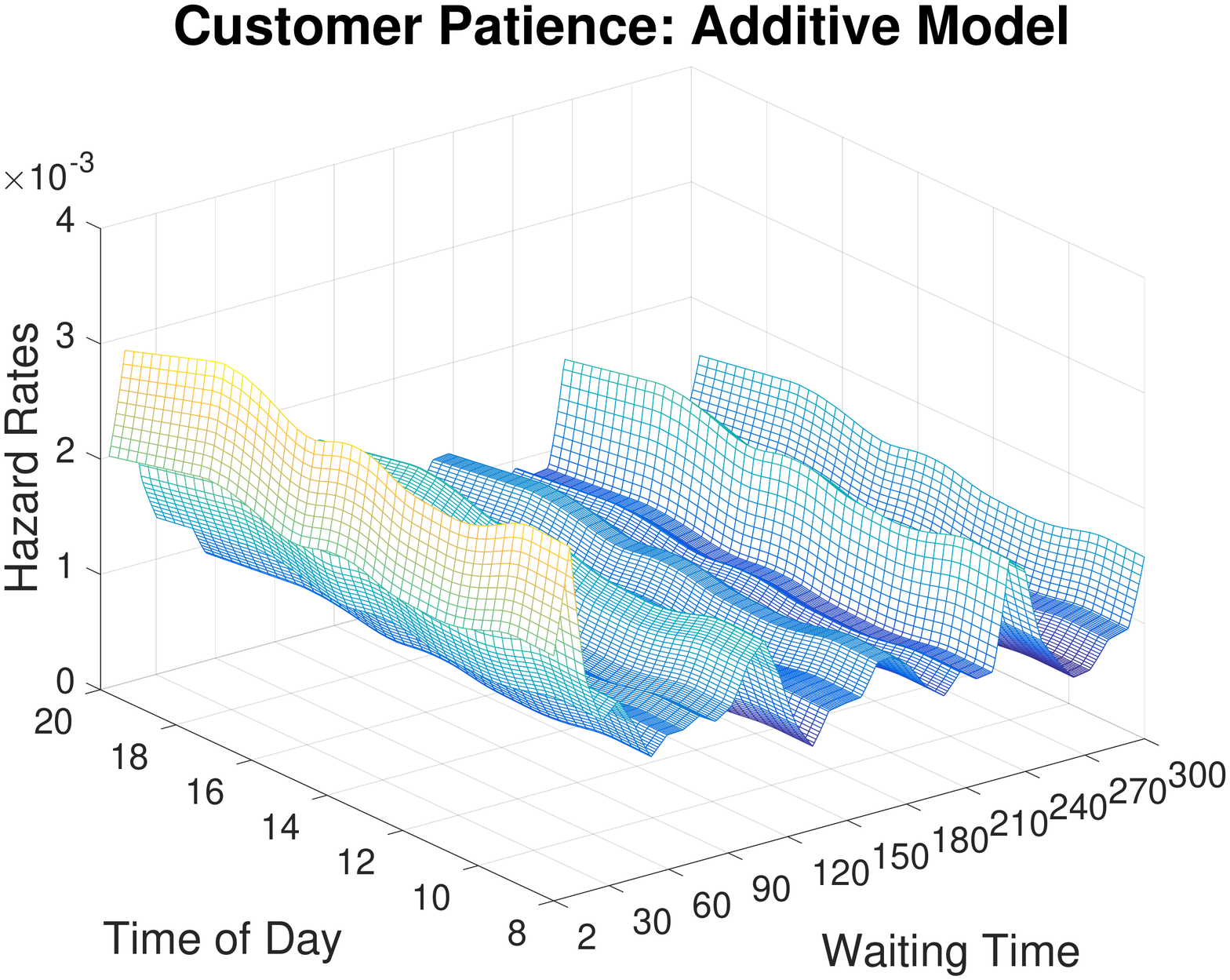}
\includegraphics[width=3in]{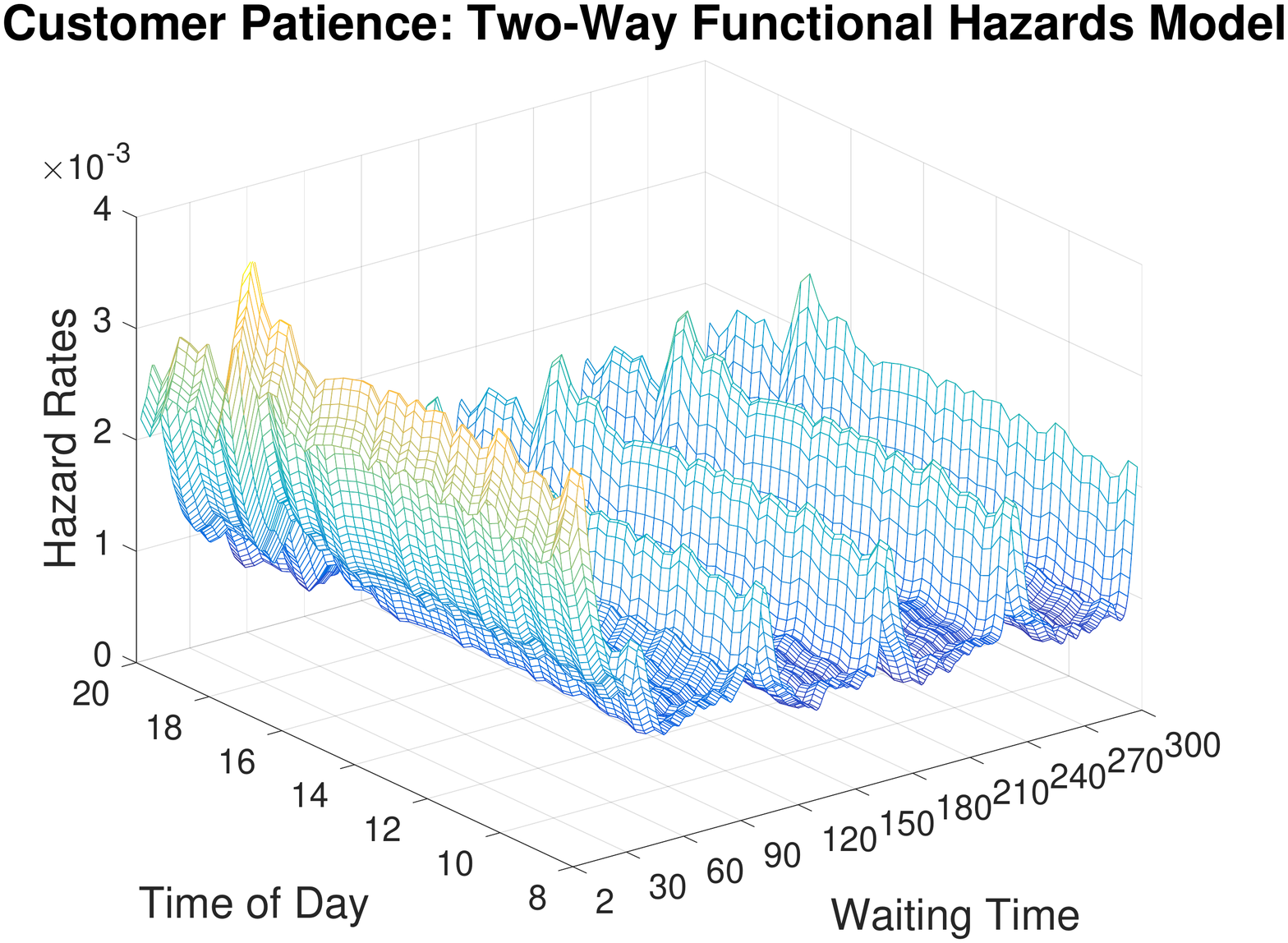}
\end{center}
\caption{Call Center Quick\&Reilly Data: Hazard Surfaces of Customer Patience. Left: the two-way additive model \citep{kauermann2006additive}; Right: the tF-Hazards model.}
\label{fig:psurf}
\end{figure}

To better depict the time-varying patterns, we decompose the hazard surfaces along both timescales.
The results are presented in Figure \ref{fig:puv}. For model identification and simpler interpretation, we fix the mean of the time-of-day component to be one for both methods.
The resulting hazard functions along the waiting time (left panel) can be interpreted as the daily average, and are comparable to the aggregated hazard function estimated from the pooled data (the dotted-dashed line) \citep{mandelbaum2013data}.
We also provide the $95\%$ pointwise confidence band for the tF-Hazards estimate from bootstrap.
More specifically, we randomly sample the same number of calls with replacement from the original observations, and fit the tF-Hazards to the bootstrapped samples to get a new estimate.
We repeat this procedure 100 times, and get 100 estimates of the same quantity.
Then we obtain the empirical confidence interval at each time point of the estimate.
For the additive model, we obtain the $95\%$ pointwise confidence band from the R package.
{However, since we only use one tenth of the total observations, the resulting confidence band is much wider than that from the tF-Hazards method (even wider than the current y-axis limits). Thus we omit it in Figure \ref{fig:puv}.}

\begin{figure}[h]
\begin{center}
\includegraphics[width=3in]{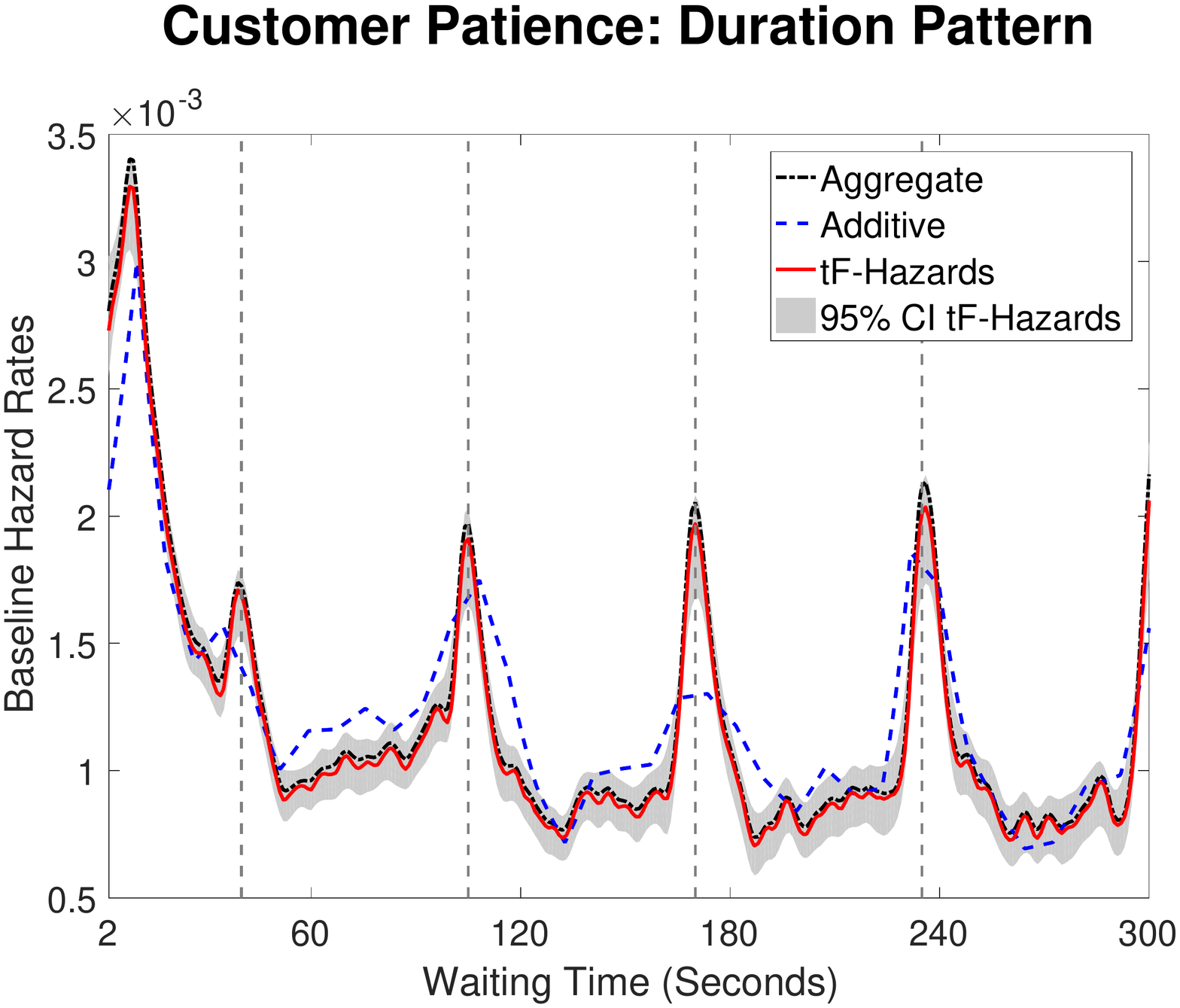}
\includegraphics[width=3in]{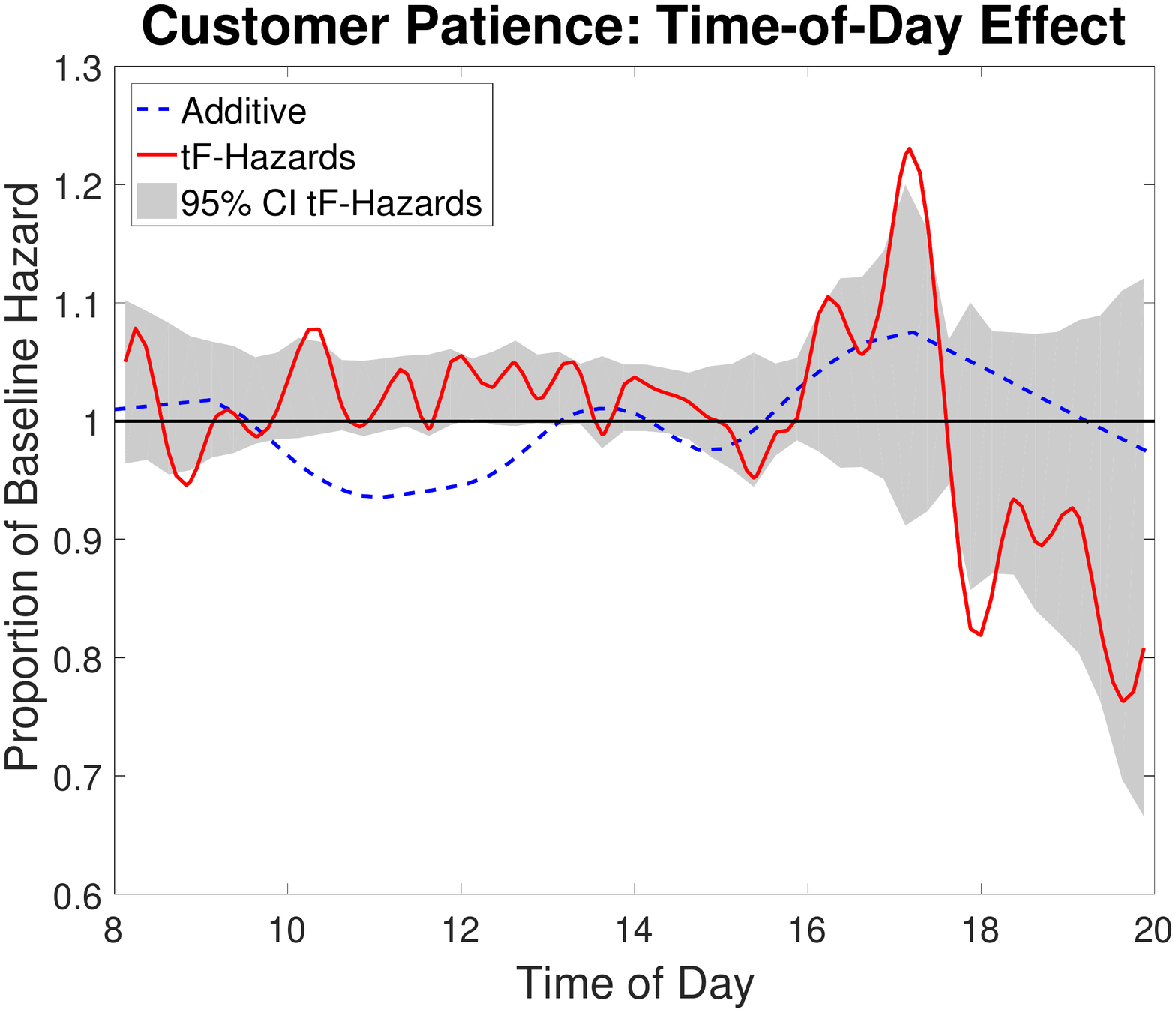}
\end{center}
\vskip-.1in
\caption{Call Center Quick\&Reilly Data: customer patience patterns along duration (left) and time of day (right). Periodic peaks of hazard rates along duration are marked by vertical dashed lines. The $95\%$ pointwise confidence bands for tF-Hazards are from 100 bootstraps; the confidence bands for the additive model are omitted due to excessive width.}
\label{fig:puv}
\end{figure}

In the left panel of Figure \ref{fig:puv}, the three hazard functions have concordant patterns in general.
The aggregated hazard function  and the tF-Hazards hazard function almost always reside in the confidence band, while the additive hazard function is less so.
{The hazard rates of abandonment decrease within the first minute of waiting, indicating customers become more patient as they wait longer in the queue.
This pattern is consistent with discoveries from previous empirical studies on an independent dataset \citep{brown2005statistical,aktekin2014bayesian}.
We also notice that there are periodic peaks about 65 seconds apart (marked by vertical dashed lines). 
The peaks turn out to be triggered by automatic announcements (e.g., ``Your call is important to us").
Recall that a large hazard rate corresponds to a high chance of abandonment. 
Thus, the peaks indicate that the announcements, designed to encourage customers to remain waiting, actually achieve the opposite of the original intent by stimulating abandonment.
Similar conclusions were drawn in \cite{brown2005statistical} and \cite{mandelbaum2013data}.}

The right panel of Figure \ref{fig:puv} is unique to our study, which presents the time-of-day patterns estimated by the additive model and the tF-Hazards model.
This offers a novel perspective of the dynamics of customer patience across different times of a day. As mentioned above, for model identification and intuitive interpretation, the time-of-day component is normalized to have mean 1 (indicated on the y-axis); as such, for a fixed time point within the day, the y-value of the component can be interpreted as the proportion of the patience hazard function relative to the average hazard function (shown in the left panel).

{Due to the high censoring rate (about 80\%, either answered or truncated at 5 minutes), the confidence band from the tF-Hazards model is a bit wide. In particular, at the end of the day when there are much fewer calls, the confidence band is even wider. Nonetheless, we can still discern some interesting time-of-day features from the tF-Hazards result. Firstly, the chance of abandonment is relatively high during the daytime, which indicates a low customer patience level.
Secondly, the hazard rates reach a peak around 5:00 pm, when people are about to get off work.
Finally, the hazard rates are relatively low after 6:00 pm, indicating that customers are generally more patient in the evening. The estimate from the additive model shows a slightly different pattern, probably due to the huge reduction of data; its confidence band is again omitted due to the excessive width.}

Understanding the waiting-time and time-of-day patterns of customer patience can help practitioners improve call center operations and better manage abandonment rates.
For one thing, the managers may explore new strategies of ``actively managing" patience, other than ``passively" periodically playing recorded messages to keep customers waiting.
Many companies nowadays have provided callback services, for example.
On the time-varying aspect, managers can use time-of-day specific hazard function to better staff call centers, and adaptively schedule the work shift.

\textcolor{black}{We also apply tF-Hazards to estimate the hazard rate surface including the calls with waiting times less than 2 seconds. The results are contained in Section B of the supplementary material. In summary, there is a steep peak in the estimated hazard function for waiting time less than 2 seconds; the new estimate has more wiggles between the periodic peaks than the current estimate in the left panel of Figure~\ref{fig:puv}, due to the fact that our method uses one single smoothing parameter to control the overall roughness. Nevertheless, the overall patterns of the estimated hazard functions are the same as in Figure \ref{fig:puv}. More detailed comparisons are included in Section B of the supplementary material.}

\subsection{Offered Wait}\label{subsec:w}
Similar to the patience study, we also exploit the waiting time data to investigate the dynamic patterns of offered wait in the call center. Because offered wait is a result of interplay among service demand, agent capacity and system protocol, studying offered wait allows the managers to better understand system status and performance evaluation.

{We use the same grids and domains for waiting time (2 seconds to 300 seconds by second) and time of the day (8:00 am to 8:00 pm by 15 minutes) as in the patience study.
Namely, we consider the same set of calls. The only difference is that now the abandoned calls are censored observations and the answered calls are complete observations.
As a result, the censoring rate (about 50\%, either abandoned or truncated at 5 minutes) is much lower than that in the patience study.}
For the same reasons mentioned in Section~\ref{subsec:p}, we set the rank of the tF-Hazards model to be one, and fit the additive model only with one tenth of all observations.
Figure \ref{fig:wsurf} shows the hazard surfaces estimated by the two models respectively.

\begin{figure}[h]
\begin{center}
\includegraphics[width=3in]{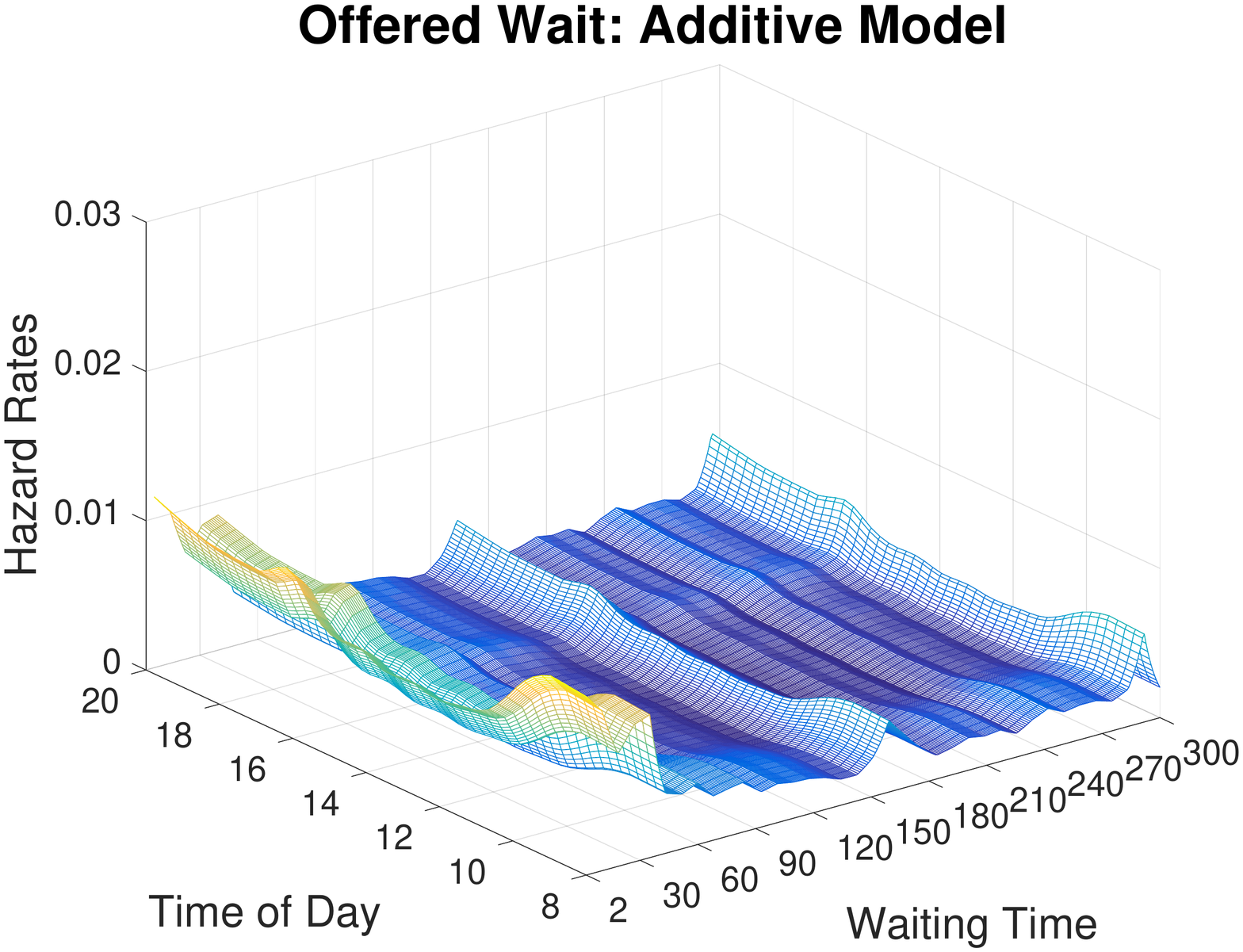}
\includegraphics[width=3in]{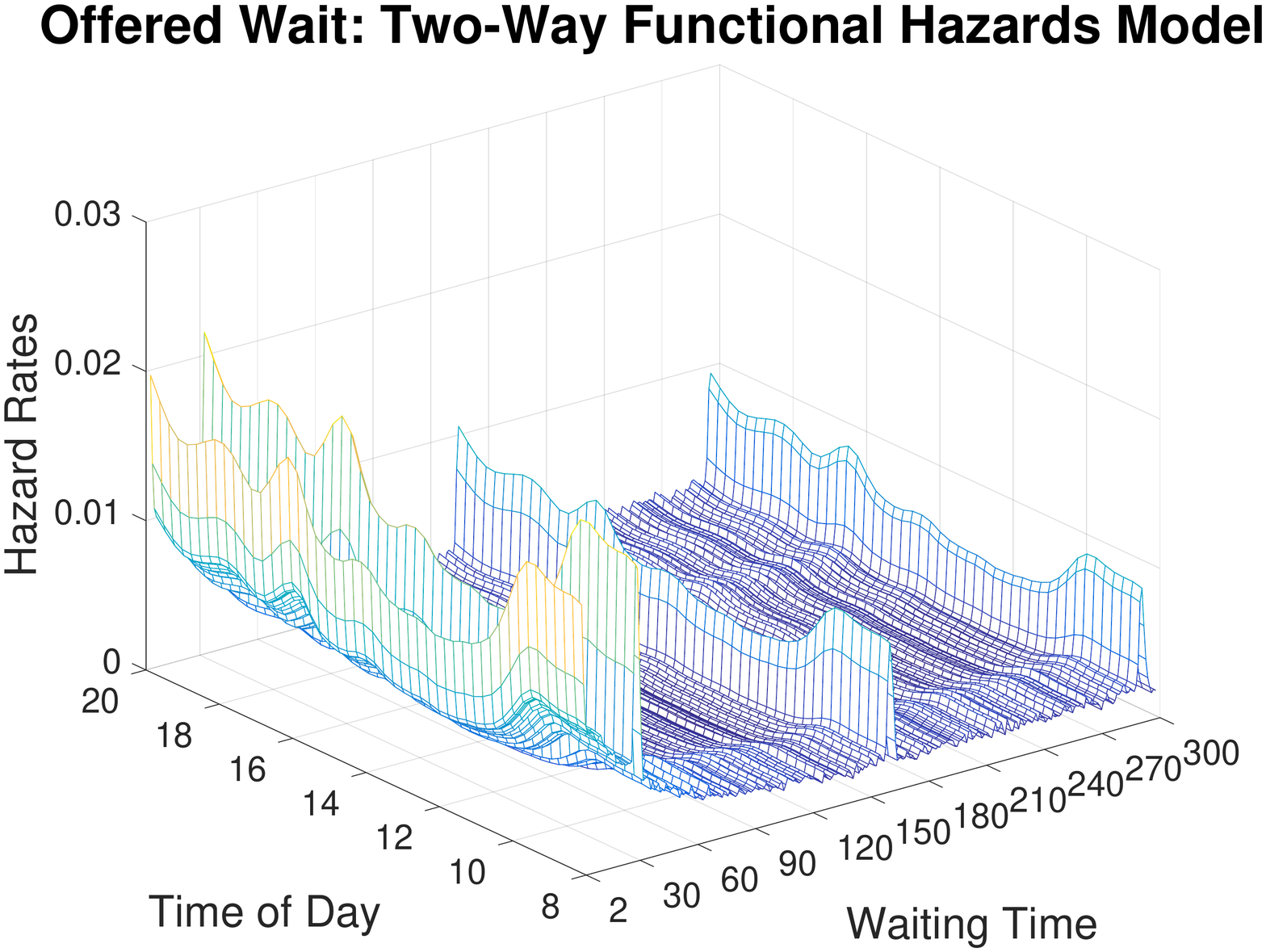}
\end{center}
\vskip-.1in
\caption{Call Center Quick\&Reilly Data: Hazard surfaces of offered wait estimated from different methods. Left: the additive model; Right: the proposed tF-Hazards model.}
\label{fig:wsurf}
\end{figure}

Figure \ref{fig:wuv} shows the duration component and the time-of-day component estimated from each method. Again, we add the aggregated hazard function estimated from the pooled data \citep{mandelbaum2013data} to the left panel. {We also provide the 95\% pointwise confidence intervals for the tF-Hazards method. The confidence bands are much narrower (barely visible for the duration component in the left panel) than the counterparts in the patience study, mainly due to the much lower censoring rate.}


\begin{figure}[h]
\begin{center}
\includegraphics[width=3in]{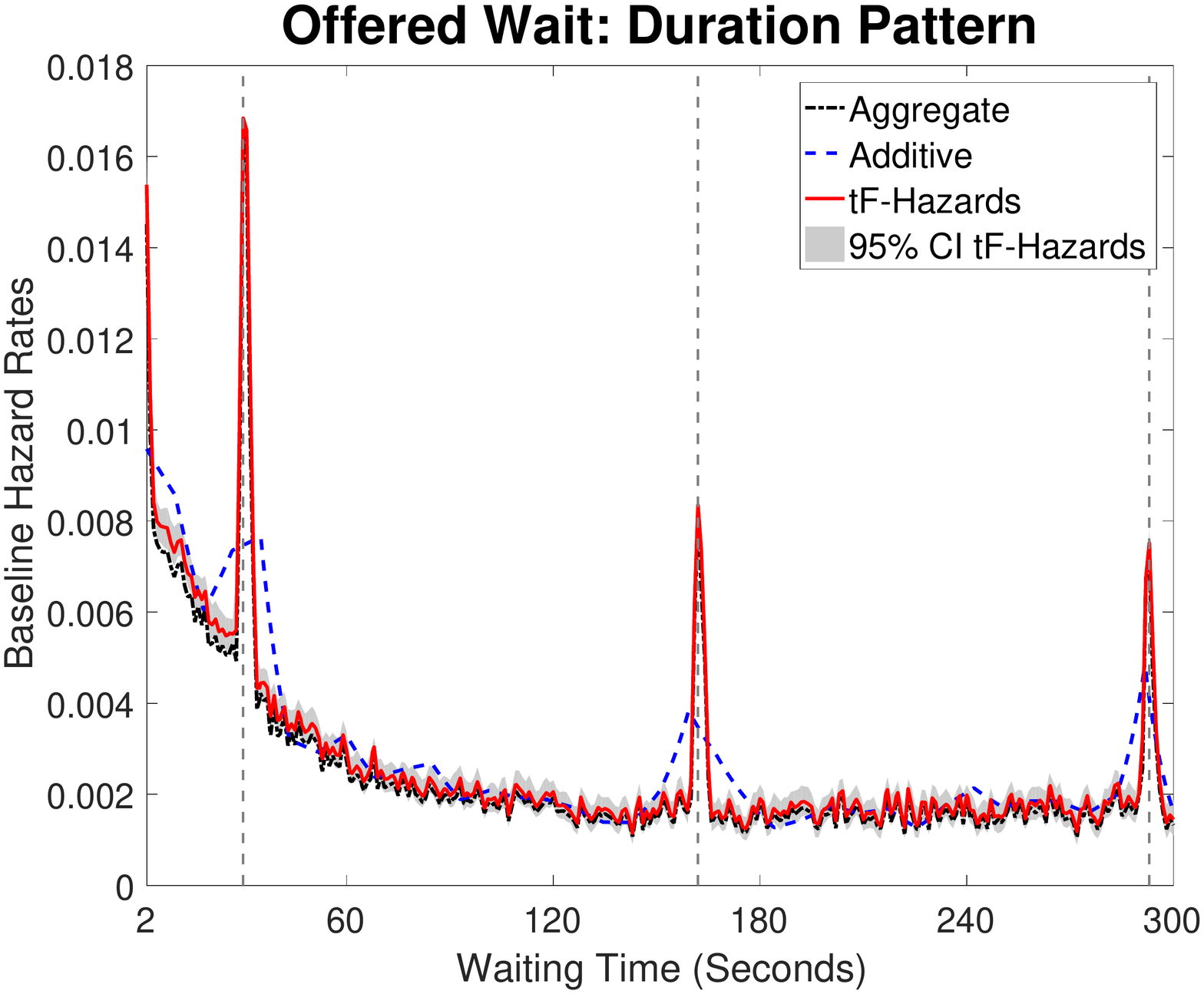}
\includegraphics[width=3in]{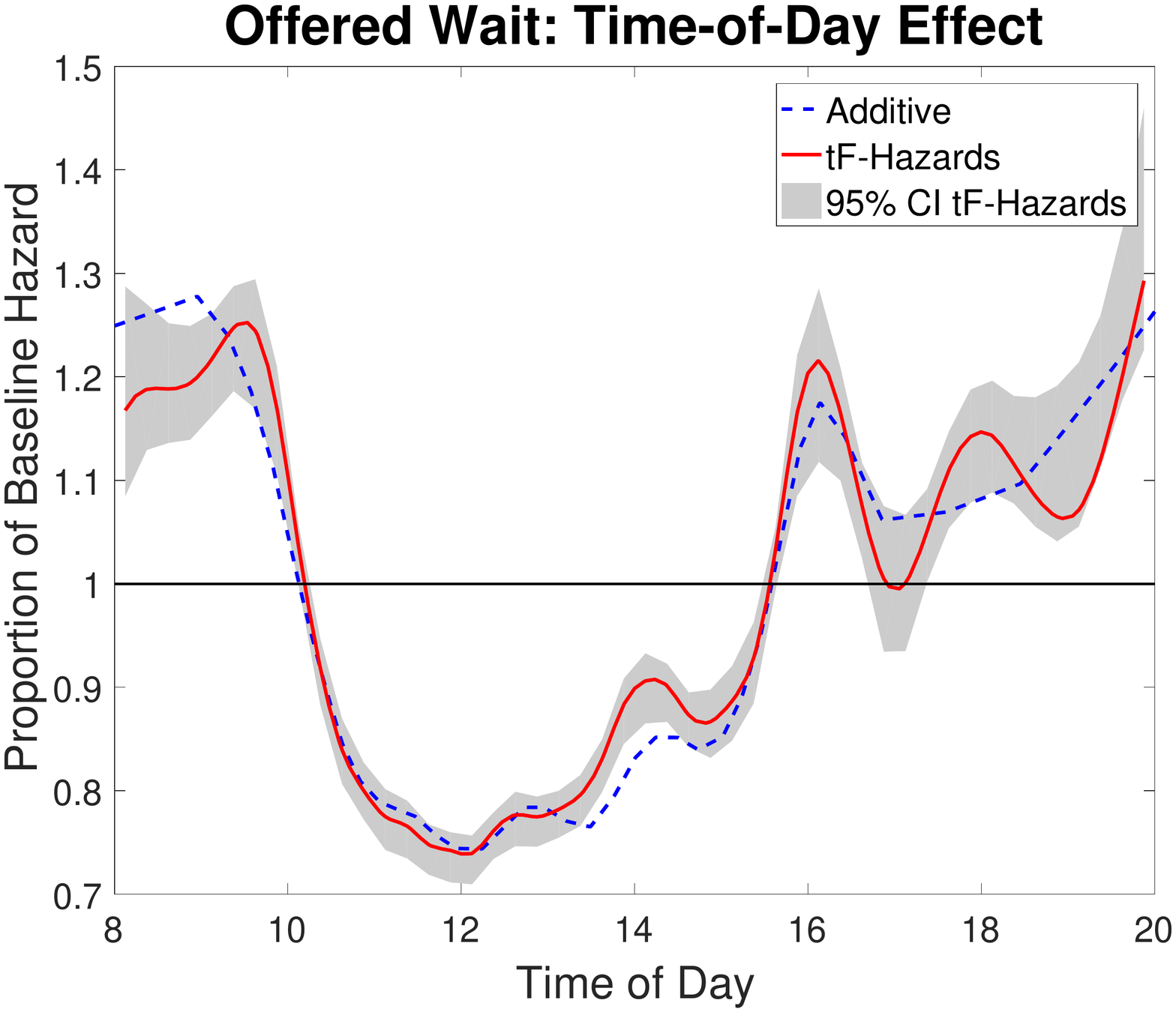}
\end{center}
\vskip-.1in
\caption{Call Center Quick\&Reilly Data: offered wait patterns along duration (left) and time of day (right). Periodic peaks of hazard rates along duration are marked by vertical dashed lines. The $95\%$ pointwise confidence bands for tF-Hazards are from 100 bootstraps; the confidence bands for the additive model are omitted due to excessive width. }
\label{fig:wuv}
\end{figure}

{In the left panel of Figure \ref{fig:wuv}, we observe that the hazard function estimated from tF-Hazards is more similar to the aggregated hazard function than the additive method.
Both tF-Hazards and the aggregated approach capture significant peaks about 130 seconds apart (marked by vertical dashed lines); yet the additive model has much lower peaks.
According to \cite{mandelbaum2013data}, the peaks are operationally meaningful and are ``associated with times at which customers enjoy an upgrade in their priority status."}

The right panel of Figure \ref{fig:wuv} presents the time-of-day patterns estimated from the additive model and the tF-Hazards model.
{Both methods give consistent patterns.
The hazard rates are relatively high before 10:00 am, and then go down and stay low during the day with a dip at 12:00 pm, before bouncing up after 4:00 pm and maintaining a high level in the evening.
Recall that in the offered wait study, higher hazard rates loosely connect to better service quality (i.e., a call is more likely to be answered).
The time-of-day pattern reflects that the service quality is generally low between 10:00 am and 4:00 pm, and high otherwise.
A plausible explanation is that: in early morning the call volume is moderate, resulting in a good service quality; then the call volume goes up and remains high during business hours, especially around the lunch break at noon, making the service quality relatively low; the system load drops dramatically as people get off work, so the quality bounces up again.}

The above insights provide useful guidance to managers for understanding system protocol impact and evaluating system performance across time-of-day. The peaks at different waiting times reflect the impact of the periodic priority status updating. The valleys and humps at different time-of-day reveal the system performance, which may indicate opportunities for better matching demand and capacity.

\textcolor{black}{Similarly, we run a separate analysis to include all those calls with waiting time less than 2 seconds, and compare the results with those in Figure~\ref{fig:wuv} in Section C of the supplementary material. The overall messages can be summarized as follows. First of all, the baseline hazard rate function (left panel) remains about the same for waiting time above 2 seconds, after including the short-wait calls. One can see the same general decay and the peaks caused by the waiting status update. Secondly, due to the extremely high percentage of calls with short wait durations, there is a huge spike (approximately 0.15 in magnitude) in the hazard rate near the origin, which obscures the detailed patterns shown in the left panel for waiting times longer than 2 seconds. Finally, as mentioned before, over 98\% of the calls with waiting times less than 2 seconds were answered immediately. This extreme amount of short waiting calls has distinct hazard rates from the rest of the data. After incorporating these calls, the intra-day pattern (right panel) is driven by the percentage of such calls at different time of the day, and hence is very similar to the intra-day pattern of the percentage of calls answered within 2 seconds. More details are provided in Section C of the supplementary material.}

{
\subsection{M/M/n+G Queue}
To demonstrate how to take advantage of the tF-Hazards model in a call center queueing system, we consider the M/M/n+G queue \citep{zeltyn2005call}.
The queue is characterized by Poisson arrivals (M), exponentially distributed service times (M), and generally distributed customer patience times (G).
It is a generalization of the M/M/n+M (or Erlang-A) model, where the customer patience time is assumed to be exponentially distributed.
\cite{zeltyn2005call} studied the asymptotics of the M/M/n+G model under different staffing regimes.

The tF-Hazards method improves the standard approach by allowing the estimated hazard functions to be different at different times of a day.
In particular, the waiting behavior in the morning may be different from that in the afternoon.
By taking into account the time-of-day variation, call center managers can evaluate the system performance more accurately, and adjust the staffing regime dynamically.

Our simulation study fixes the arrival rate to be 100, the service rate to be 1, and the number of agents to be 100. The hazard function for patience is set as the piecewise constant function estimated from the Quick\&Reilly data. We particularly consider three levels of patience (i.e., low, medium, high) presumably corresponding to different times of a day, through multipling the baseline hazard function by 1.5, 1, and 0.5, respectively. Based on~\cite{zeltyn2005call} and~\cite{aktekin2014bayesian}, we numerically evaluate the steady-state probability of abandonment, given a customer is not served immediately upon arrival. As a result, we obtain the probability of abandonment as 0.0095 (for the low patience setting), 0.0068 (for the medium patience setting), 0.0042 (for the high patience setting), respectively. Apparently, different patience levels lead to distinct system performances, which can be viewed as corresponding to different times of a day.
Therefore, the proposed tF-Hazards model will be helpful in practice as it can provide a more detailed characterization of the varying waiting behavior at different times of a day.
}

\section{Discussion}\label{sec:dis}
Our research is motivated by efficient management of call center operations, in particular agent staffing. A crucial (but understudied) issue is empirical modeling of customer waiting behavior -- estimation of hazard functions of customer patience time and/or system offered wait time.
For the first time, we study the hazard functions as a two-way surface of both wait duration and time of day, to provide inputs to time-varying queueing models. We develop a two-way functional hazards model by considering a smooth and low-rank hazard function of two timescales, namely duration and time of day.
To fit the model, we employ a penalized likelihood framework where the likelihood is derived from a two-way piecewise constant hazards model, together with a penalty term to enforce smoothness of the hazard function in both timescales.
We derive an ADMM algorithm to estimate model parameters. 
The method naturally accommodates missing values and leverages information in all observations to obtain an accurate and efficient estimate of the underlying hazard surface.
Application of the method to a US Bank call center data set reveals interesting insights about customer waiting behavior. We demonstrate via simulation how our model can be used for estimating system performance. 
The proposed method is generally applicable to other real data problems with competing risks.

There are a few open questions that require further investigations.
One future research direction is to incorporate covariate adjustment into the two-way hazards model.
One may extend the idea of the Cox proportional hazards model \citep{cox1972regression} and assume the hazard responds to the covariate effect exponentially.
In the context of call center waiting time studies, possible covariates include different days of a week, different customer types, and so on.
The two-way proportional hazards model can be used to study how different factors affect customer patience and call center efficiency. {Another direction is to accommodate dependent observations.
Incorporating dependency may further improve the inference of customer waiting dynamics.}

%
%
%
%
%

%
\if1\blind
{
\section*{Acknowledgement}
The research was partially supported by NSF grants DMS-1106912, DMS-1208952, and DMS-1407655, the Xerox UAC Foundation, and the University of Hong Kong Stanley Ho Alumni Challenge Fund. The authors also thank Technion SEELab for providing the call center data.
}\fi



\clearpage
\appendix
{\noindent \bf Appendix: Tuning Parameter Selection}\label{subsec:tuning}

There are multiple penalty parameters in the optimization problem \eqref{penlik}.
Simultaneous selection of all the tuning parameters in a data-driven fashion can be a huge computational burden.
As a remedy, we nest the selection of different parameters inside the ADMM algorithm.
We will discuss in more detail below.

The roughness penalty parameters $\lambda_{\bu_i}$ and $\lambda_{\bv_i}$ control the smoothness of $\bu_i$ and $\bv_i$ when estimating $\U$ and $\V$.
Since we estimate $\U$ and $\V$ in \eqref{UV1} through tF-SVD, we adopt the adaptive tuning selection procedure as in \cite{huang2009analysis}.
In particular, the selection of $\lambda_{\bu_i}$ and $\lambda_{\bv_i}$ is embedded in the alternating algorithm of tF-SVD and use the generalized cross-validation (GCV) criteria.
Closed-form GCV scores have been derived from leaving out rows and columns for fast computation \citep{huang2009analysis}.

Another tuning parameter is $\rho$.
Since $\rho$ as a penalty parameter is left out in \eqref{UV1}, it primarily serves as the step size in the dual update of \eqref{Lambda}.
Although the step size in the ADMM algorithm can be fixed as any positive value with little effect on the final solution theoretically, it does affect  convergence rate in practice \citep{boyd2011distributed}.
To improve the convergence rate of the algorithm and alleviate the dependence on the initial choice of the penalty parameter, we follow \cite{he2000alternating} and select $\rho$ adaptively in each iteration.

In particular, we first define two residuals in each iteration as
\begin{align*}
\mbox{Primal residual:    }\quad& z^{(l)}\triangleq \|\bH^{(l)}-\U^{(l-1)}{\V^{(l-1)}}^T \|^2_F,\\
\mbox{Dual residual:    }\quad& s^{(l)}\triangleq\|\bH^{(l)}-\bH^{(l-1)}\|^2_{F}.
\end{align*}
We remark that a small value of $\rho$ leads to $z^{(l)}\gg s^{(l)}$, while a large value of $\rho$ leads to $z^{(l)}\ll s^{(l)}$.
To see this, we temporarily set $\bTheta=\0$, and examine \eqref{thetasol}.
From the {l'Hospital's} rule and the limit theory, it is easy to see that
\bes
\lim_{\rho\rightarrow0} \widetilde{h}_{jk}={d_{\cdot jk}\over t_{\cdot jk}} =\widehat{h}_{jk}, \quad\mbox{ and }\quad \lim_{\rho\rightarrow\infty} \widetilde{h}_{jk}=\bu^T_{(j)}\bv_{(k)}.
\ees
Namely, when $\rho$ approaches zero, the entries of $\bH^{(l)}$ always have similar values to the corresponding MLE, and thus $s^{(l)}\approx0$;
when $\rho$ approaches infinity, the entries of $\bH^{(l)}$ have similar values to the corresponding entries of $\U^{(l-1)}{\V^{(l-1)}}^T$, and thus $z^{(l)}\approx0$.

The idea of the adaptive selection of $\rho$ in each iteration is to keep both residuals within a reasonable factor of one another as they both converge to zero.
More specifically, we propose to use the following updating scheme
\bes\label{rho}
\rho^{(l)}=\left\{
\begin{aligned}
&10\rho^{(l-1)},\quad  &100s^{(l)}&<z^{(l)};\\
&2\rho^{(l-1)},\quad &10s^{(l)}&<z^{(l)}\leq100s^{(l)};\\
&\rho^{(l-1)},\quad &s^{(l)}/10&\leq z^{(l)}\leq 10s^{(l)};\\
&\rho^{(l-1)}/2,\quad &s^{(l)}/100&\leq z^{(l)}<s^{(l)}/10;\\
&\rho^{(l-1)}/10,\quad && \quad\,\, z^{(l)}<s^{(l)}/100.
\end{aligned}
\right.
\ees
We recommend initializing the algorithm with a small value of $\rho$ (e.g., $\rho^{(0)}=0.1$) in practice.
The scheme works well in all our numerical studies and the value of $\rho$ becomes fixed after a few iterations.
See discussions of the convergence property of the ADMM algorithm with a variable penalty parameter in \cite{he2000alternating}.

Finally, we remark that the rank $r$ of Model \eqref{model_d} also needs to be determined from data.
One approach is to check the scree plot of the MLE of the hazard rate matrix, as a common practice in the principal component analysis literature.
Other methods for estimating the rank exist as well \citep[see][for example]{owen2009bi}.
In most real applications, a small rank number is usually sufficient, since the first few rank-one layers capture the most important patterns of a hazard  function that are highly interpretable. 
Further investigation of rank selection for the tF-Hazards model is a future research direction and beyond the scope of this paper.

\clearpage
\setcounter{figure}{0}
\renewcommand{\thefigure}{S.\arabic{figure}}
\renewcommand{\thetable}{S.\arabic{table}}
\renewcommand\theequation{S.\arabic{equation}}
\setcounter{equation}{0}

\begin{center}
\if1\blind
{
{\Large\bf Supplementary Materials for \\``To Wait or Not to Wait: Two-way Functional Hazards Model for Understanding Waiting in Call Centers'' \\ by Gen Li, Jianhua Z.\ Huang and Haipeng Shen}
}\fi

\if0\blind
{
{\Large\bf Supplementary Materials of \\``To Wait or Not to Wait: Two-way Functional Hazards Model for Understanding Waiting in Call Centers''}
}\fi
\end{center}
\appendix

\section{Simulation}
We demonstrate the efficacy of the proposed tF-Hazards method via a comprehensive simulation study.
As a comparison, we also consider the MLE of the standard two-way piecewise constant model.
We consider 6 different simulation settings. In all the settings, we assume the underlying hazard functions are piecewise constant in both timescales, but some are smoother than others. We fix the number of arrival time intervals to be $n=30$, and the number of waiting time intervals to be $p=30$ as well. For simplicity, we set the numbers of observations in different arrival time intervals to be the same, and any observation exceeding a prefixed time-to-event threshold is right censored. More specifically,
\bi
\item {\bf Setting 1}: the generating hazard rate matrix has unit rank; the left and right singular vectors of the hazard rate matrix are obtained by discretizing some underlying smooth functions. The number of observations in each arrival time interval is 100.
\item {\bf Setting 2}: the hazard rate matrix is the same as that in {\bf Setting 1}.  The number of observations in each arrival time interval is 500.
\item {\bf Setting 3}: the generating hazard rate matrix has unit rank; the left and right singular vectors are filled with uniform random numbers between 0 and 0.8. The number of observations in each arrival time interval is 100.
\item {\bf Setting 4}: the hazard rate matrix is the same as that in {\bf Setting 3}. The number of observations in each arrival time interval is 500.
\item {\bf Setting 5}: the generating hazard rate matrix is full-rank and filled by uniform random numbers from 0 to 0.5. The number of arrivals in each interval is 100.
\item {\bf Setting 6}: the hazard rate matrix is the same as that in {\bf Setting 5}. The number of arrivals in each interval is 500.
\ei

For each setting, we generate time-to-event observations from the given hazard rate matrix, 
apply the MLE piecewise constant method and the proposed method to the data, and evaluate the estimation accuracy of the two methods.
In particular, we set the rank of the proposed method to be 1 in all settings (for {\bf Setting 1 -- 4}, this is the correct rank; for {\bf Setting 5 and 6}, the rank is underestimated.)
We calculate the Frobenius norm of the difference between the estimated and the true hazard rate matrices to compare different methods.
We repeat the procedure 100 times for each simulation setting.
The results are summarized in Figure \ref{fig:box}.

\begin{figure}
  \centering
  \includegraphics[width=6in]{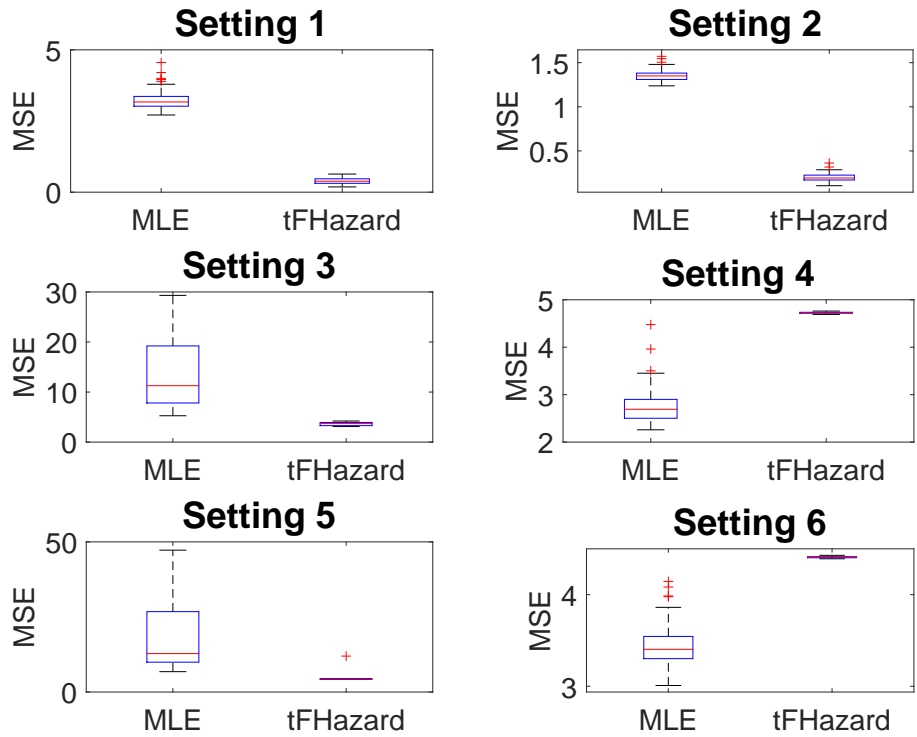}
  \caption{Simulation results. The box-plots show the Frobenius norm of the difference between the estimated and the true hazard rate matrices. Each plot corresponds to one simulation setting.}\label{fig:box}
\end{figure}

When the generating hazard rate matrix has low-rank smooth structure ({\bf Setting 1 and 2}), the tF-Hazards estimate is much more accurate than the MLE method.
When the true hazard rates are non-smooth ({\bf Setting 3 -- 6}), the roughness penalty in tF-Hazard introduces bias to the estimate, but at the same time reduces the variance of the estimate. The overall performance reflects the bias-variance tradeoff.
We observe that the comparison between the two methods largely depends on the sample size.
In particular, when the sample size is small ({\bf Settings 3 and 5}), the tF-Hazards method still outperforms the piecewise constant model due to the low variance; when the sample size is large enough ({\bf Settings 4 and 6}), the MLE is more accurate than the tF-Hazards estimate.
We also note that {\bf Settings 5 and 6} represent situations where the low-rank assumption is violated, which suggest that tF-Hazards remains useful when the sample size is insufficient.

\section{Customer Patience Study with Short Waiting Calls}
{\color{black}We apply the tF-Hazards method to estimate the underlying hazard rate surface for the Quick\&Reilly group with and without those calls with waiting times less than 2 seconds.
The results are shown in Figure \ref{fig:patience}. }
Panels (a) and (b) plot the intra-day pattern as proportion of the baseline hazard under the two scenarios. Panels (c) and (d) plot the baseline hazard rate as a function of waiting duration $\ge 2$ seconds, while Panel (e) shows the baseline hazard function for waiting time $\ge 0$  second.

\begin{figure}[htbp]
  \centering
  \includegraphics[width=2in]{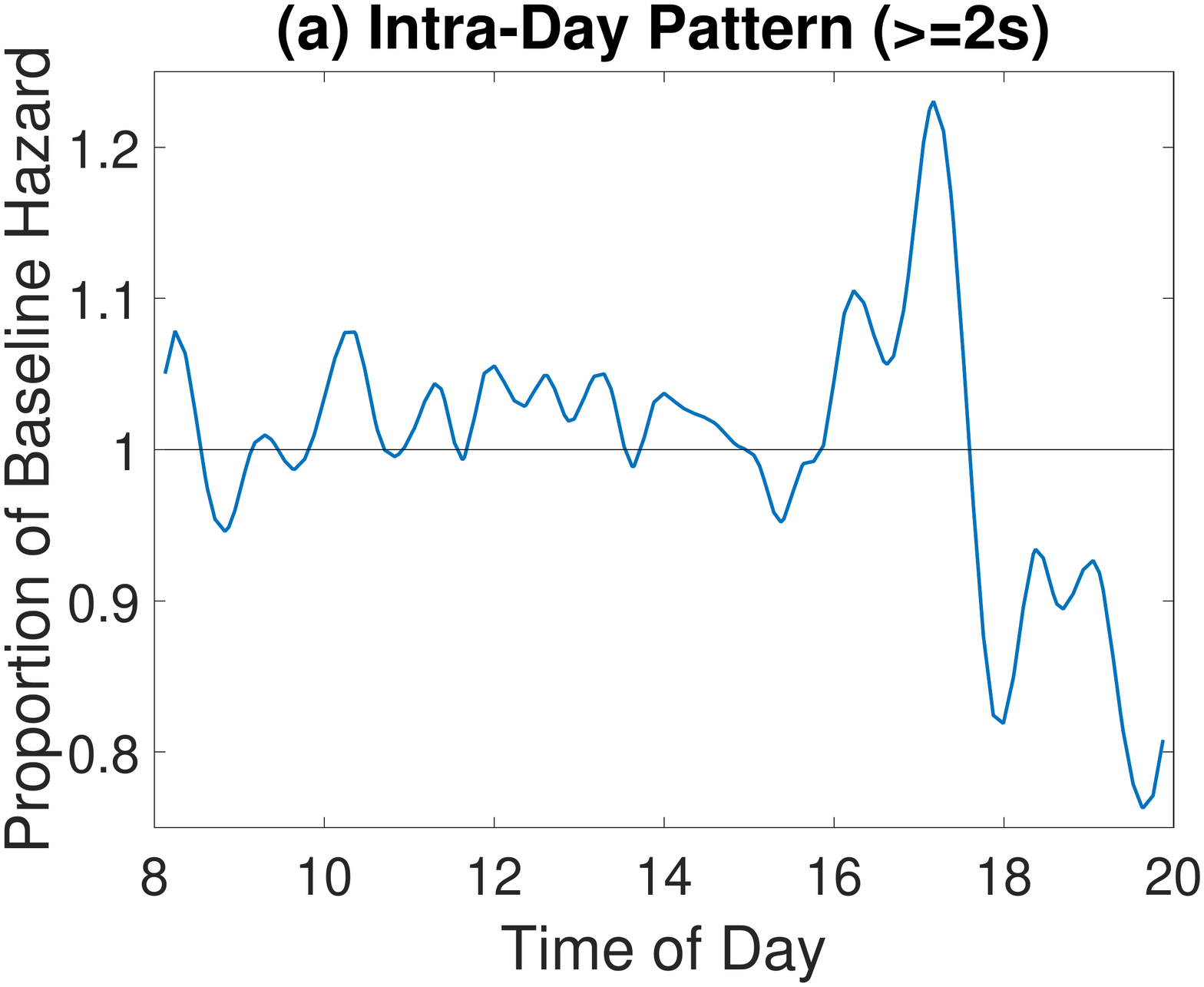}\hskip2.1in
  \includegraphics[width=2in]{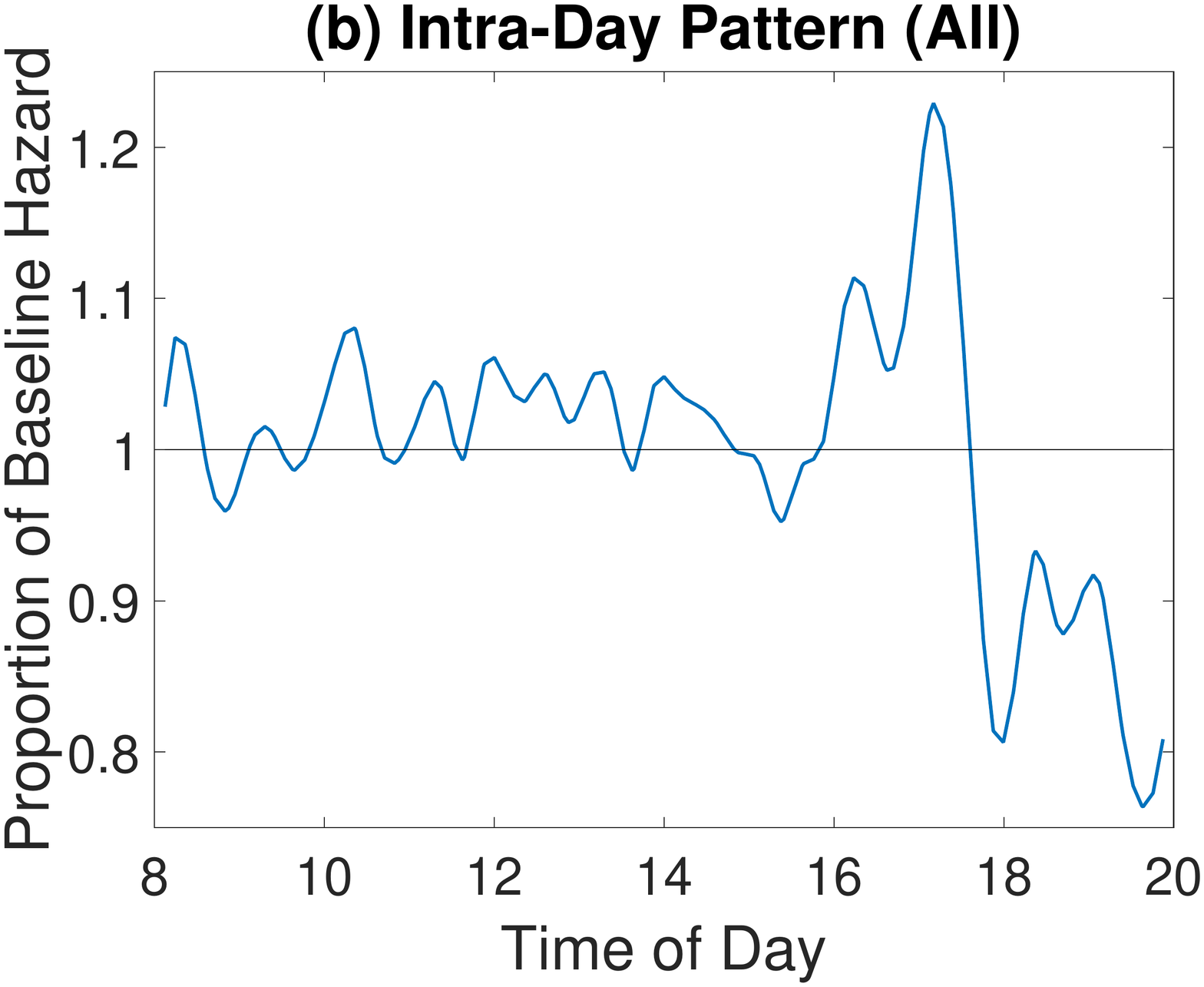}\\
  \includegraphics[width=2in]{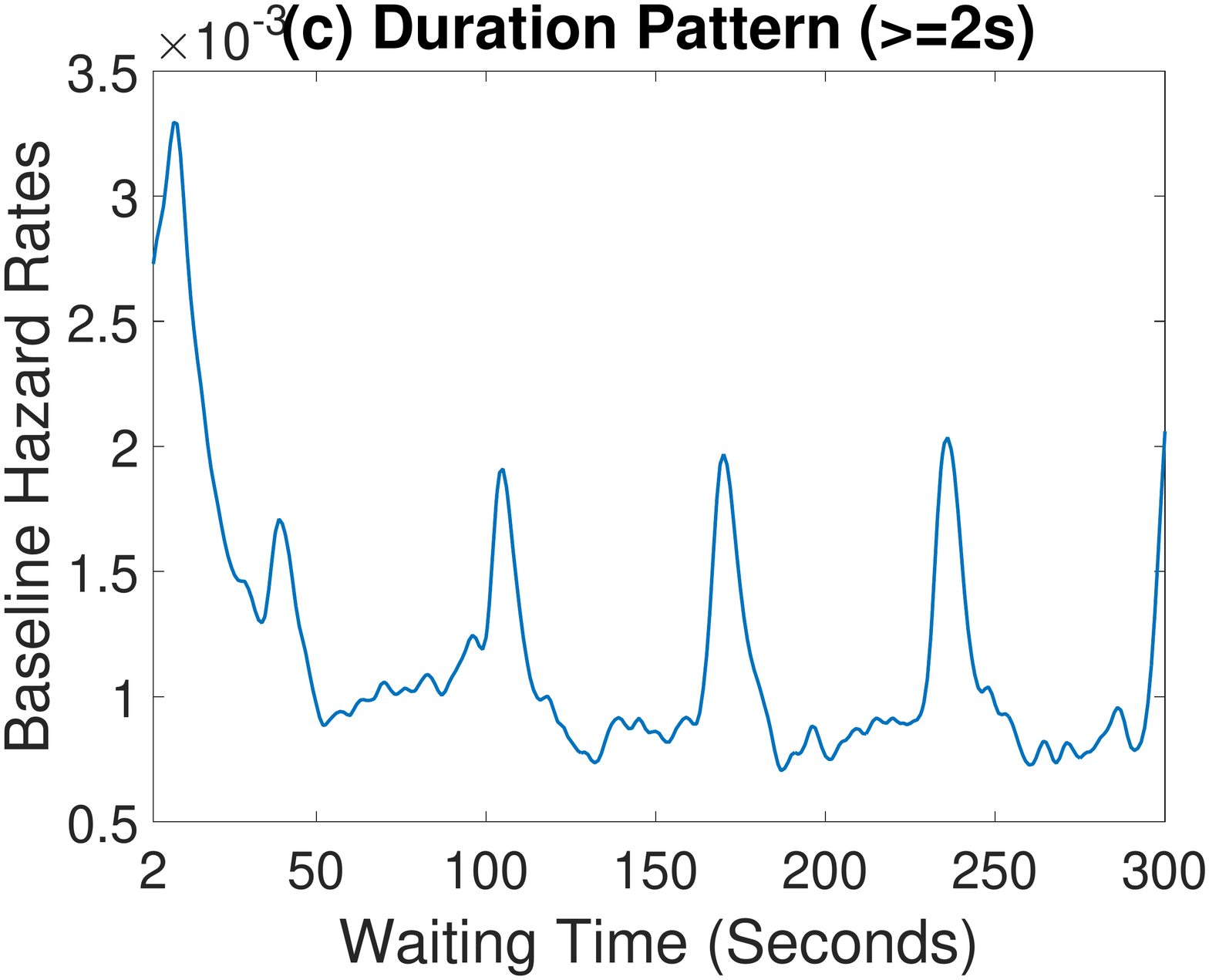}
  \includegraphics[width=2in]{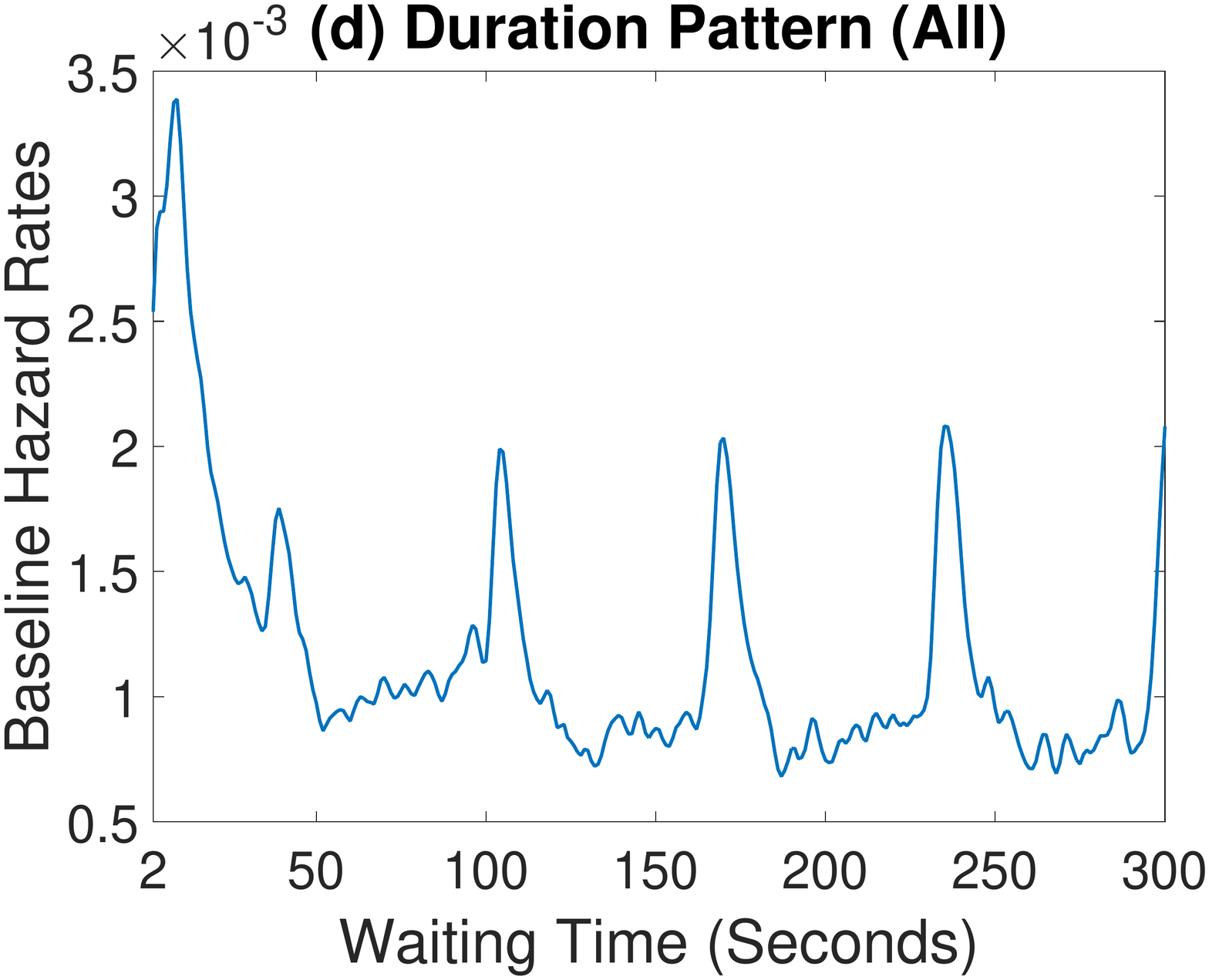}
  \includegraphics[width=2in]{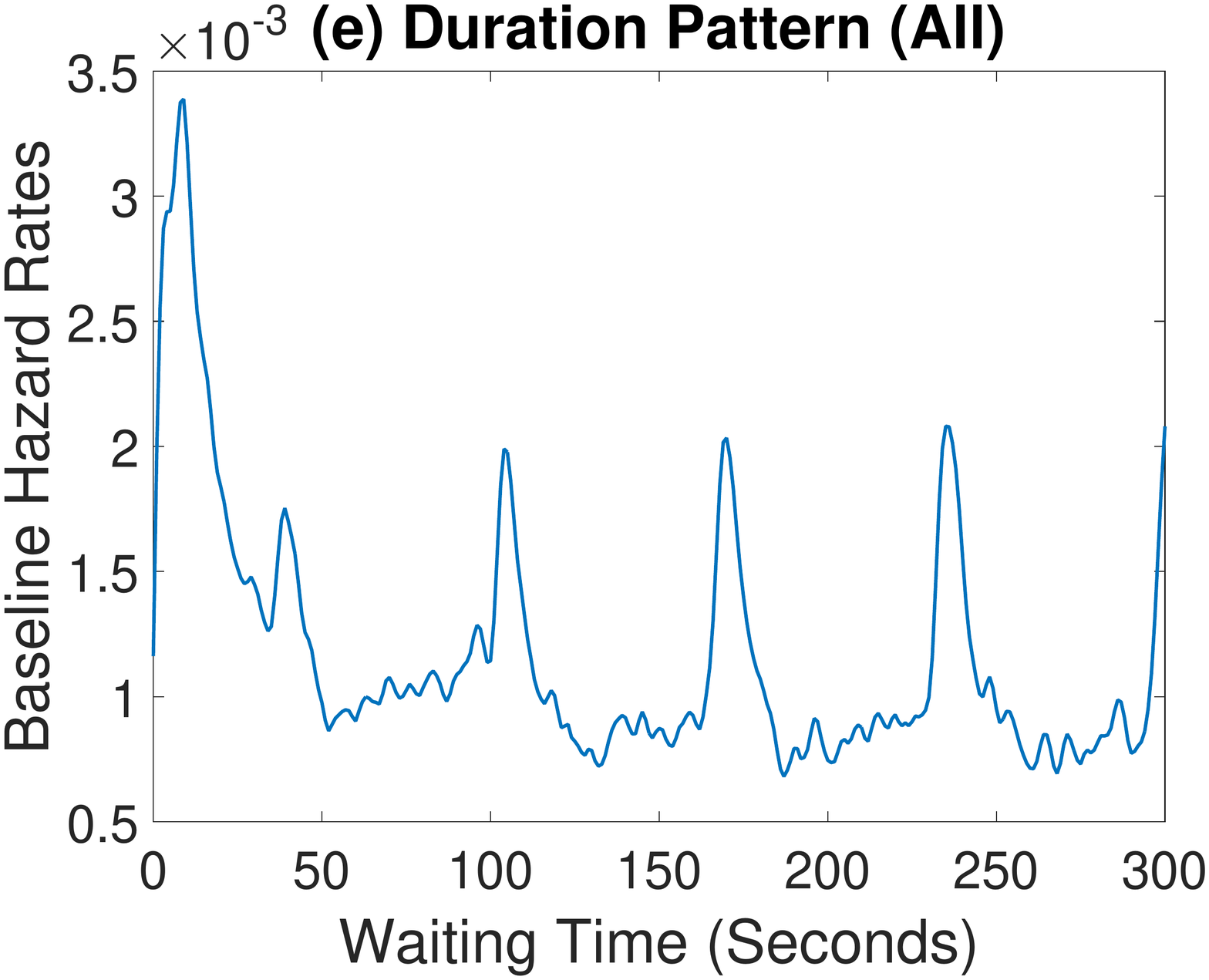}
  \caption{Customer patience study with short waiting calls for the Quick\&Reilly customer group. (a) the time-of-day pattern estimated from the $\geq 2$s data; (b) the time-of-day pattern estimated from the complete data;
   (c) the waiting pattern estimated from the $\geq 2$s data; (d) the waiting pattern estimated from the complete data (but only showing the pattern starting from 2s); (e) the waiting pattern estimated from the complete data (starting from 0s).}\label{fig:patience}
\end{figure}

We can make the following observations:
\begin{itemize}
\item Panels (a) vs. (b): Including the short-wait calls has no obvious effect on the intra-day pattern.

\item Panels (c)-(e): Again, the two baseline hazard functions in Panel (c) and Panel (d) look similar in general, with periodic peaks every minute or so.  The new estimate (with the short-waiting data) in Panel (e) has a sharp spike in the first couple of seconds (due to the high-answering-rate) and contains more wiggles (Panel (d)). This is because our method uses a single tuning parameter to control the degree of smoothness in the loading estimation. The inclusion of the short-waiting data leads to the selection of a smaller roughness penalty tuning parameter and thus results in a more wiggly estimate, without losing the general shape of the baseline function.
\end{itemize}

\section{Offered Wait Study with Short Waiting Calls}
{\color{black}We apply the tF-Hazards method to estimate the underlying hazard rate surface for the Quick\&Reilly group with and without those calls with waiting times less than 2 seconds.
The results are shown in  Figure \ref{fig:offered}.} Panels (a) and (b) plot the intra-day pattern as proportion of the baseline hazard under the two scenarios. Panel (c) plots the immediate answering rates (i.e., percentages of calls answered within 2 seconds) at different times of a day. Panels (d) and (e) plot the baseline hazard rate as a function of waiting duration $\ge 2$ seconds, while Panel (f) shows the baseline hazard function for waiting time $\ge 0$ second.

\begin{figure}[htb]
  \centering
  \includegraphics[width=2in]{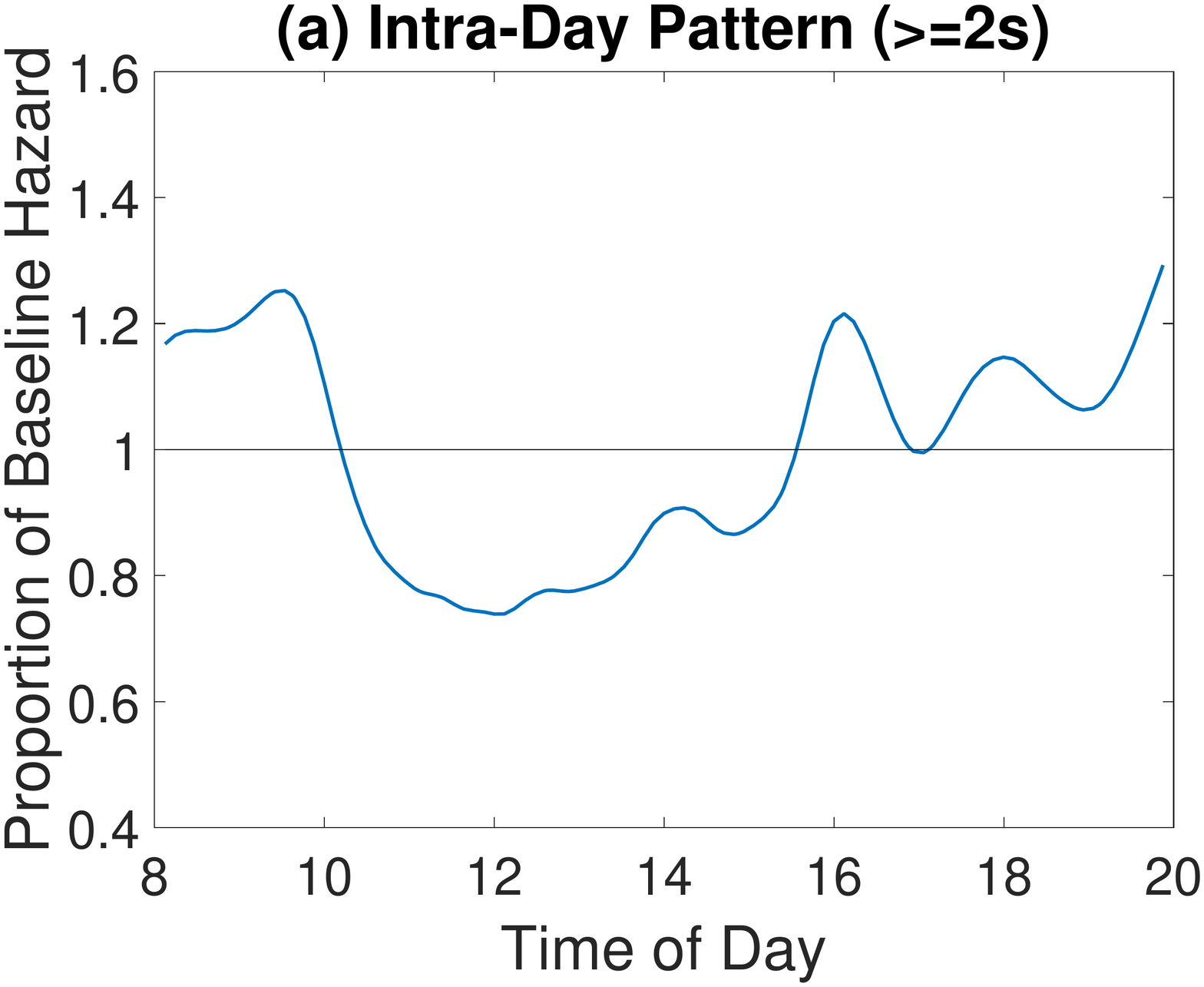}
  \includegraphics[width=2in]{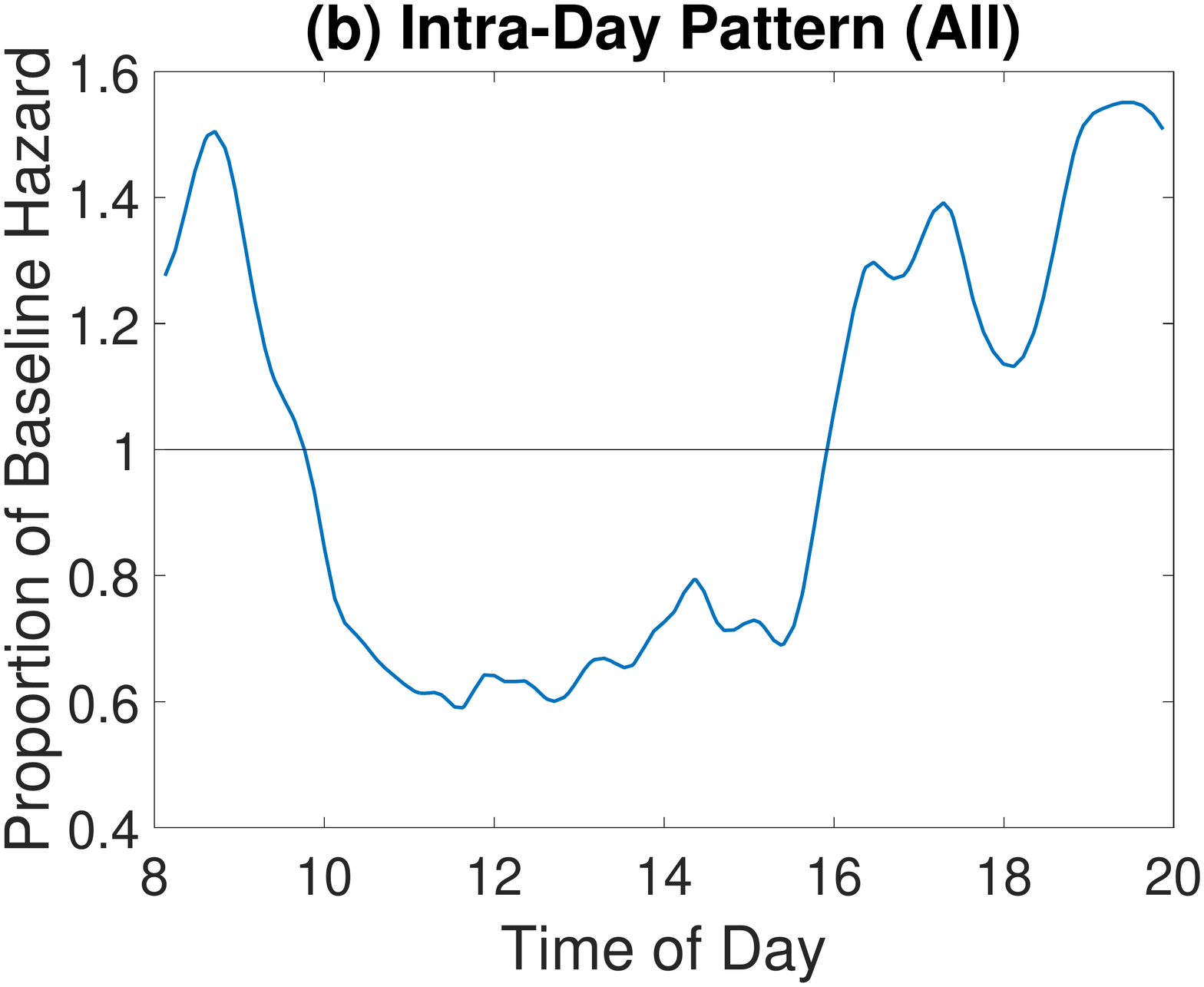}
  \includegraphics[width=2in]{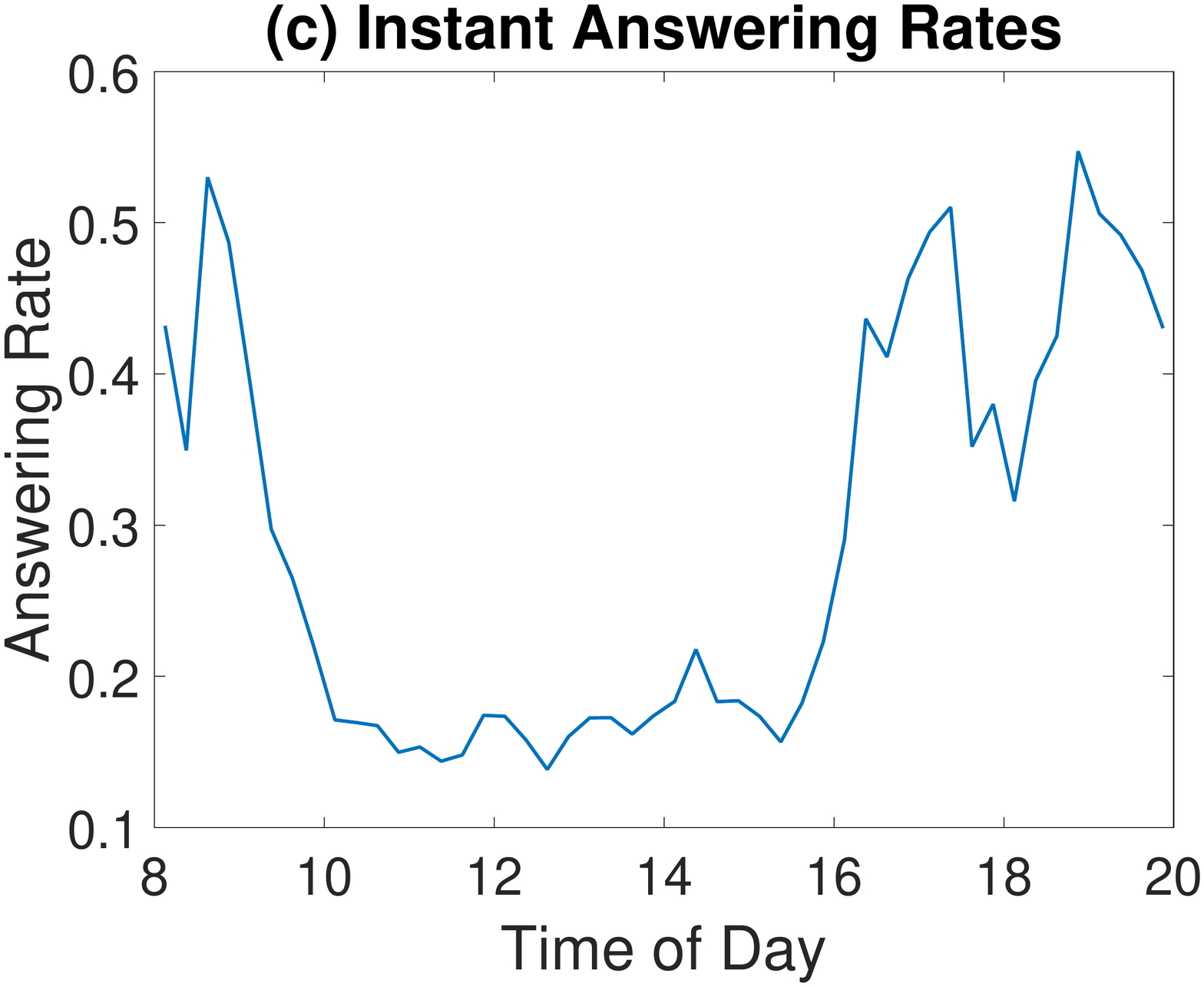}\\
  \includegraphics[width=2in]{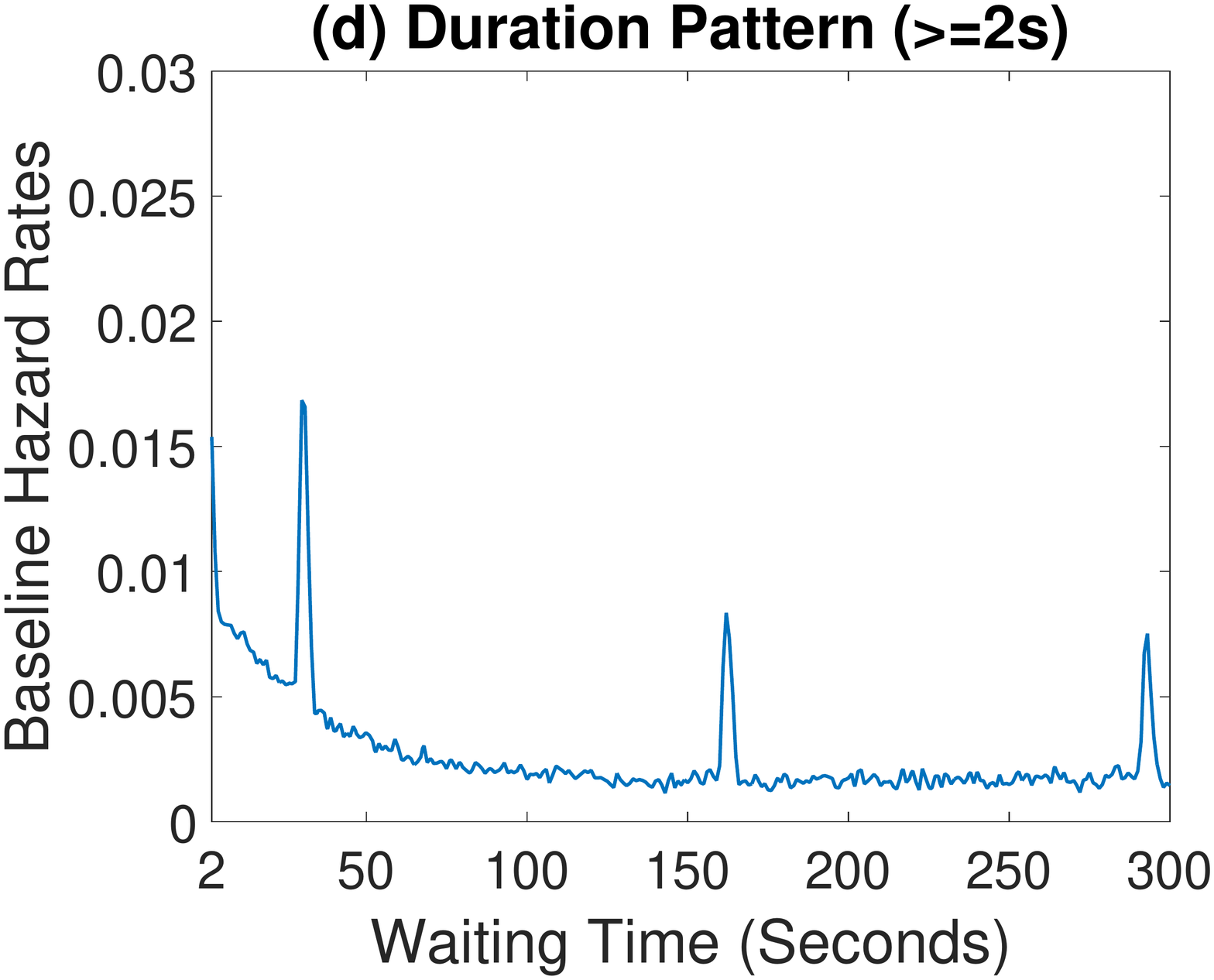}
  \includegraphics[width=2in]{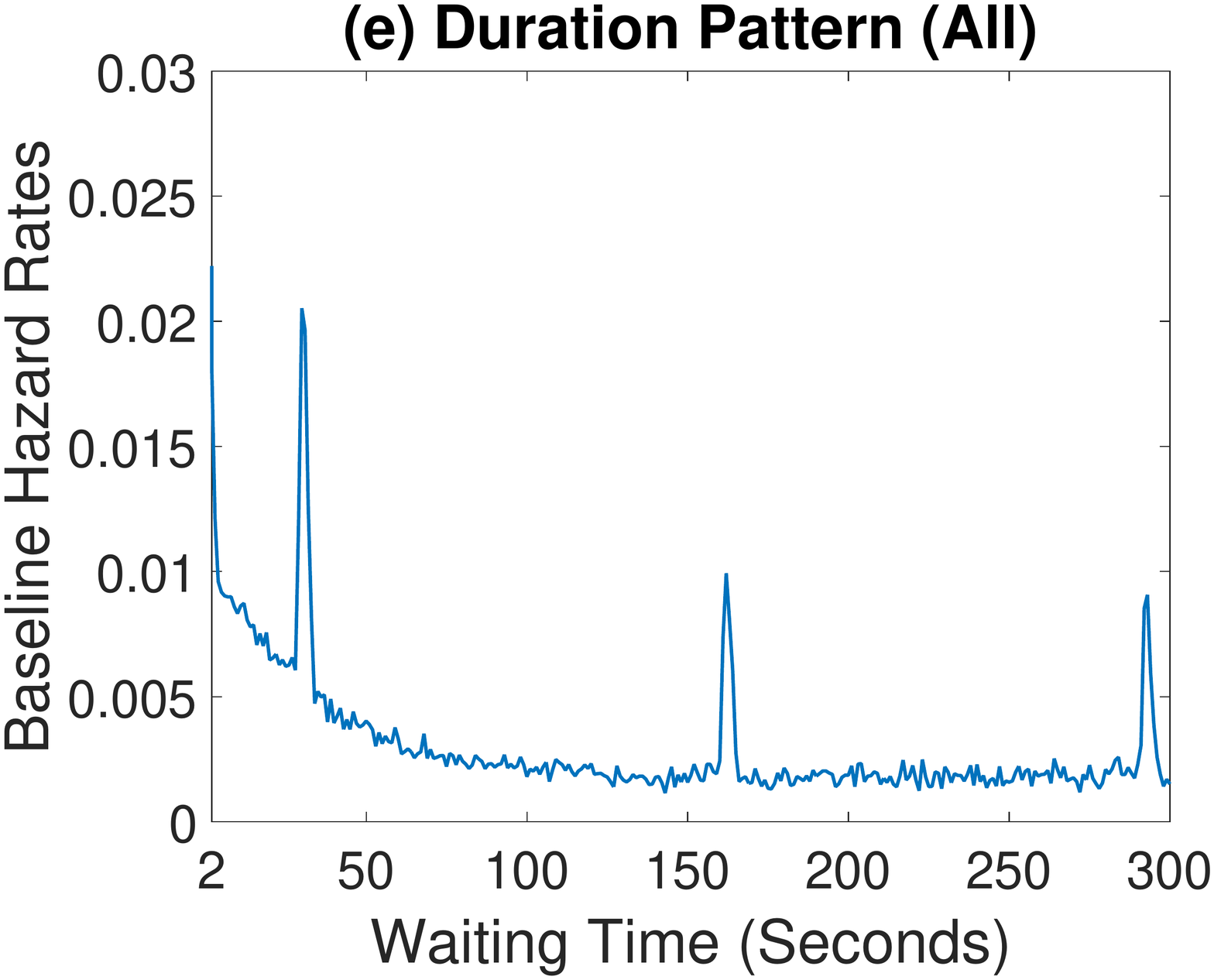}
  \includegraphics[width=2in]{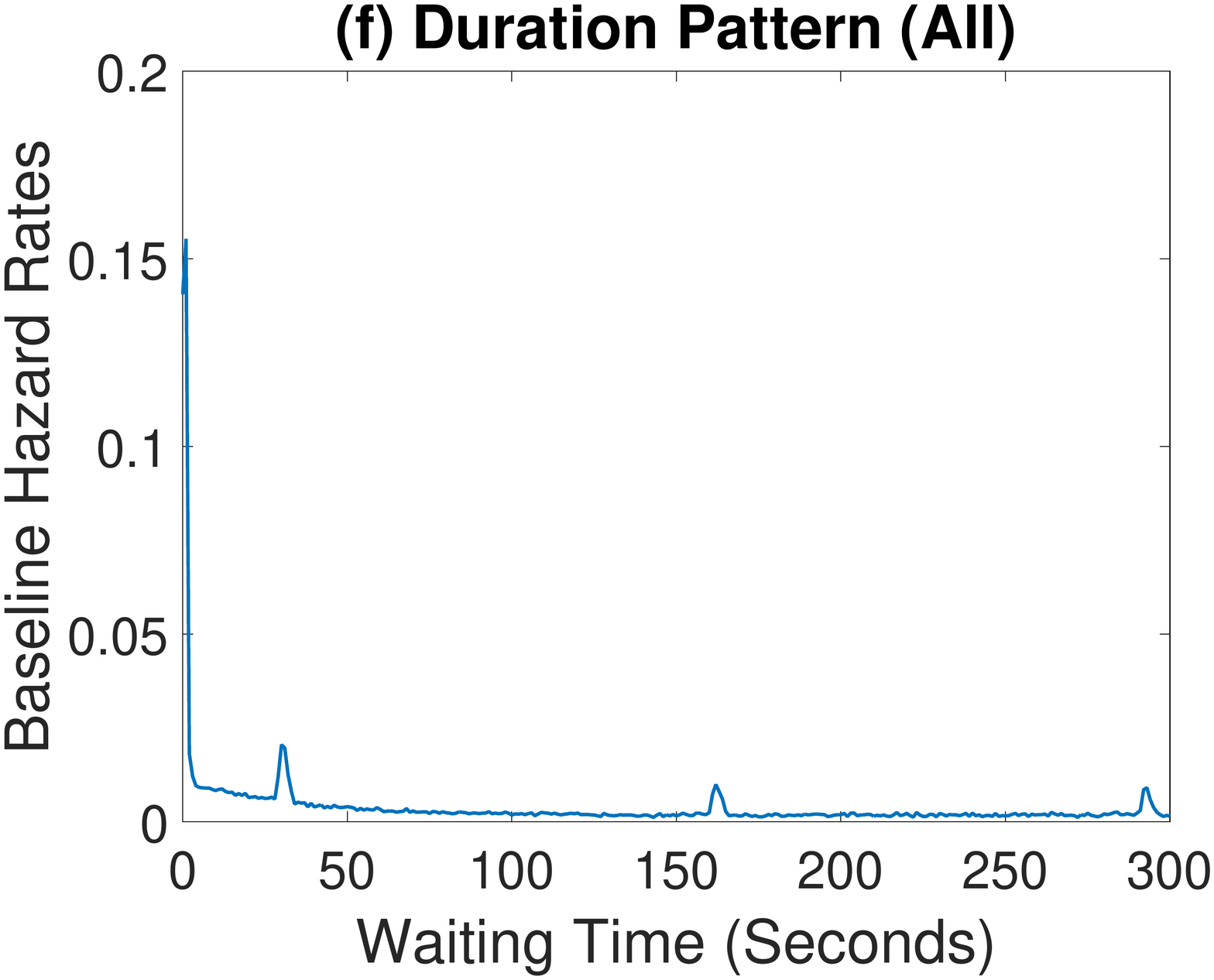}
    \caption{Offered wait study with short waiting calls for the Quick\&Reilly customer group. (a) the time-of-day pattern estimated from the $\geq 2$s data; (b) the time-of-day pattern estimated from the complete data; (c) the proportions of calls answered within 2s at different times of a day;   (d) the waiting pattern estimated from the $\geq 2$s data; (e) the waiting pattern estimated from the complete data (but only showing the pattern starting from 2s); (f) the waiting pattern estimated from the complete data (starting from 0s).}\label{fig:offered}
\end{figure}

We can make the following observations:
\begin{itemize}
\item Panels (a)-(c): Including the calls with short waiting times does affect the intra-day pattern (Panel (a) vs. (b)). However, the pattern shown in Panel (b) is very similar to the pattern of the percentage of calls answered within 2 seconds. As mentioned in the main paper, over 98\% of the calls with waiting times less than 2 seconds were answered immediately. This extreme amount of short waiting calls has distinct hazard rates from the rest of the data. If we incorporate these calls, the intra-day pattern is driven by the percentage of such calls during the day.

\item Panels (d) vs. (e): The baseline hazard rate function remains about the same for waiting time $\ge$ 2 seconds, after including these calls. One can see the same general decay and the peaks caused by the waiting status update.

\item Panel (f): Due to the extremely high percentage of calls with short wait durations, there is a huge spike (approximately 0.15 in magnitude) in the hazard rate near the origin, which obscures the detailed patterns shown in Panel (e).

\end{itemize}

In summary, the proposed method can incorporate those calls with waiting times less than 2 seconds. However, due to the special nature of the data (i.e. almost all these calls are answered immediately), the estimated hazard function for the patience will have small wiggles around the general trend, and the hazard function for the offered wait will have a huge spike near the origin, which makes it difficult to reveal the detailed pattern. As such, we chose to leave out these short-wait calls when estimating the hazard functions in the main paper.

{\color{black}
\section{Analysis of Different Service Groups}
Different service groups typically have different offered wait and customer patience patterns.
In the US bank call center example, there are 17 service types, where 5 groups are dominant with more than 5\% of the total calls, respectively. In particular, the 5 groups are Retail (55.18\%), Quick\&Reilly (11.09\%), CCO (5.79\%), Consumer Loans (5.44\%), and Business (5.23\%). They make up over 82\% of the total calls.
We focus on the Quick\&Reilly group in the main paper. In this section, we further study the remain 4 major groups.

We need to specify the domains and intervals for waiting time and time of day, respectively, in the study of each service group. Similar to the Quick\&Reilly study, we set the waiting-time interval length to be 1 second, and the time-of-day interval length to be 15 minutes. To avoid the effect of short waiting calls, we always set the start waiting time to be 2 seconds. The end waiting time, and the start and end of the time-of-day intervals are chosen on a case-by-case basis. Figures \ref{fig:retail}--\ref{fig:business} show the histograms of observed waiting times (in seconds) and arrival rates (in 15-minute intervals) of calls with waiting time $\ge 2$ seconds.
We particularly set 200 seconds to be the end waiting time for the Retail group, 300 seconds for the CCO group, and 120 seconds for the Consumer Loans and Business groups.
In addition, we concentrate on the 7am--8pm time-of-day interval for the Retail group, 8am--5pm for the Business group, and 8am--8pm for the CCO and Consumer Loans groups.

\begin{figure}[htbp]
\begin{center}
\includegraphics[width=2.4in]{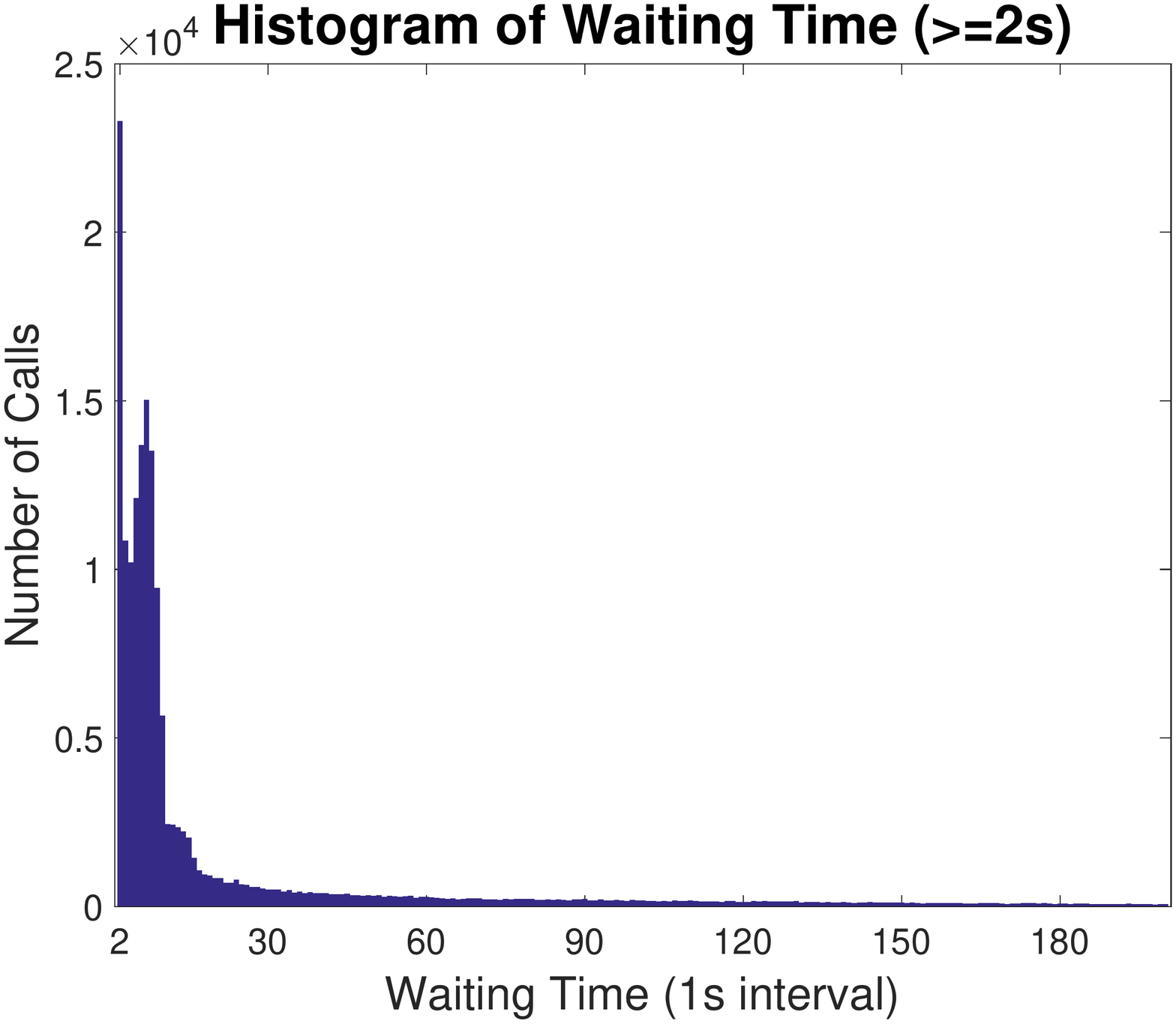}
\includegraphics[width=2.4in]{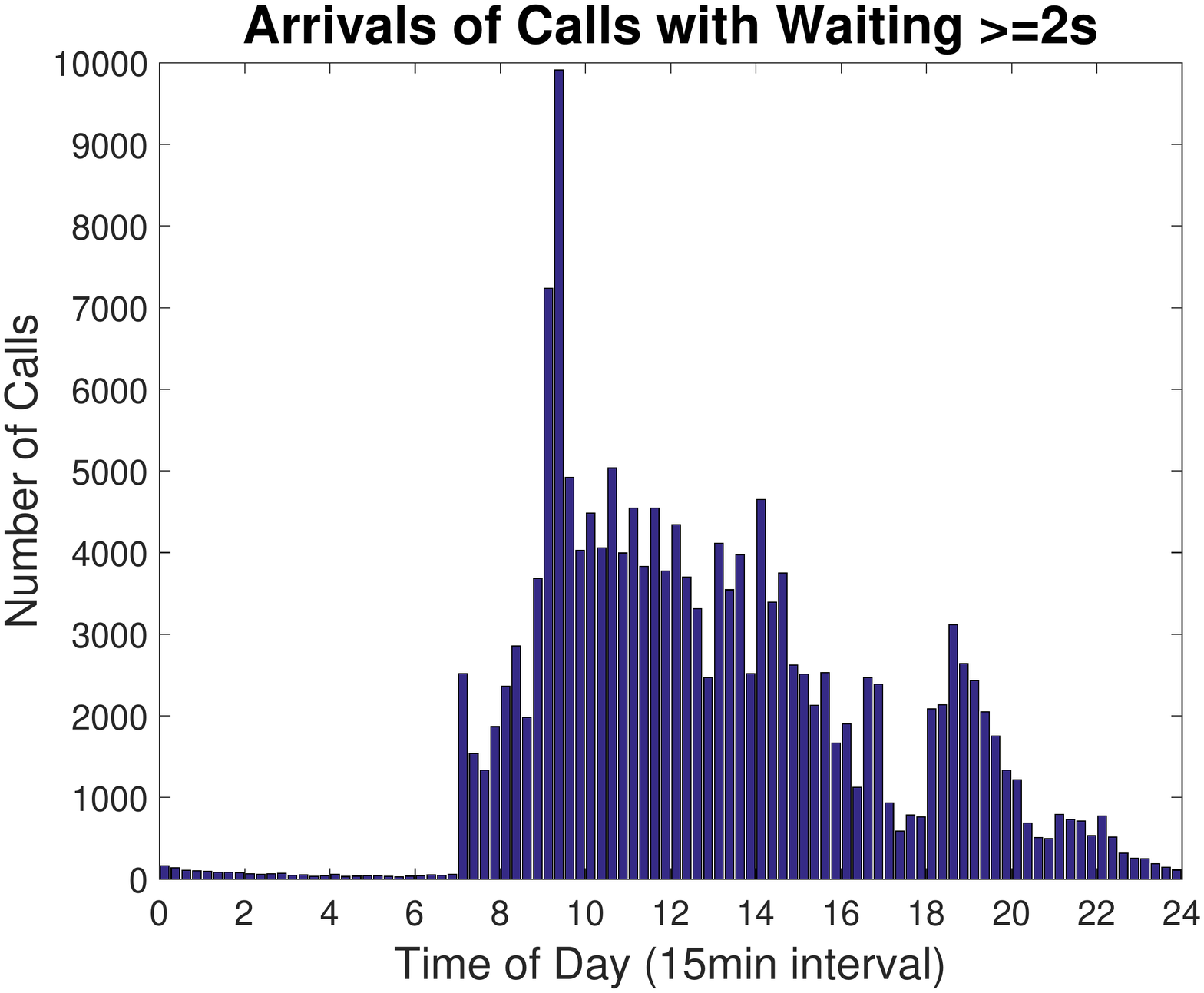}
\end{center}
\vskip-.1in
\caption{Retail service group. Left: the histogram of observed waiting times (in seconds) for calls with waiting time between 2 and 200 seconds. Right: the histogram of arrival rates (in 15-minute intervals) of calls with waiting time $\ge 2$ seconds.}
\label{fig:retail}
\end{figure}

\begin{figure}[htbp]
\begin{center}
\includegraphics[width=2.4in]{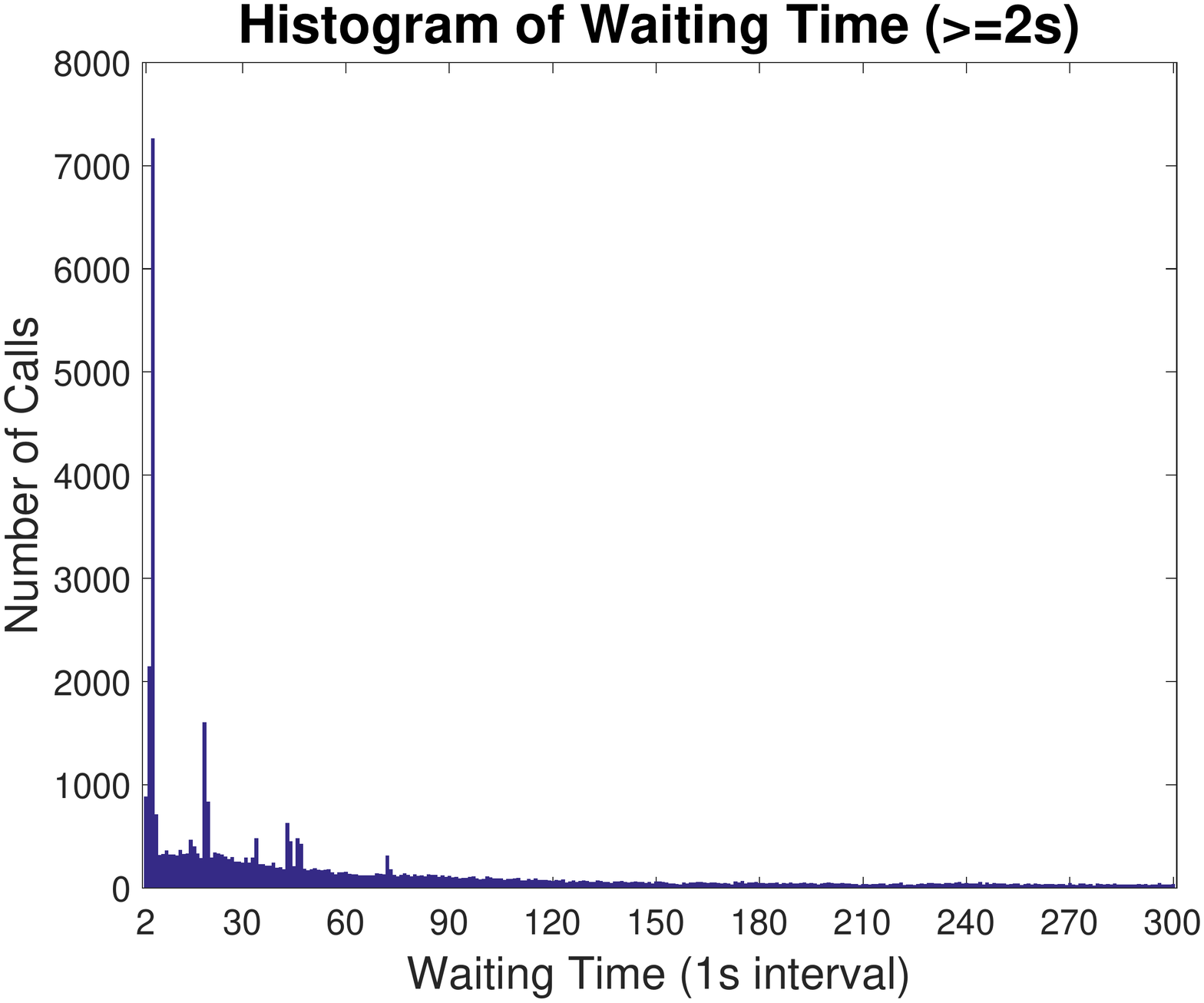}
\includegraphics[width=2.4in]{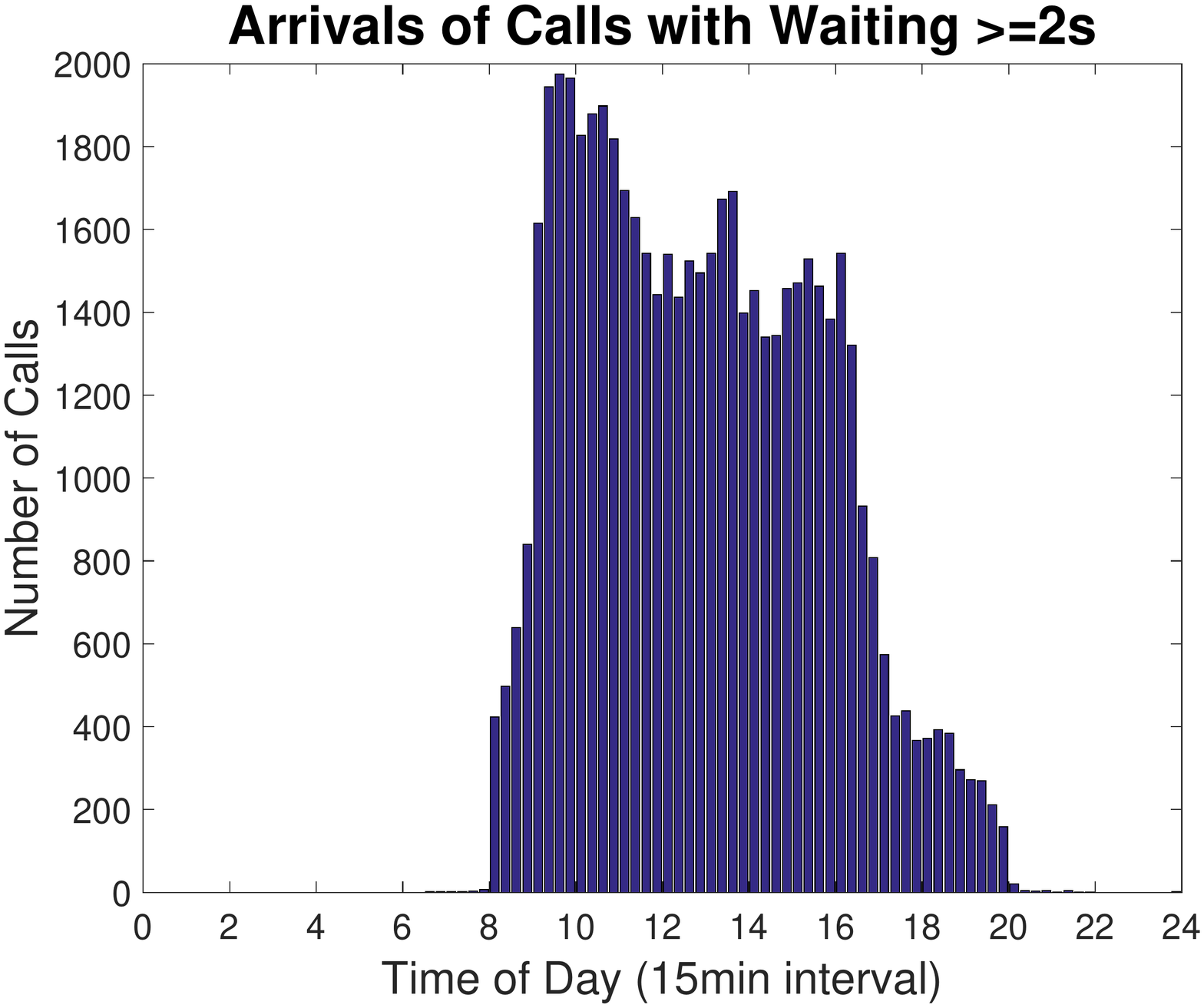}
\end{center}
\vskip-.1in
\caption{CCO service group. Left: the histogram of observed waiting times (in seconds) for calls with waiting time between 2 and 300 seconds. Right: the histogram of arrival rates (in 15-minute intervals) of calls with waiting time $\ge 2$ seconds.}
\label{fig:cco}
\end{figure}

\begin{figure}[htbp]
\begin{center}
\includegraphics[width=2.4in]{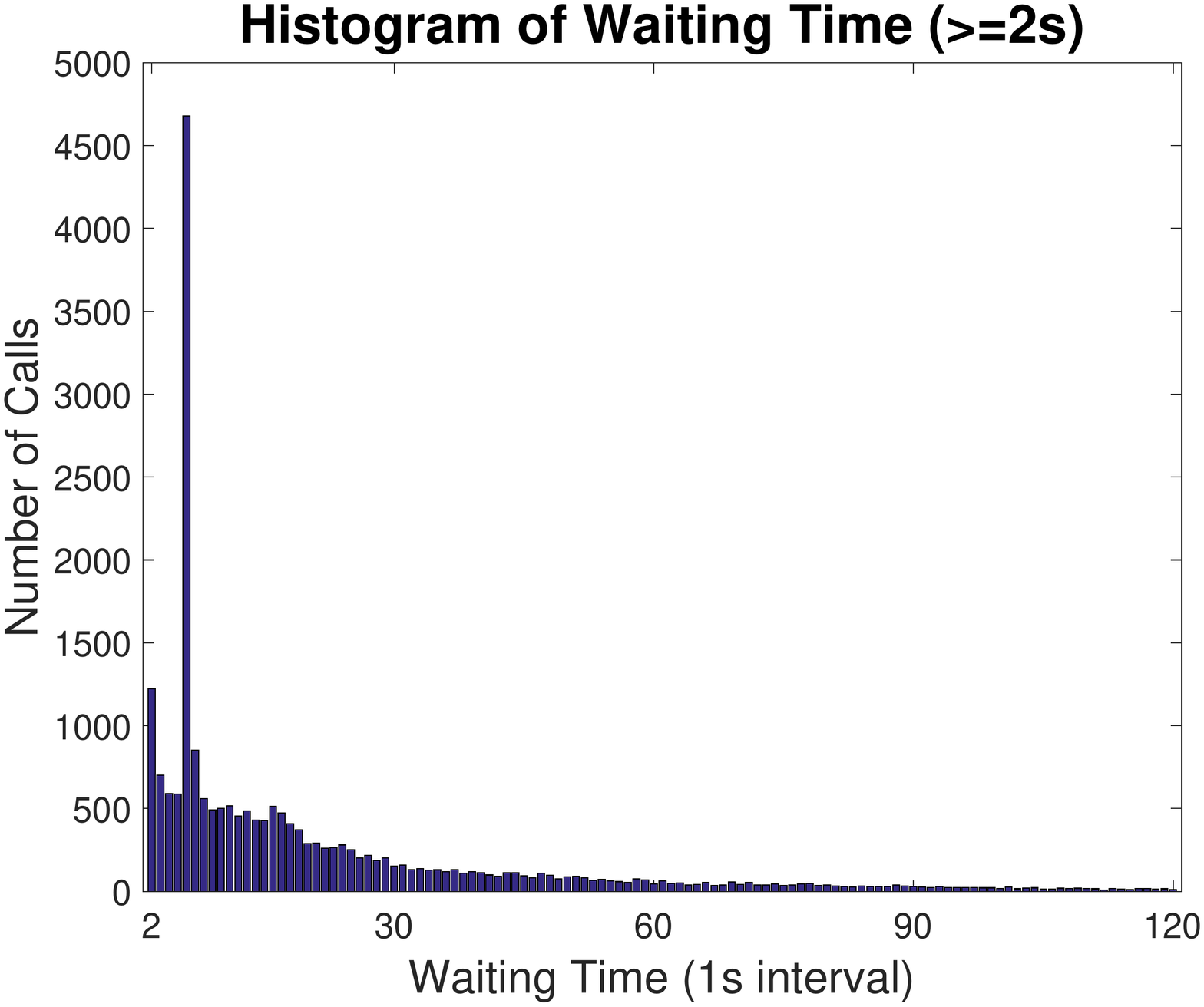}
\includegraphics[width=2.4in]{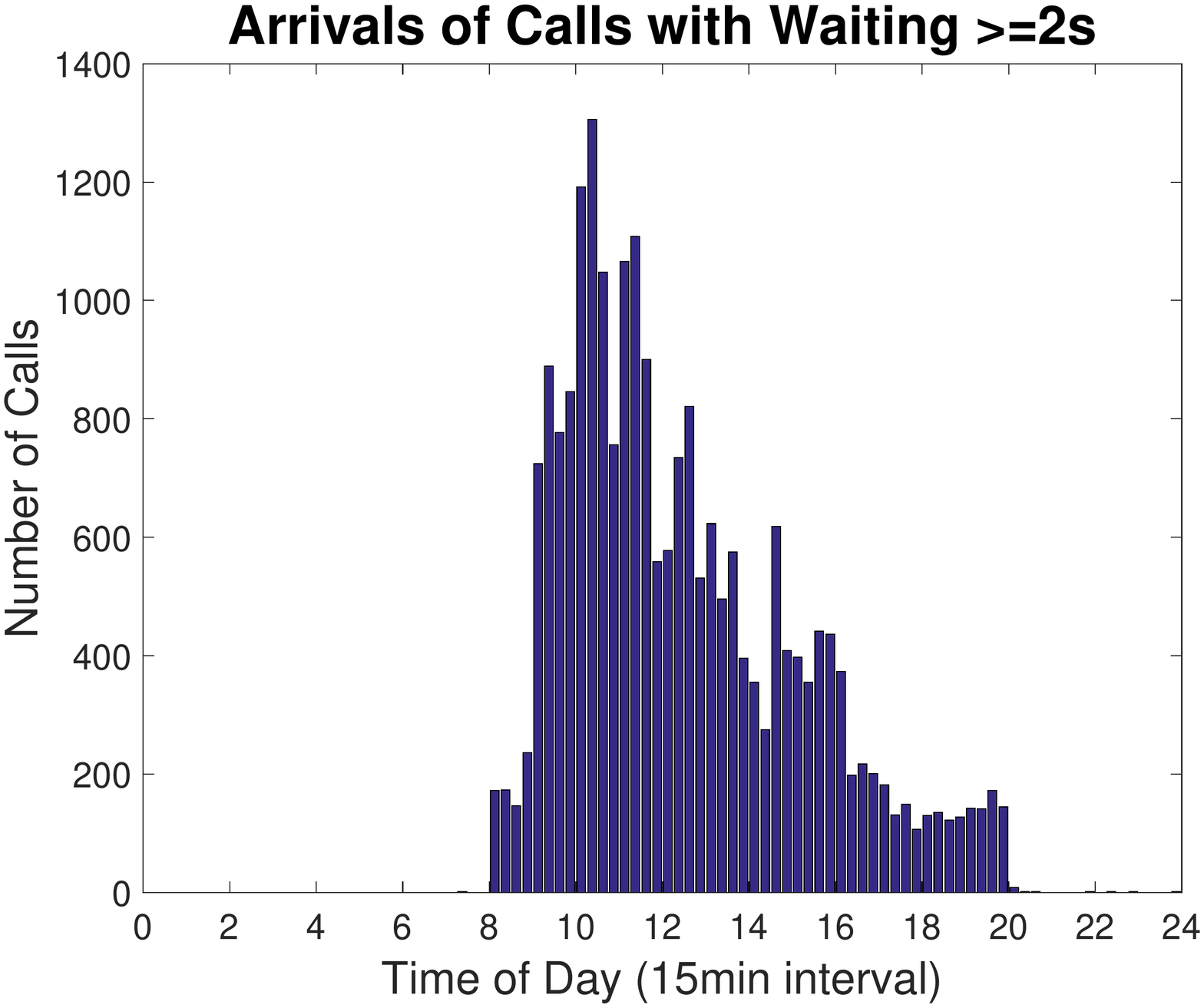}
\end{center}
\vskip-.1in
\caption{Consumer Loans service group. Left: the histogram of observed waiting times (in seconds) for calls with waiting time between 2 and 120 seconds. Right: the histogram of arrival rates (in 15-minute intervals) of calls with waiting time $\ge 2$ seconds.}
\label{fig:loan}
\end{figure}

\begin{figure}[htbp]
\begin{center}
\includegraphics[width=2.4in]{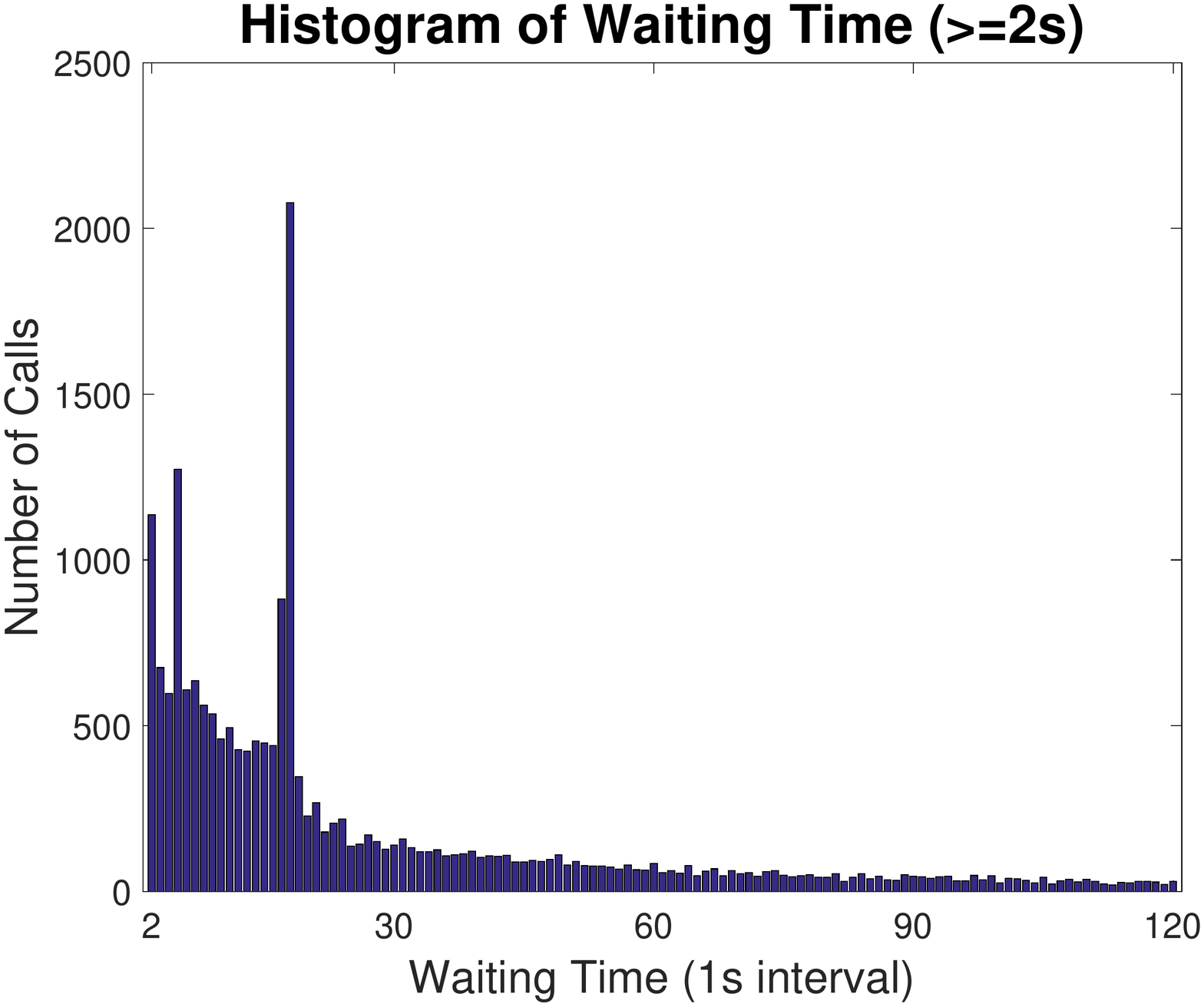}
\includegraphics[width=2.4in]{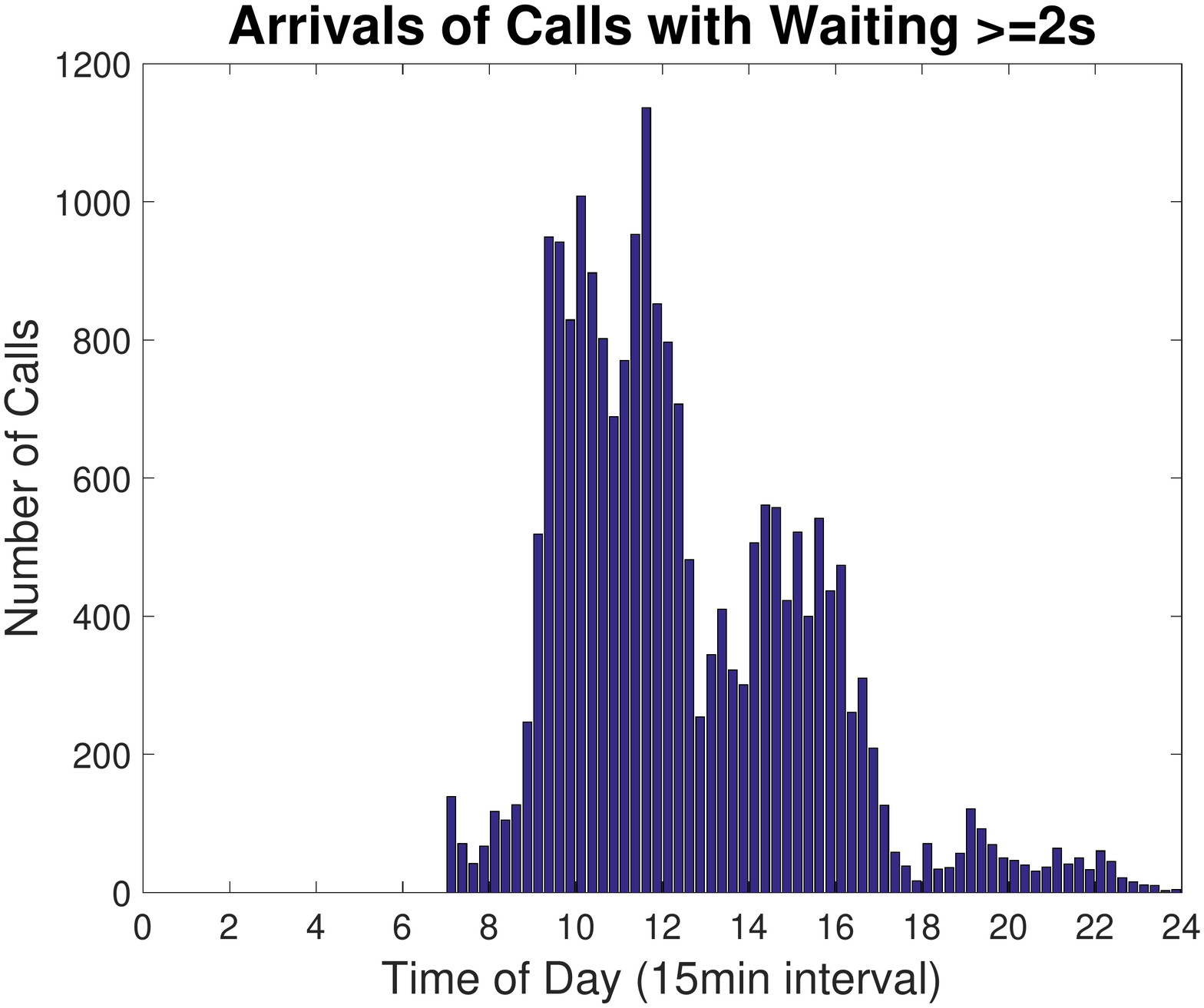}
\end{center}
\vskip-.1in
\caption{Business service group. Left: the histogram of observed waiting times (in seconds) for calls with waiting time between 2 and 120 seconds. Right: the histogram of arrival rates (in 15-minute intervals) of calls with waiting time $\ge 2$ seconds.}
\label{fig:business}
\end{figure}

We first apply the tF-Hazards method (with unit rank) to analyze the customer patience patterns in different groups.
The duration patterns are shown in Figure \ref{fig:dur_p}.
As a reference, we also provide hazard functions estimated from the pooled data across different times of a data. The estimates from both methods are large consistent within each service group.
The time-of-day patterns are shown in Figure \ref{fig:tod_p}.
Apparently, different service groups have distinct patience patterns both along the waiting time direction and across different times of a day.
In particular, along the waiting-time direction, we generally do not observe periodic peaks in hazard functions as in the Quick\&Reilly study, possibly due to different on-hold strategies.
The specific patterns for each service group call for further investigation.

\begin{figure}[htbp]
\begin{center}
\includegraphics[width=3in]{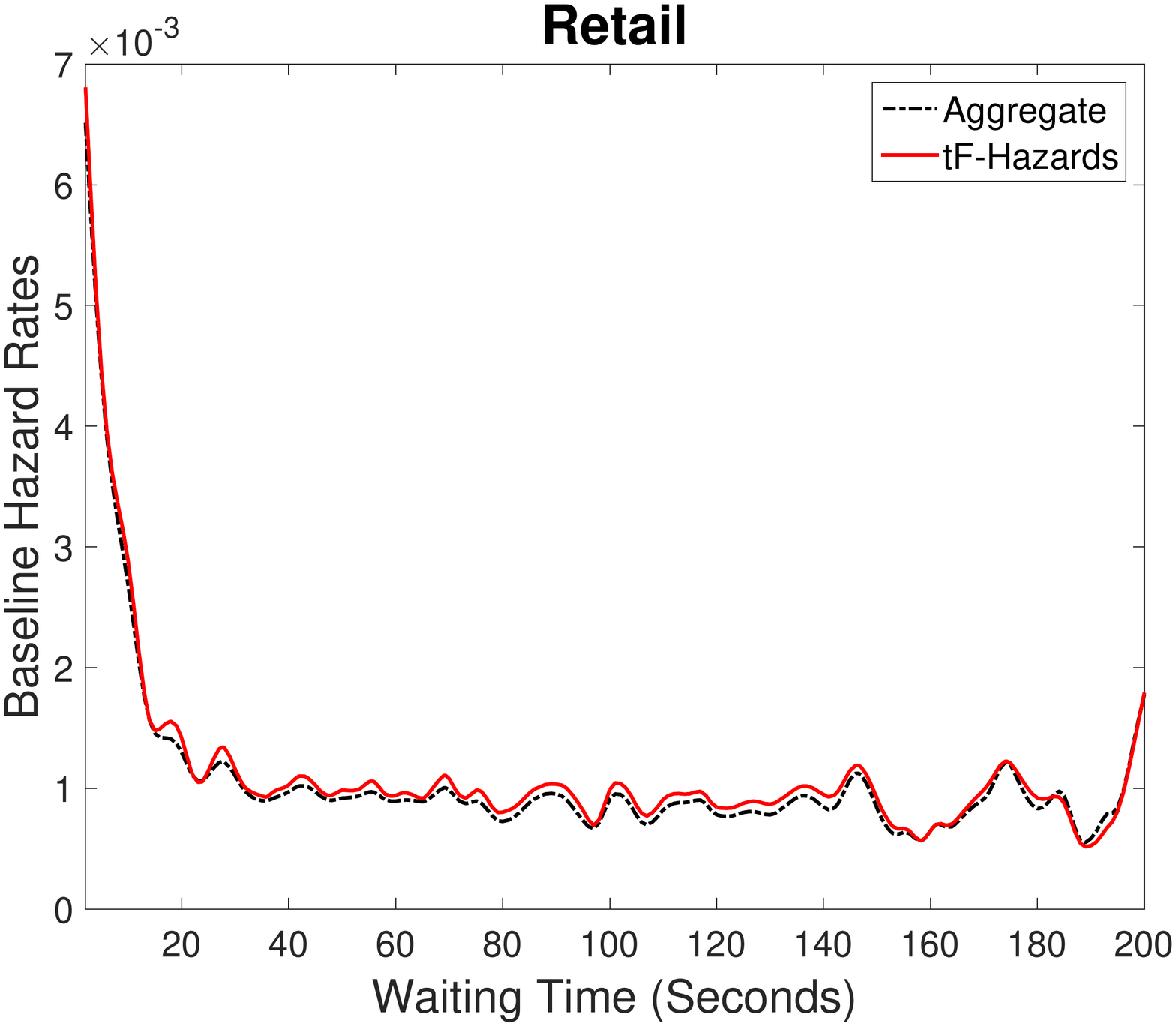}
\includegraphics[width=3in]{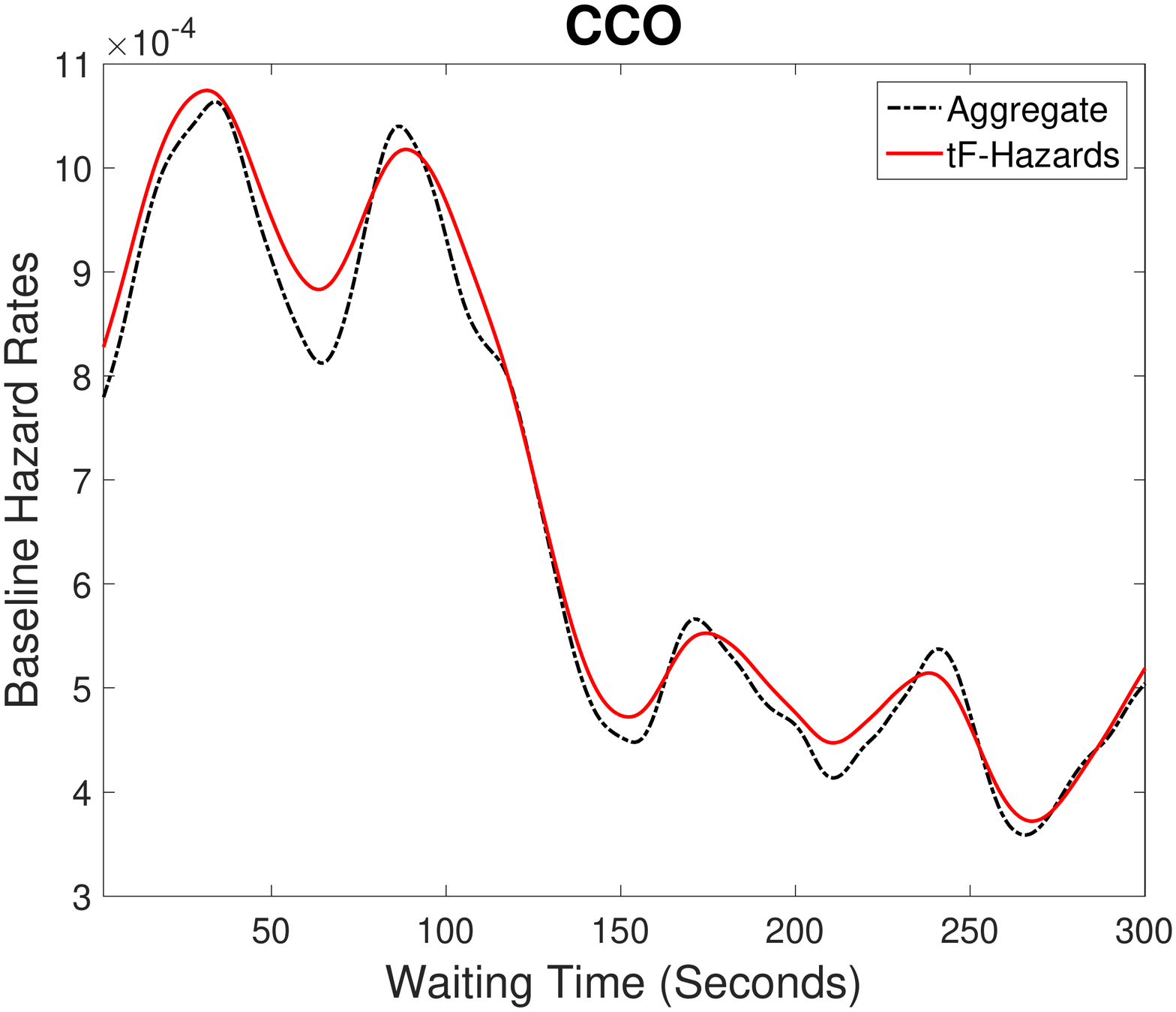}
\includegraphics[width=3in]{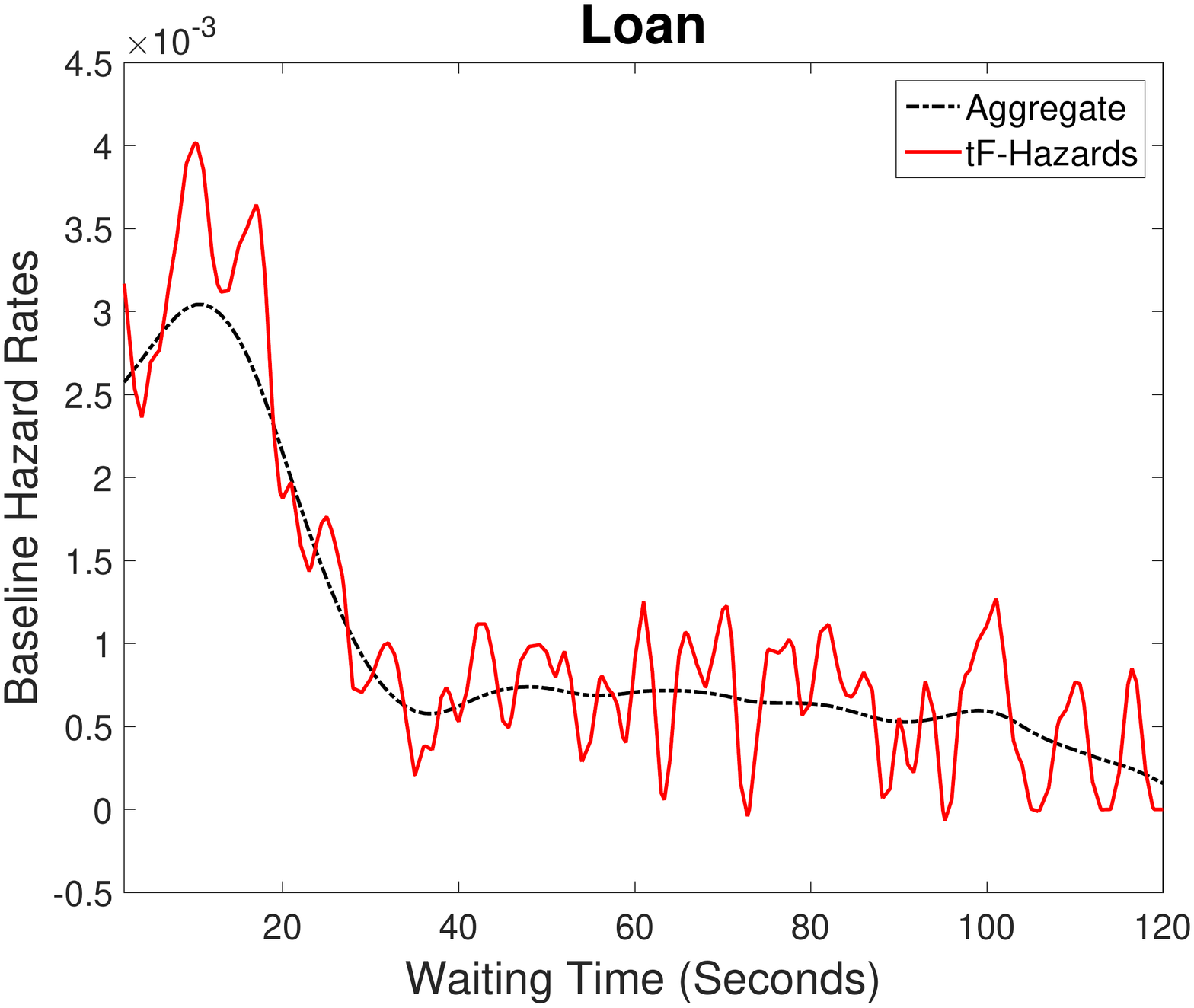}
\includegraphics[width=3in]{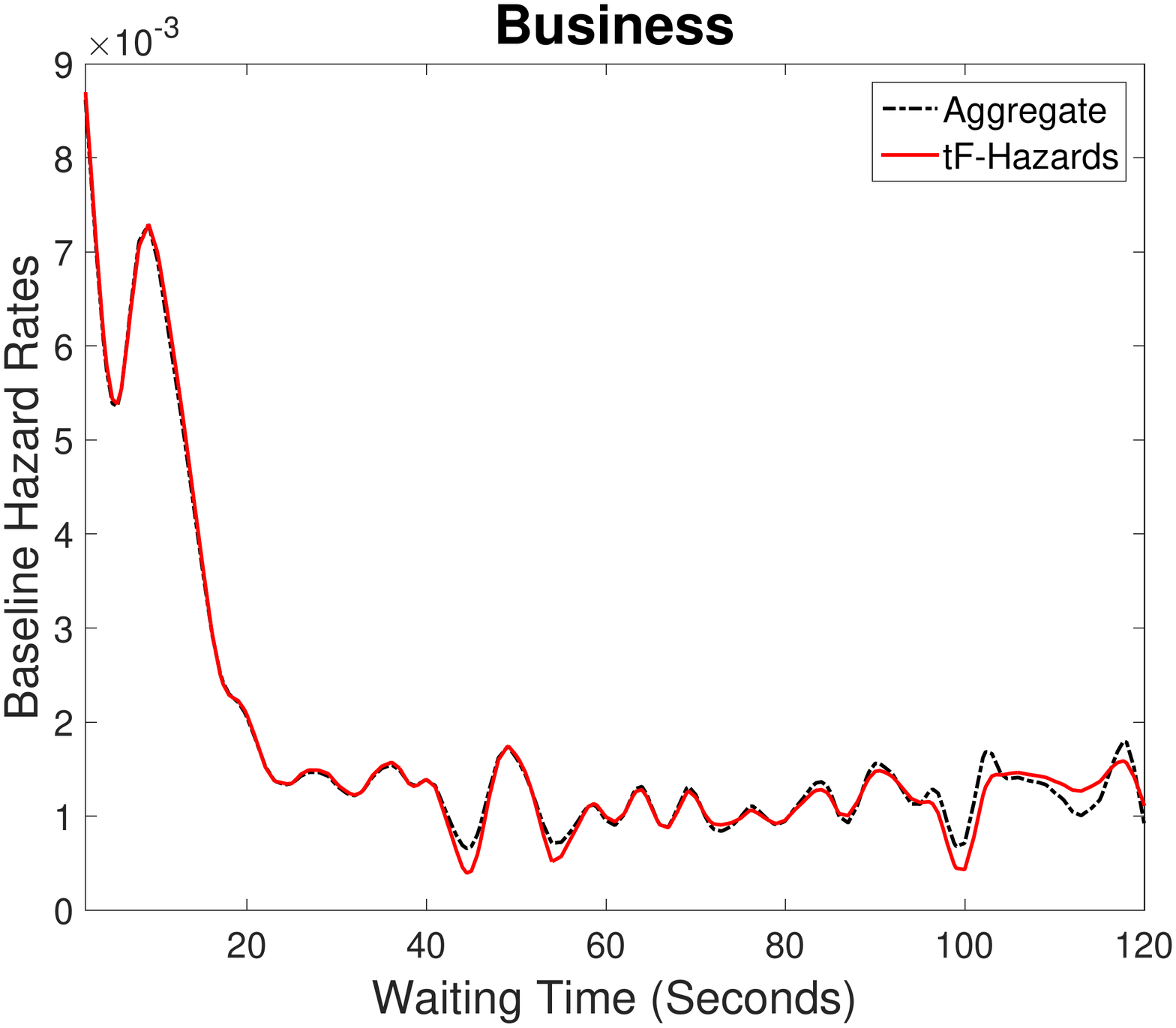}
\end{center}
\vskip-.1in
\caption{Duration patterns of customer patience in different service groups. Red lines are hazard functions estimated from the tF-Hazards method; black dashed lines are aggregated hazard functions estimated from the pooled data.}
\label{fig:dur_p}
\end{figure}

\begin{figure}[htbp]
\begin{center}
\includegraphics[width=3in]{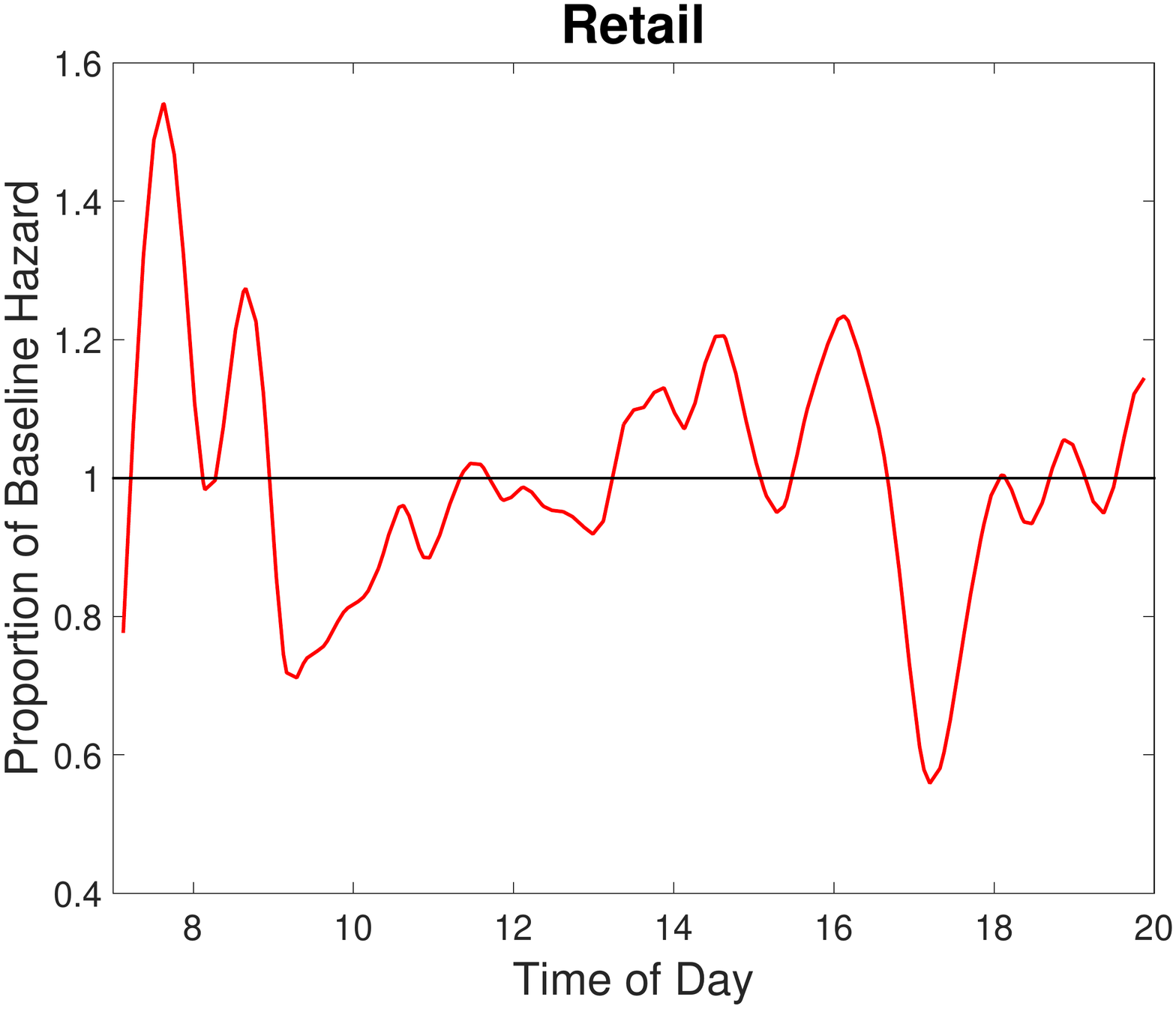}
\includegraphics[width=3in]{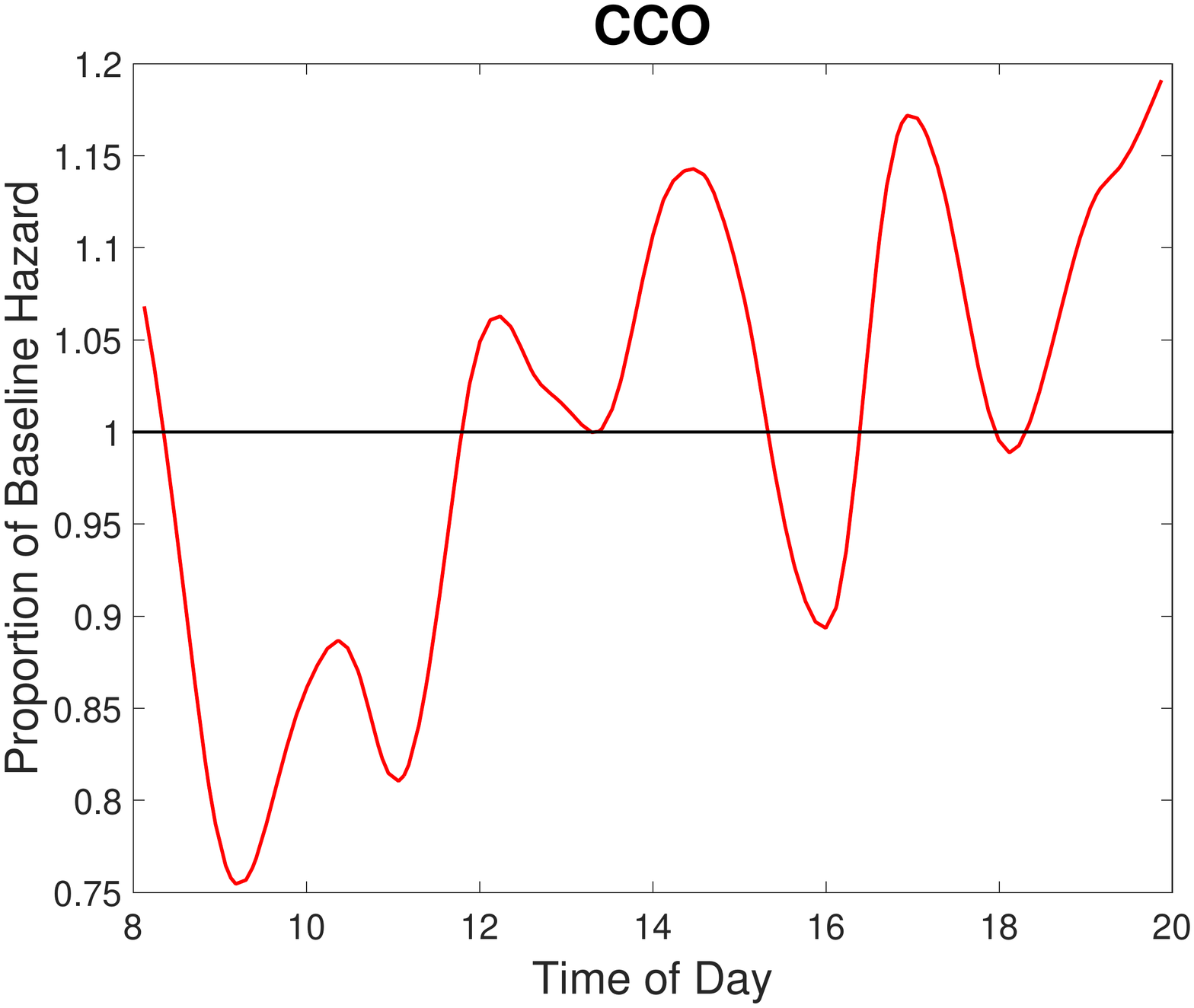}
\includegraphics[width=3in]{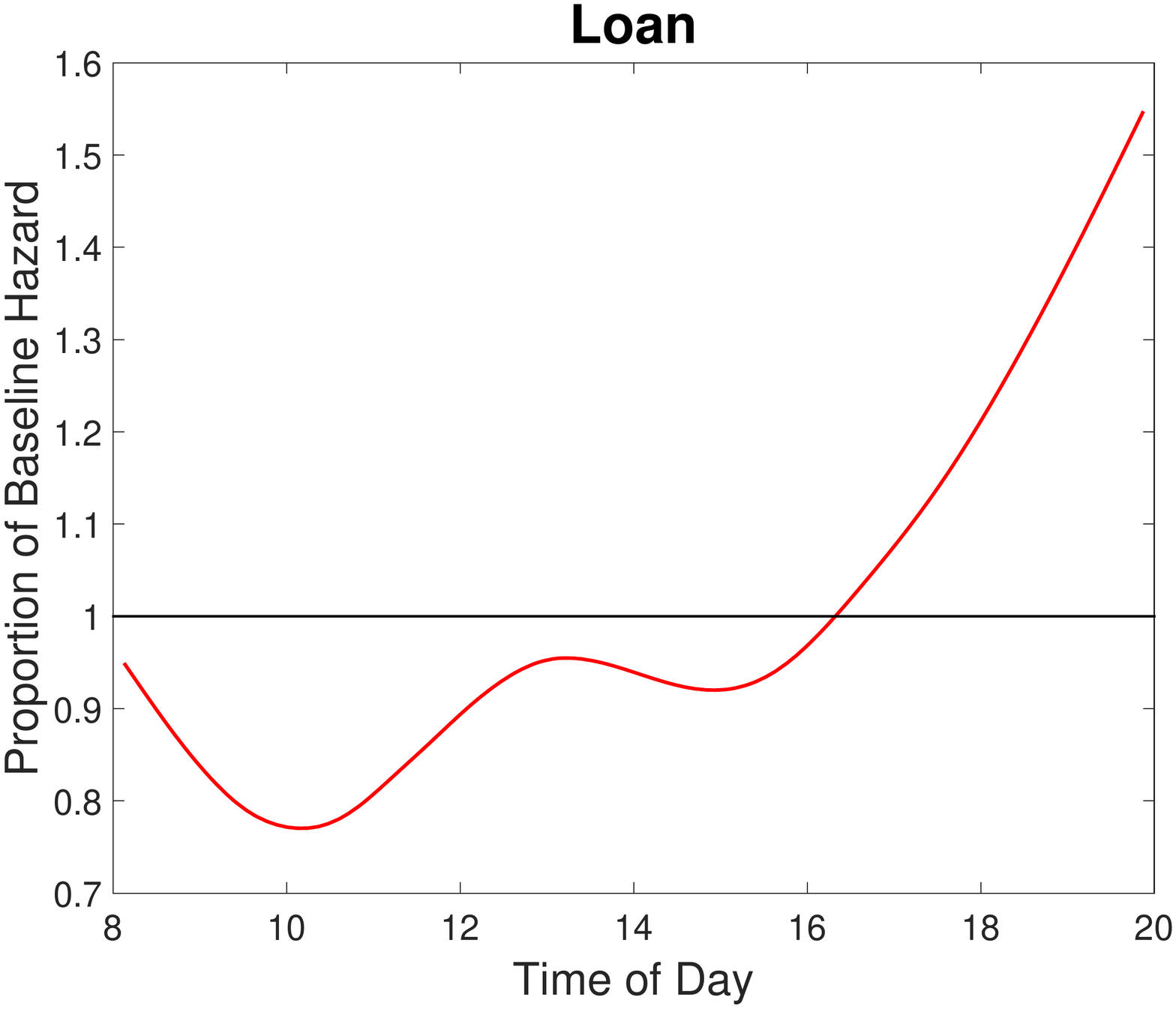}
\includegraphics[width=3in]{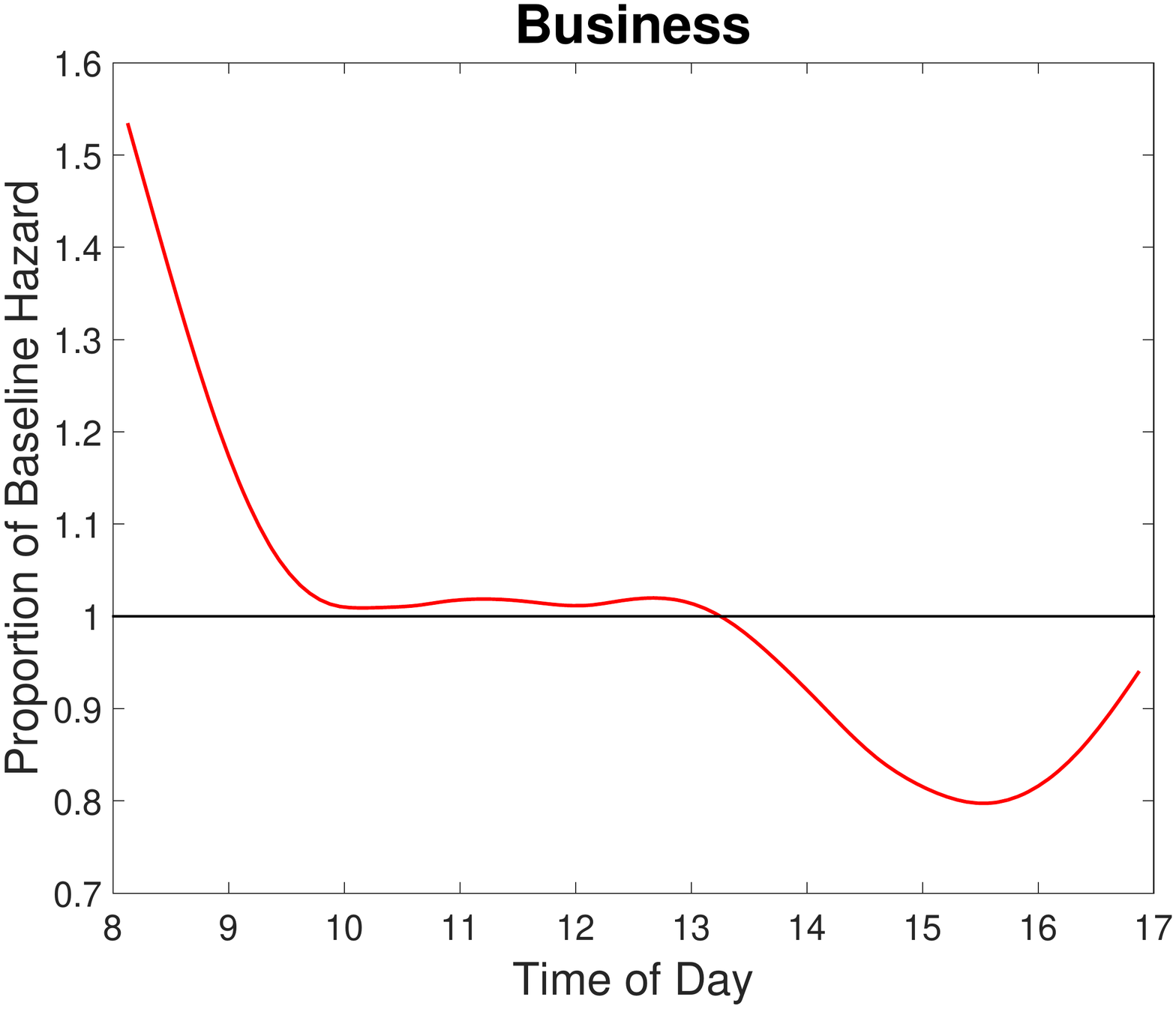}
\end{center}
\vskip-.1in
\caption{Time-of-day patterns of customer patience in different service groups.}
\label{fig:tod_p}
\end{figure}

We also apply the proposed method with unit rank to analyze the offered wait in different groups.
The duration and time-of-day patterns are shown in Figures \ref{fig:dur_w} and \ref{fig:tod_w}.
Similarly, different service groups present distinct features that require further studies.

\begin{figure}[htbp]
\begin{center}
\includegraphics[width=3in]{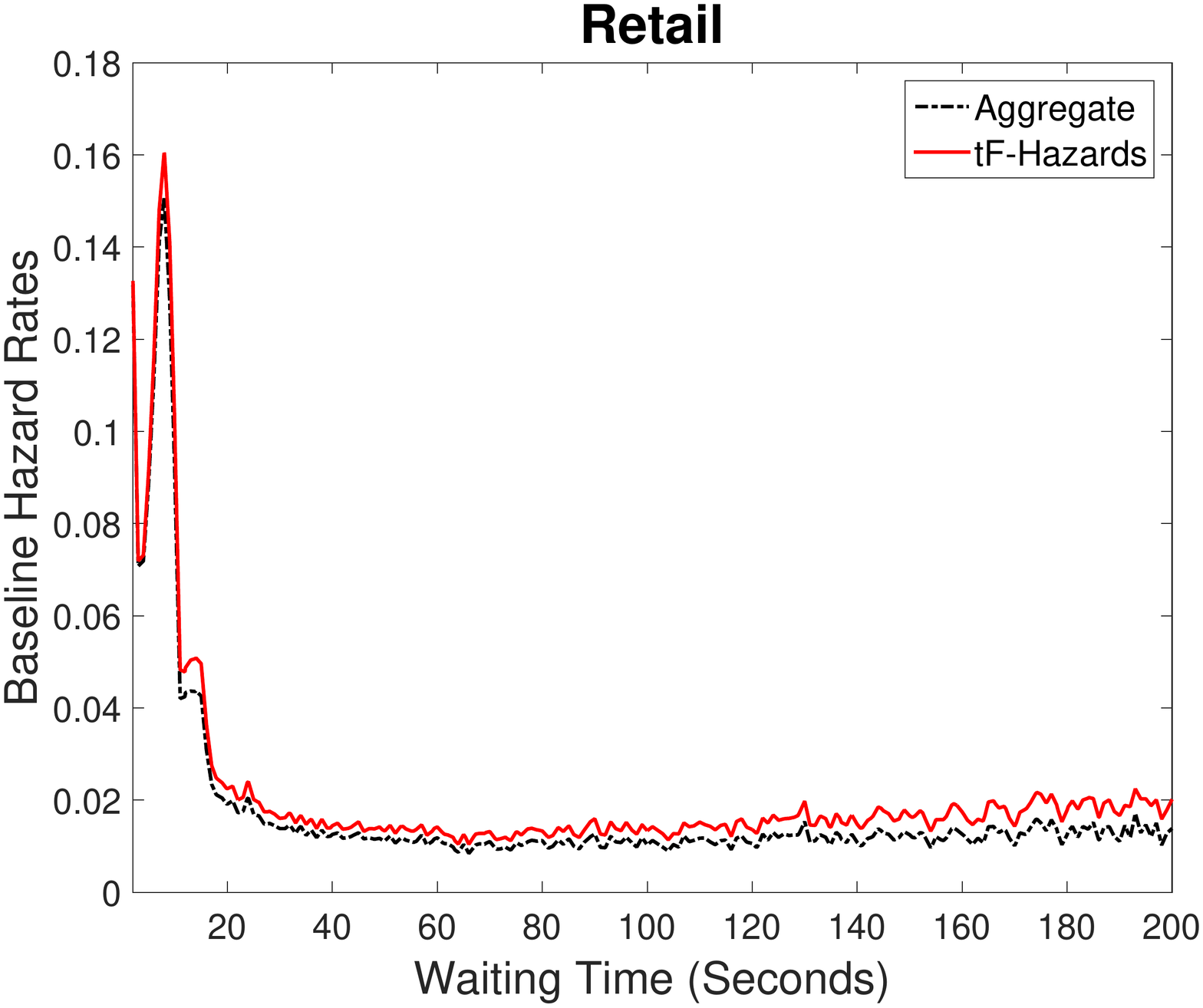}
\includegraphics[width=3in]{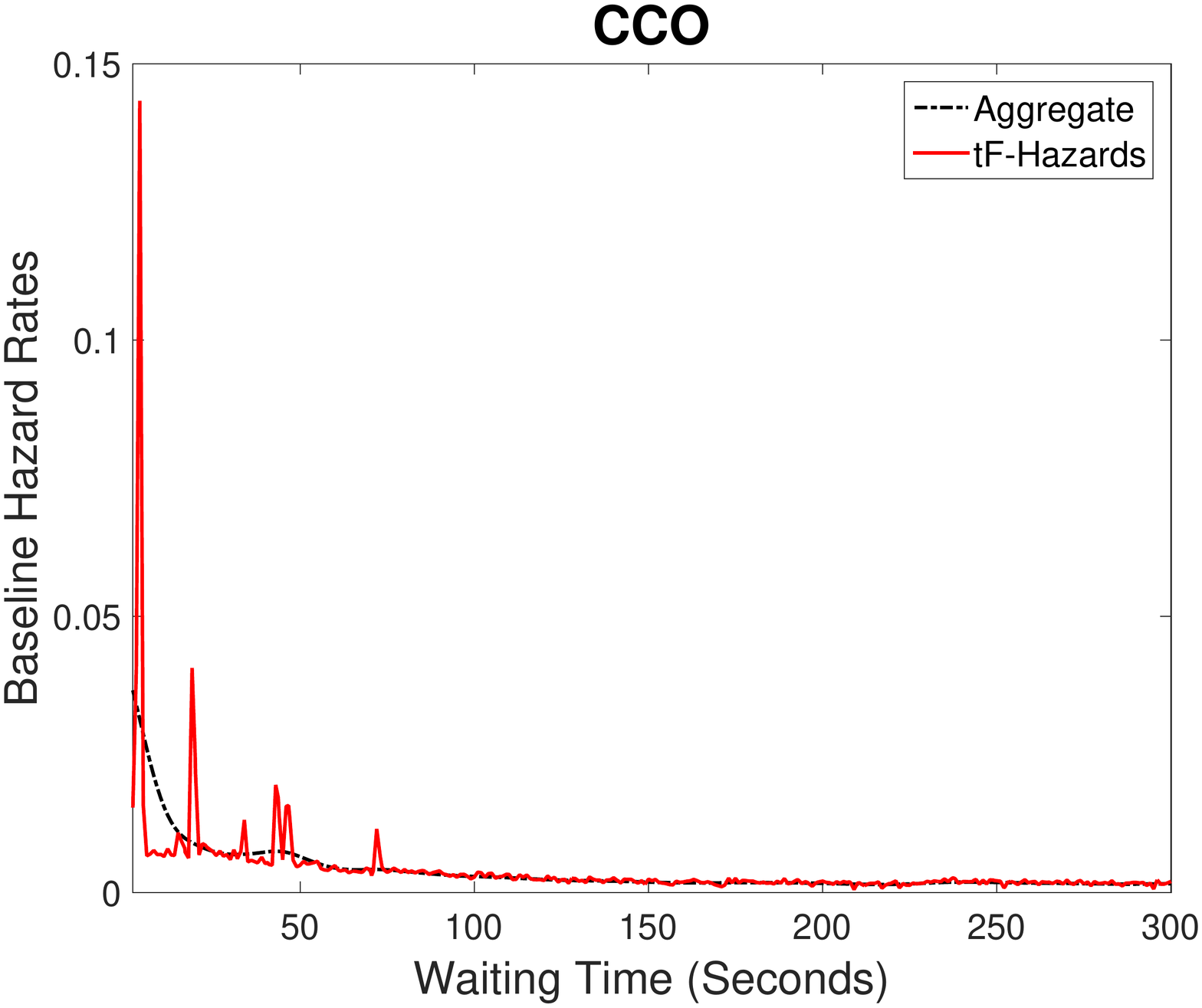}
\includegraphics[width=3in]{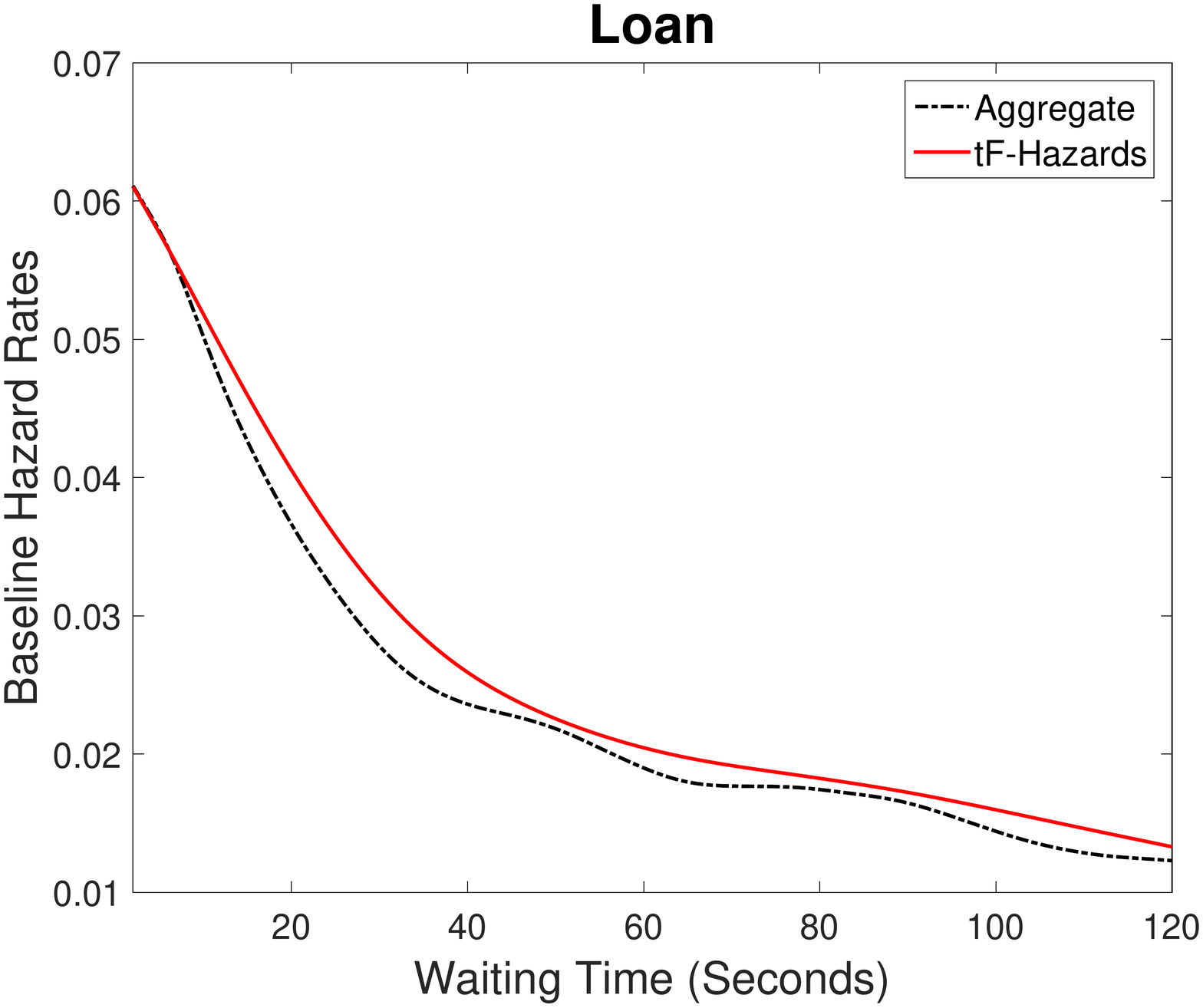}
\includegraphics[width=3in]{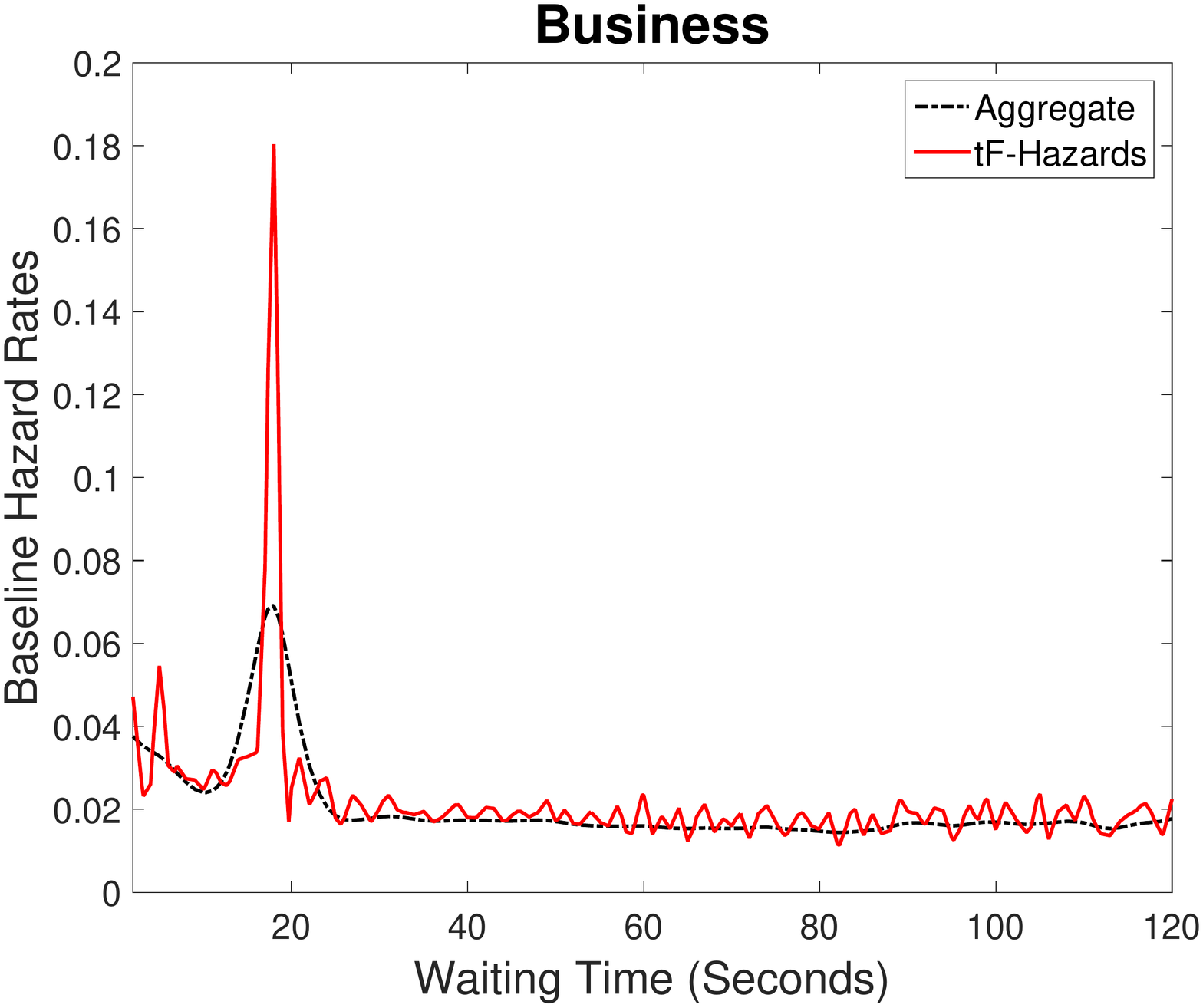}
\end{center}
\vskip-.1in
\caption{Duration patterns of offered wait in different service groups. Red lines are hazard functions estimated from the tF-Hazards method; black dashed lines are aggregated hazard functions estimated from the pooled data.}
\label{fig:dur_w}
\end{figure}

\begin{figure}[htbp]
\begin{center}
\includegraphics[width=3in]{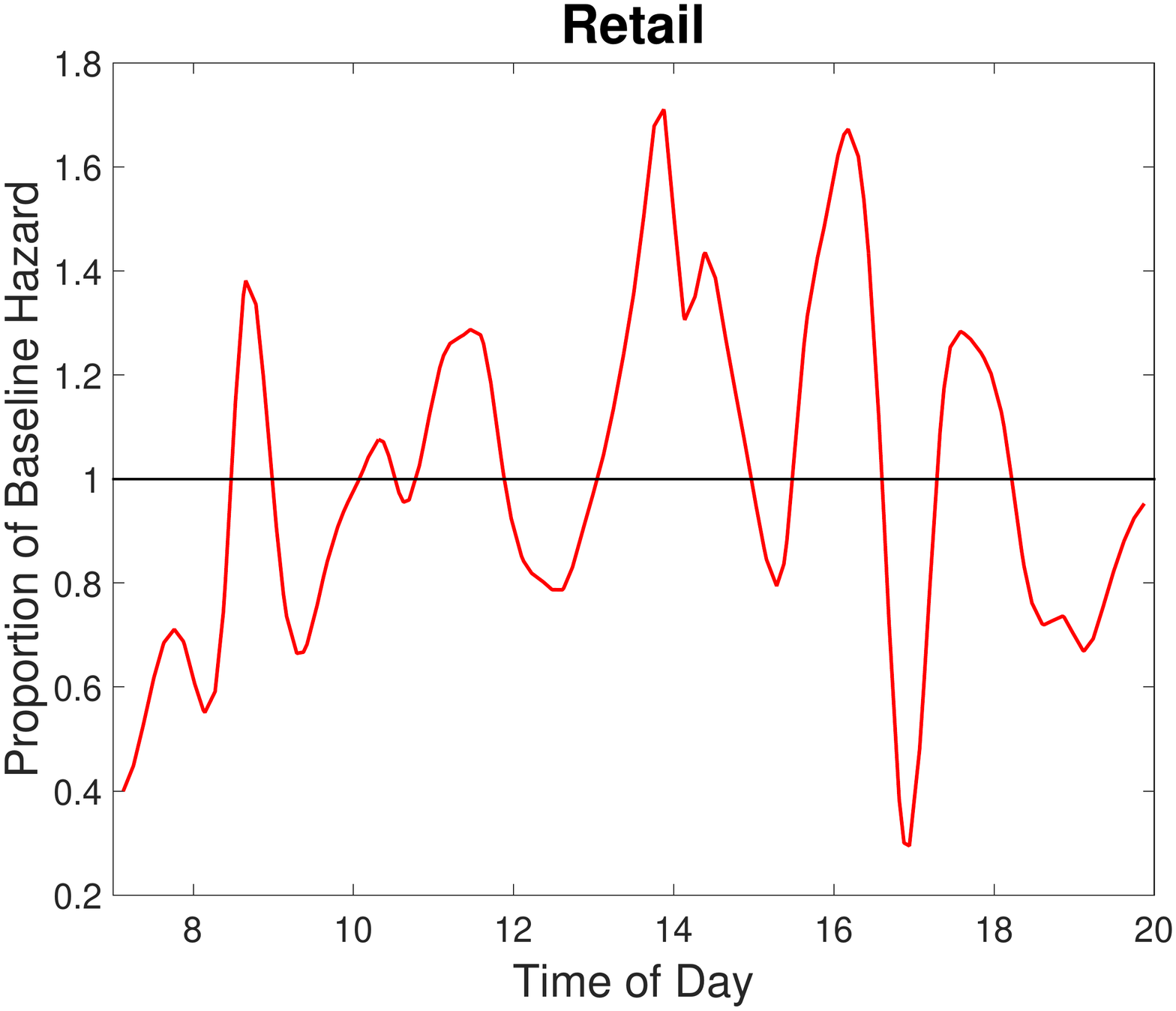}
\includegraphics[width=3in]{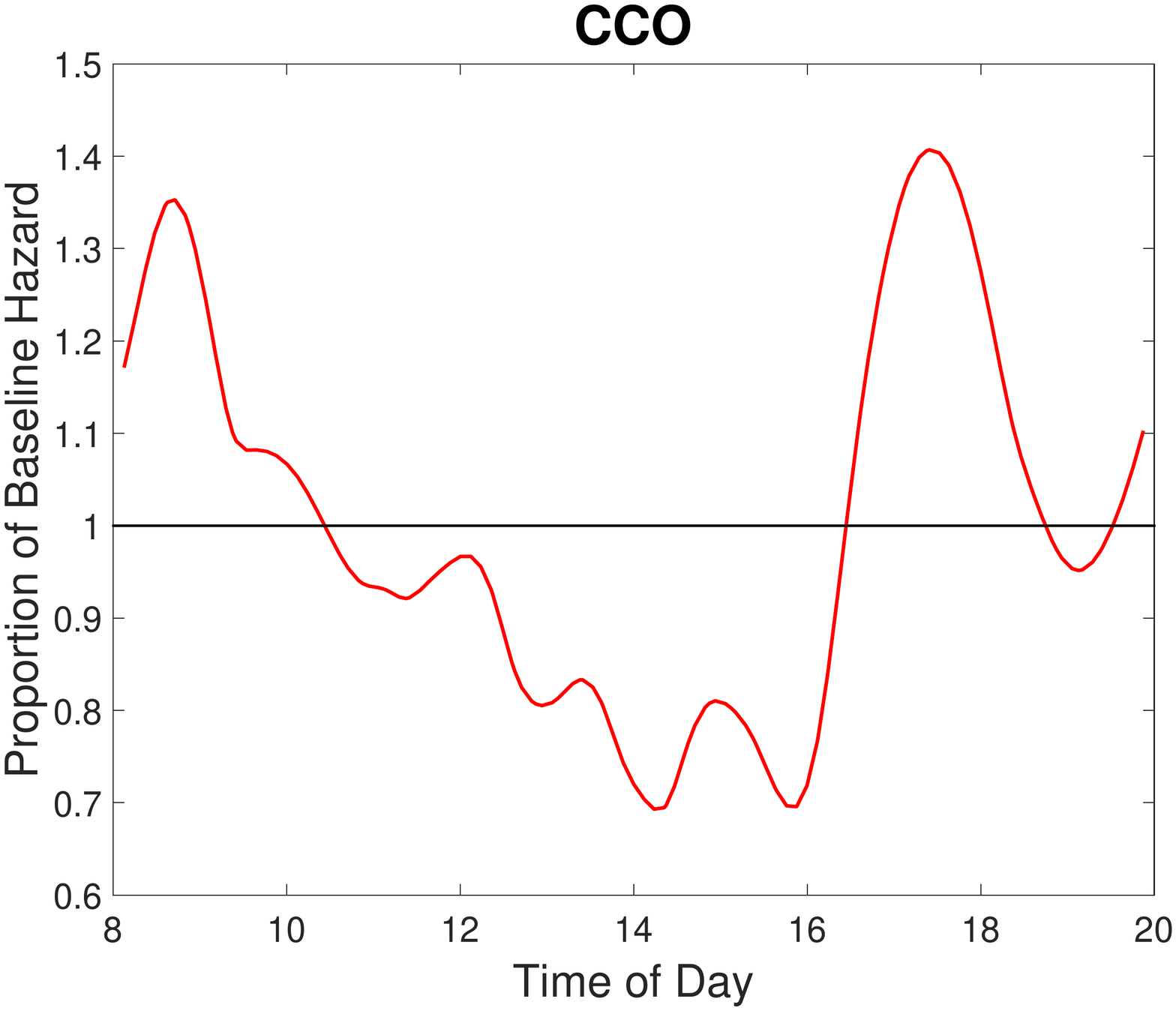}
\includegraphics[width=3in]{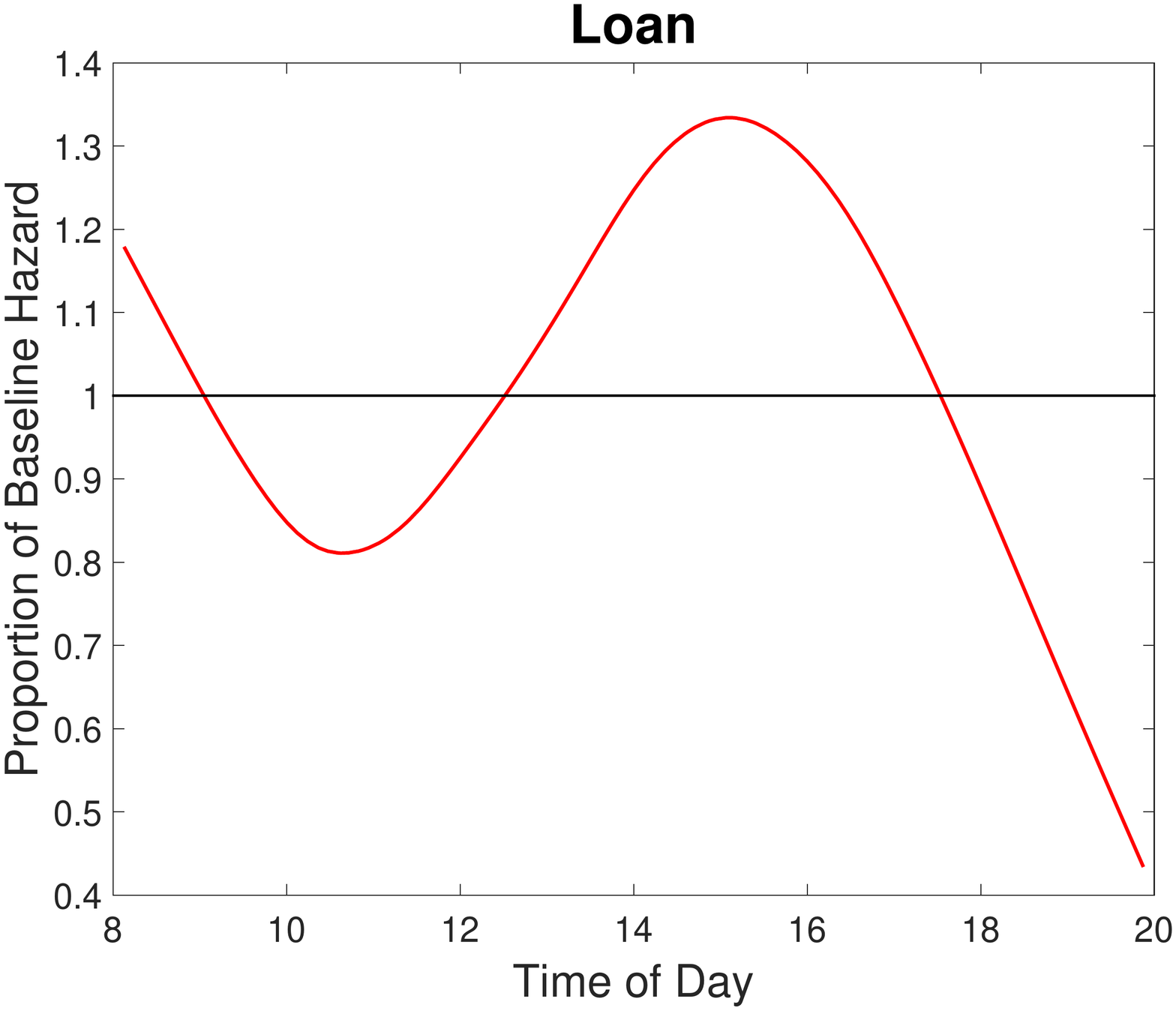}
\includegraphics[width=3in]{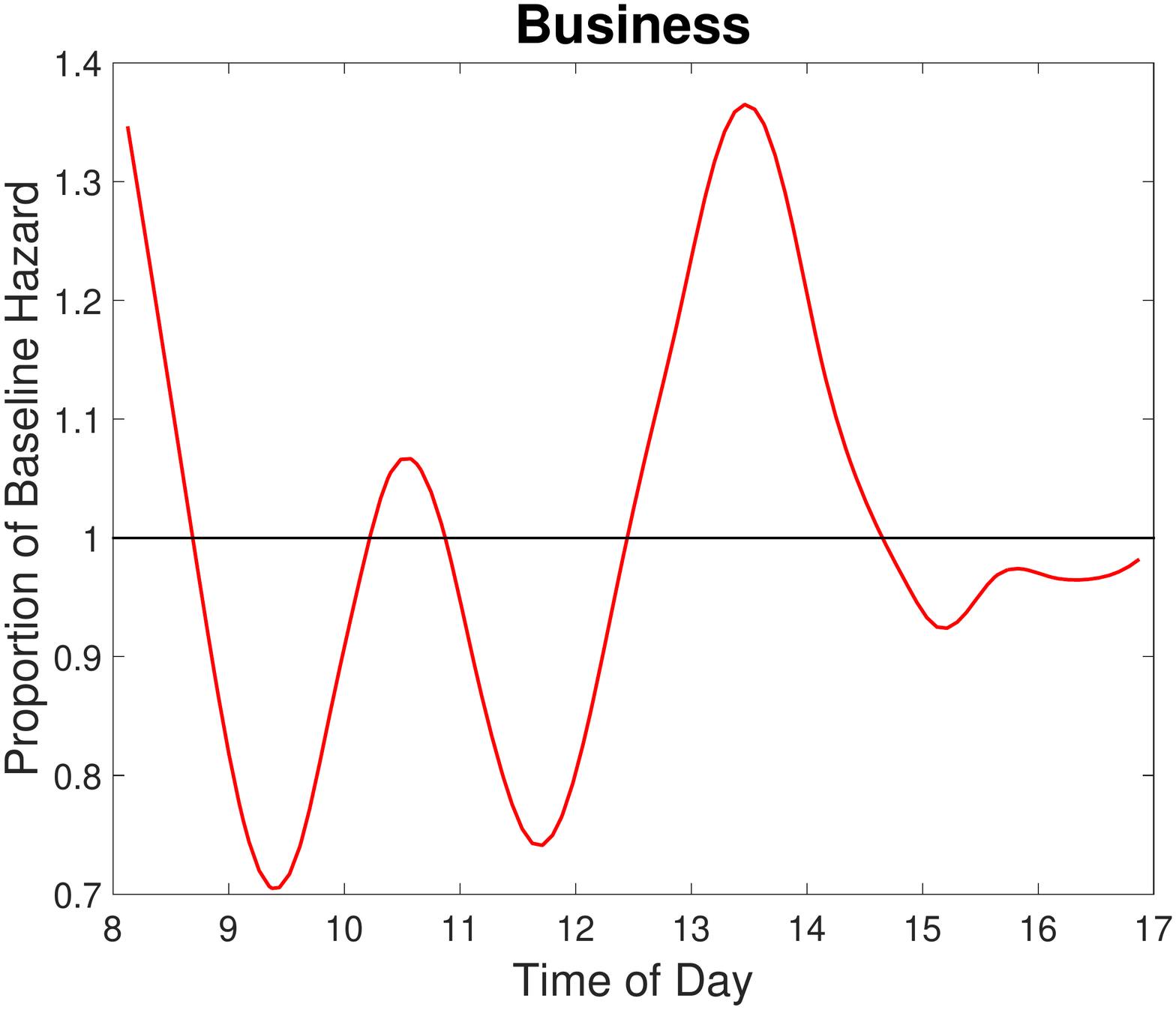}
\end{center}
\vskip-.1in
\caption{Time-of-day patterns of offered wait in different service groups.}
\label{fig:tod_w}
\end{figure}

}

\end{document}